\documentclass[aps,reprint,prx,longbibliography]{revtex4-2}

\usepackage{amssymb,amsfonts,amsmath}
\usepackage{subcaption}
\usepackage{graphicx}\usepackage{xcolor}
\usepackage{dcolumn}\usepackage{bm}\usepackage[normalem]{ulem}
\usepackage{comment} 
\usepackage{booktabs}
\graphicspath{{./}{../}}

\usepackage{float}
\restylefloat{table}

\newcommand{\lstad}{\ell_{core}}
\newcommand{\Tstad}{T_{core}}
\newcommand{\ellcut}{\ell_{cut}}
\newcommand{\ellcore}{\ell_{core}}

\newcommand{\pd}{\partial}
\newcommand{\nn}{\nonumber}
\newcommand{\GR}{\mathcal{R}}
\newcommand{\GH}{\mathcal{H}}
\newcommand{\rhodyn}{\mathcal{\rho}_{dg}}
\newcommand{\Sg}{\mathcal{S}}
\newcommand{\hL}{\hat{L}}
\newcommand{\bJG}{{\bf J_{\cal R}}}
\newcommand{\bER}{{\bf E_{\cal R}}}
\newcommand{\bPG}{{\bf P_{\cal R}}}
\newcommand{\ER}{ E_{\cal R}}

\newcommand{\taudep}{\tau_{dep}}
\newcommand{\taugrav}{\tau_{vg}}
\newcommand{\ddiv}{\text{div}}
\newcommand{\BL}{{\tiny{\mbox{BL}}}}
\newcommand{\ann}{{per}}

\newcommand{\Phil}{\Phi^{(\ell)}}

\newcommand{\dis}{dis}
\newcommand{\srr}{\sigma_{rr}}
\newcommand{\sqq}{\sigma_{\theta\theta}}

\begin{document}

\title{How viscous bubbles collapse: \\ topological and symmetry-breaking instabilities
in curvature-driven hydrodynamics}

\author{Benny Davidovitch}
\affiliation{Physics Department, University of Massachusetts Amherst, Amherst MA 01003}

\author{Avraham Klein} 
\affiliation{Physics Department, Ariel University, Ariel 40700, Israel}

\begin{abstract}
  The duality between deformations 
  of elastic bodies and non-inertial flows in viscous liquids has been a guiding principle in decades of research. However, this duality is broken 
  when a spheroidal or other doubly-curved liquid film is suddenly forced out of 
  mechanical equilibrium, 
  as occurs {\emph{e.g.}} when the 
  pressure inside a liquid bubble drops  
  rapidly due to 
  rupture or controlled evacuation.  
  In such cases the film 
  may evolve through a non-inertial yet geometrically-nonlinear surface dynamics, which has remained largely unexplored.
  We reveal the driver of such dynamics as 
  temporal variations in the 
  curvature of 
  the evolving surface. 
  Focusing on the prototypical example of a floating 
  bubble that undergoes  
rapid depressurization, we show that the bubble surface 
  evolves {\emph{via}} a topological instability and a subsequent front propagation, whereby a small planar zone nucleates and expands in the spherically-shaped film, bringing about hoop 
  compression 
  and triggering another, symmetry-breaking instability and radial wrinkles that grow in amplitude and invade the flattening film. 
  Our analysis reveals the dynamics as 
  a non-equilibrium branch of ``Jellium'' physics, whereby a 
  rate-of-change of surface curvature in a viscous film is akin to charge 
  in an electrostatic medium that comprises polarizable and conducting domains.   
  We explain key features underlying recent experiments and highlight a qualitative inconsistency between the prediction of linear stability analysis and the observed ``wavelength'' of surface wrinkles.
  Our analysis points to the existence of a nonlinear curvature-driven
  mechanism for pattern selection in viscous flows.
\end{abstract}

\maketitle

Whether it is a molten glass blown or drawn, a floating lava balloon, or a constituent of draining foam, 
the thermodynamic equilibrium
state of a
liquid {film} consists of tensile isotropic stress, whose value $2\gamma$, is twice 
the corresponding surface tension (force/length). Consequently,
the shape of the {film} is given by the famous Laplace law:
\begin{equation}
\Delta P = 4\gamma H  \ ,  \label{eq:Laplace-law}
\end{equation}
where $\Delta P$ is the excess pressure in the interior gas, 
and $H$ is the mean curvature of the film \cite{landau1987course}. 
Laplace's law~(\ref{eq:Laplace-law}) and its modification by viscous flow within a liquid film \cite{Trouton1906,MiddlemanBook1995,Howell1996} constitute an intriguing
mechanism for surface dynamics induced
by temporal variation of pressure or boundary conditions. 
However, the unique properties of such viscous film dynamics are often obscured by 
experimental and conceptual challenges.
A primary experimental hurdle is that surface dynamics are affected by external parameters, {\emph{e.g.}} puncturing of the film which generates small and rapidly expanding holes \cite{Debregeas1998}, as well as gravitational effects \cite{da2000rippling}. 
At the conceptual level, despite the profound similarity between Newtonian viscous dynamics and Hookean solid elasticity,
often termed the Stokes-Rayleigh analogy (SRA) \cite{Slim2012},
there exist also pivotal differences that complicate common attempts to draw   
parallels between the well-studied deformations of elastic sheets at mechanical equilibrium 
and a quasi-static shape evolution of viscous films. These differences stem from the existence of a preferred (``target'') 
metric in elastic sheets \cite{Efrati2009} and its absence in liquid films, yielding distinct impacts of surface tension \cite{Kumar2020} and Gaussian curvature \cite{Howell1996} on the respective stress tensors.

\begin{figure*}[t!] 
  \centering
  \begin{subfigure}{0.4\hsize}
    \includegraphics[width=\hsize,clip,trim=0 20 0 0]{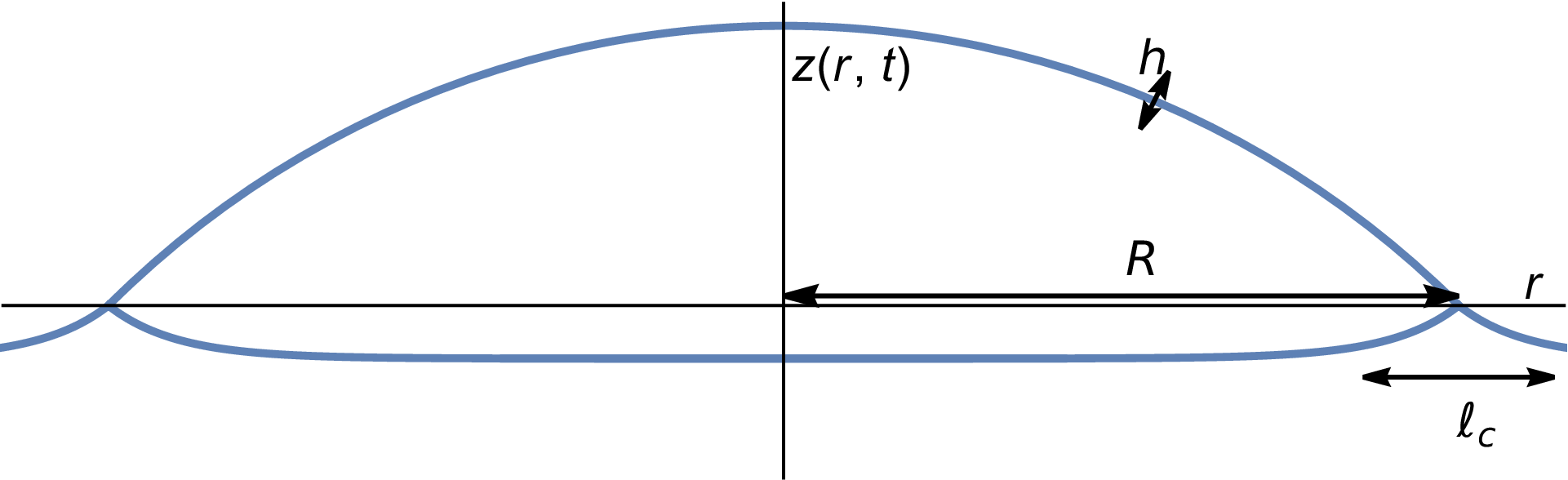}
    \caption{}
  \end{subfigure}
  \begin{subfigure}{0.43\hsize}
    \includegraphics[width=\hsize]{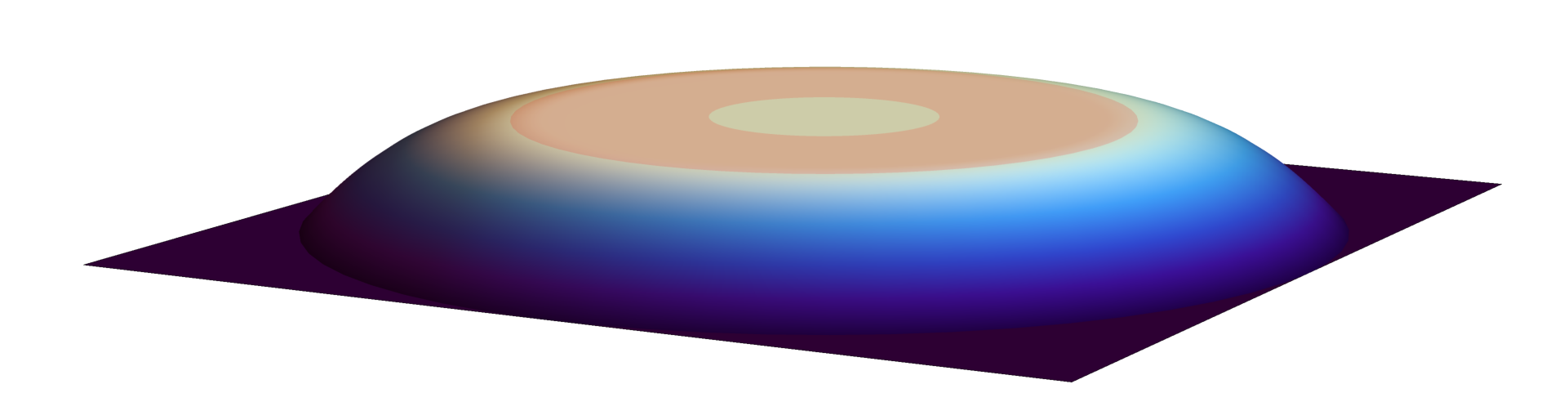}
    \caption{}
  \end{subfigure}
  \begin{subfigure}{0.12\hsize}
    \includegraphics[width=\hsize,clip,trim=20 20 20 20]{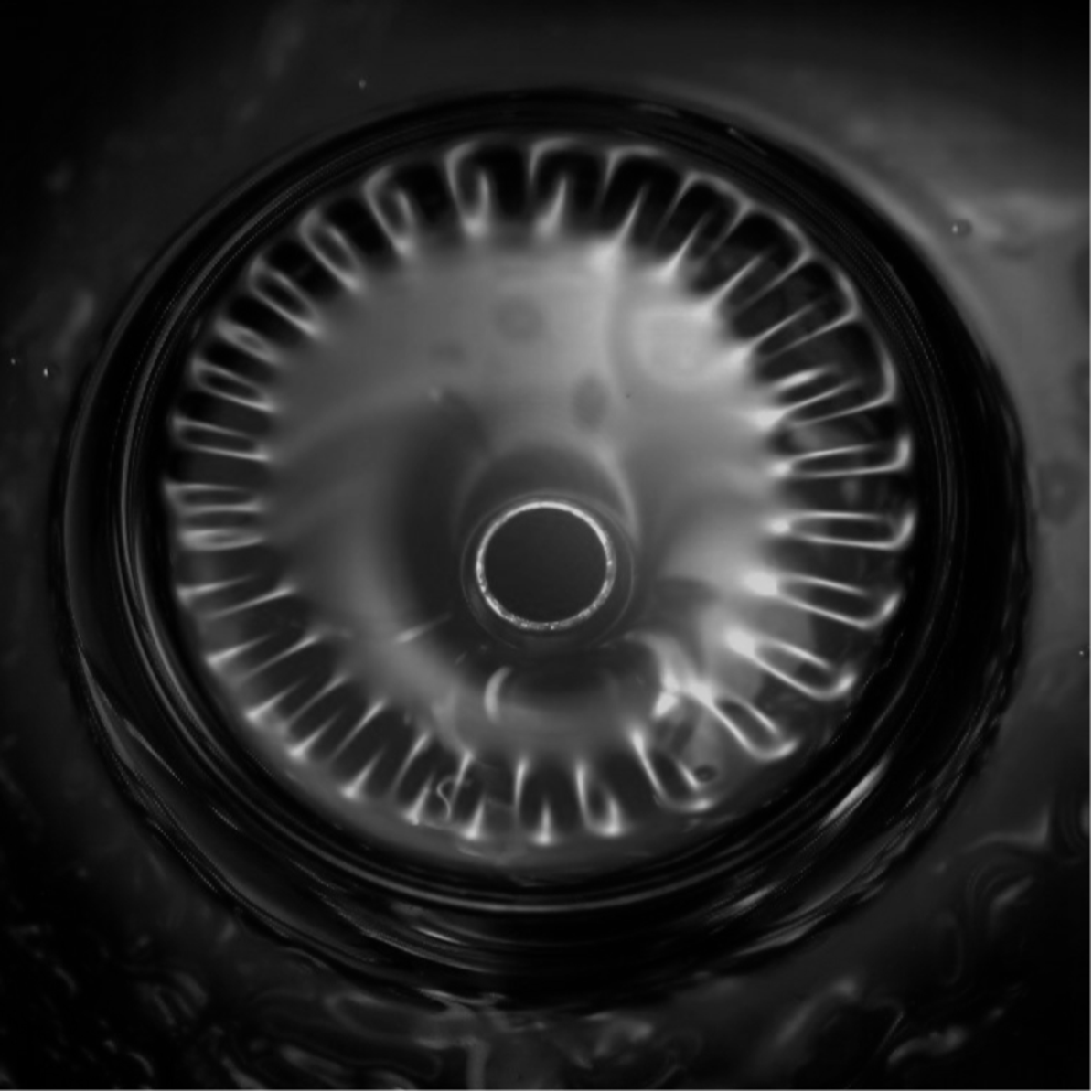}
    \caption{}
  \end{subfigure}
          \caption{Wrinkling instability of a collapsing bubble. (a) An illustration of the experimental setup. Depicted are a hemispherical bubble of radius $R_0 \gg \ell_c$ and thickness $h \ll \ell_c$, where $\ell_c$ is the (gravity)-capillary length that characterizes both radial and vertical size of a meniscus 
    between the bubble and the liquid bath.  
    (b) A depiction of our numerical solution of the bubble evolution, showing a moving front separating a central flat core from a curved periphery. The bubble dynamics are entirely dictated by the motion of the front. The red shaded area highlights a region with negative hoop stress, which triggers the wrinkling instability (c) A snapshot of an experimental realization of the collapsing bubble (top view), similar to what was presented in Ref.~\cite{Oratis2020}. (Courtesy of A.T. Oratis and J.C. Bird)     .   
        }
    \label{fig:expSetup}
\end{figure*}

A recent experiment by Oratis {\emph{et al.}} \cite{Oratis2020}
offers a peephole into the
viscous dynamics of curved liquid films. Revisiting the rapid depressurization of floating bubbles \cite{Debregeas1998}, 
the
authors demonstrated that the emergence of highly ordered pattern of radial wrinkles, which 
invade the flattening liquid film (see Fig.~\ref{fig:expSetup}c), {is almost unaffected}
by the presence of an expanding hole or gravitational field. This discovery reveals the surface dynamics 
as being governed solely by an interplay of viscous and capillary forces,
in contradiction with previous proposals \cite{da2000rippling}.
On the other hand,
although SRA appears 
as a natural 
scenario,  
prompted by the   
seeming resemblance to wrinkling instabilities in azimuthally-compressed elastic sheets \cite{Huang2007,King2012,Paulsen2016,Box2019}
and corresponding proposals of scaling rules for the 
wrinkles' ``wavelength'' in terms of 
parameters of the liquid film \cite{Debregeas1998,da2000rippling,Oratis2020},
the dynamical origin of compression 
remains elusive.
In the absence of a theory that explains the 
emergence 
of a preferentially hoop (azimuthal) compression 
that 
\textit{exceeds} the stabilizing 
surface tension, 
the 
adequacy of such models remains dubious. 
Thus, from the perspective of pattern formation theory, a desired objective is a    
theoretical framework that 
addresses the origin and spatial extent of net azimuthal (hoop) compression 
in the evolving film, 
describes the corresponding axially-symmetric surface dynamics, and provides a quantitative basis for its stability analysis.  


In this paper, we show that SRA is generally valid only if the shape of a liquid film
evolves slowly in comparison to a visco-capillary rate, $\tau_{vc}^{-1} \!\sim\! \gamma /\eta h$, where $\eta,h$ are, respectively, the dynamic viscosity and film thickness. At faster (yet non-inertial) rates, the shape evolution of liquid films 
may not be captured by SRA, but rather 
by universal, \emph{curvature-driven} surface dynamics. Such a dynamics is imparted by
viscous resistance to temporal variation of the  
intrinsic (Gaussian) curvature, $\GR$, and   
extrinsic (mean) curvature, $\GH$, of the surface, and is expressed through a \textit{dynamo-geometric} ``charge'' density,   
$\rhodyn$, 
that is determined by their rates of change. 
This scalar field 
acts as source  
of visco-capillary stress tensor within the liquid film, much like the vector
fields that are  generated by electrostatic charge in vacuum or polaraziable media. 
We reveal the experimental observations of Ref.~\cite{Oratis2020} as a ramification of such a surface dynamics.
There, a rapid depressurization of a liquid film that is effectively pinned by 
its meniscus with the bath (Fig.~\ref{fig:expSetup}a),
gives rise to  
a radially moving front, $r = r_f(t)$, 
that localizes temporally-varying 
curvature, $\rhodyn (r_f) \neq 0$, between 
a flat core 
and a curved periphery (Fig.~\ref{fig:expSetup}b). 
Crucially, the front emerges
in tandem with
a point-like 
(disclination) charge 
at the center of the film, such that the vicso-capillary stress in the   
flattening film 
is governed by a 
\textit{neutral} 
dynamo-geometric charge density: 
\begin{equation}
\rhodyn (r,t) \sim  q \cdot [r^{-1} \delta(r) - r_f^{-1} \delta (r-r_f)] \ . 
\label{eq:Dyndensity}
\end{equation}  
This topological instability, in which 
a front-disclination pair emerges to govern the 
flattening of a 
spherically-shaped film, gives rise to a hoop-compressive stress 
and thereby a symmetry-breaking instability and consequent 
growth of radial wrinkles.

The above scenario is similar to the classic Thomson problem in 2D elasticity \cite{Thomson1904}, where crystalline defects emerge to ``screen'' an imposed geometric charge density $\propto \GR$ and suppress its associated stress \cite{Bowick2000,Bausch2003}. 
The profound link to electrostatics further extends the classic analogy between 
Wigner crystals, the Abrikosov lattice in type-II superconductors, and the   
2D elasticity of curved crystals \cite{NelsonBook2002}, to non-equilibrium 2D viscous hydrodynamics.
Nevertheless, in this non-equilibrium process 
the dynamo-geometric charges 
$q(t),-q(t)$ in ~\eqref{eq:Dyndensity},  
and their respective locations, $r=0$ and  $r_f(t)$, 
are 
determined by 
the first law of thermodynamics, 
through a balance of 
the rate of change of surface energy with the heat production in the viscous flow. 
More broadly, our theory 
reveals 
a geometrically-nonlinear branch of viscous two-dimensional (2D) hydrodynamics, governed by momentum conservation %
and the $1^{st}$ law of thermodynamics, which may be
applied also to 
2D strongly-correlated electronic liquids, such as the hydrodynamic flow in graphene
\cite{BandurinD.2016,Jesse2016,MollPhilipJ.2016}.


We commence by introducing the essential elements of our model, after which we address the axisymmetric surface dynamics
of the collapsing bubble and the consequent emergence of wrinkles. We discuss briefly experimental signatures of our predictions and 
conclude with a forward looking overview. To keep our exposition concise we delegate many technical details to \textit{Supplementary Information (SI)}.

\label{sec:setup-mainresults}

\section*{Model setup and equations of motion}
\label{subsec:setup}

 

Floating hemispherical bubbles of radius $R_0 \approx 4\gamma/\Delta P$ form when gas of pressure $\Delta P$ 
above ambient rises to the surface of a liquid bath of density $\rho$ and dynamic viscosity $\eta$, causing a thin film of thickness $h \ll \ell_C^2/R_0$ to bulge upward, where $\ell_C \equiv \sqrt{\gamma/\rho g} \ll R_0$ is the capillary length. Using the time scales, 
$\tau_{vc} \equiv \eta h/\gamma$ and $\tau_{iner} \!=\! \rho R_0^2\! /\!\eta$, at which a viscous stress generated by a flow in the film would 
balance, respectively,  
surface tension and Bernouli pressure, we define 
the dimensionless thickness, Bond number, and Ohnesorge number: 
\begin{equation}
  \epsilon \equiv h/R \ \ ; \ \ Bo \equiv (R_0/\ell_c)^2 \ , \ 
  Oh \equiv 
  \sqrt{\tau_{vc}/
  \tau_{iner}}
  \label{eq:non-dim-1}
\end{equation}
and address the parameter
regime (see Table 1):
\begin{equation}
    Oh^{-1} \ll \epsilon \ll Bo^{-1}  \ll 1 \ .
    \label{eq:regime-1}
\end{equation}
In this parameter regime 
gravity affects   
the surface shape only at the meniscus (see Fig.~\ref{fig:expSetup}), and its dynamic effect is merely a 
slow drainage 
over a characteristic time $\taugrav  \!\sim \!\eta/ \rho g R_0 \gg \tau_{vc}$. The negligibility of inertia and gravity 
means that the characteristic driving force/length and fluid velocity induced by suddenly perturbing this steady state are, respectively, $\gamma$ and $R_0/\tau_{vc}$. Furthermore, the force/length required to move the meniscus at such a 
velocity is $\sim Bo \cdot \gamma \gg \gamma$, hence it is safe to assume that the film remains effectively ``pinned'' to the bath at $r\approx R_0$. In our study, such a capillary-driven Stokes-type hydrodynamics is generated by rapidly depressurizing the gas, \textit{e.g.} $\Delta P(t) = \frac{4\gamma}{R_0} e^{-t/\taudep}$, where
\begin{equation}
    T \equiv \taudep/\tau_{vc}  \ll 1 \ . 
    \label{eq:T}
\end{equation}
As was shown in 
Ref.~\cite{Oratis2020}, this can be achieved by a controlled evacuation of the gas, without rupturing the bubble.    

\begin{table}
\label{tab:1}
\begin{center}
\begin{tabular}{|c|c|c|c|c|}
    \toprule
        $\mbox{}$ & $\taugrav$ &  $\tau_{vc}$ & $\taudep$ &  $\tau_{iner}$ \\ \midrule 
         \text{definition} & $\eta/\rho gR_0$ & $\eta h/\gamma$ &  $T \tau_{vc}$ &  $\rho R_0^2/\eta$  \\
         \bottomrule 
    \end{tabular}
\end{center}

\caption{Hierarchy of time scales. Characteristic times of drainage ($\taugrav$), visco-capillary dynamics ($\tau_{vc}$), depressurization rate ($\taudep$), and inertia ($\tau_{iner}$), expressed    
   in term of dynamic viscosity ($\eta$), mass density ($\rho$), surface tension ($\gamma$), 
    film's thickness ($h$) and radius ($R_0$), and gravitational acceleration ($g$). The order (longest time to the left) corresponds to rapid depressurization ($T \!=\! \taudep/\tau_{vc} \!\ll \!1$). 
    The text addresses also 
    depressurization processes that occur adiabatically ($ 1\!\ll\! T \!\ll\! {\taugrav}/{\tau_{vc}}$) 
    or ``superfast'' ($T/\epsilon \!\ll \! 1$) 
    .   }
\end{table}

We employ
a
Trouton method to study the hydrodynamics of thin, volumetrically-incompressible liquid films with free surfaces \cite{MiddlemanBook1995}. Specializing to dynamics that conserves the axial symmetry of the bubble, the mid-surface is given
by $\vec{X}(r,\theta,t) \!=\! r \hat{r} \!+\! z(r,t) \hat{z}$, the film thickness is $h(r,t)$, and the 
thickness-averaged components of the fluid velocity are $v_r(r,t)$ and $\partial_t z$. Following 
Howell \cite{Howell1996}, 
we employ the small-slope approximation, 
such that the  \textit{thickness-integrated} in-plane components of the stress tensor may be written via an Airy stress function $\Phi$, i.e.
\begin{subequations}
\label{eq:stress-strain-rate}
\begin{eqnarray}
\sigma_{rr} &=& r^{-1} \partial_r\Phi \approx 2\gamma + 
2\eta [h (2\dot{\varepsilon}_{rr} + \dot{\varepsilon}_{\theta\theta}) + \partial_t h]  \label{eq:stress-strain-rate-r} \\
\sigma_{\theta\theta} &=& \partial_{rr}\Phi \approx 2\gamma + 
2\eta [h (\dot{\varepsilon}_{rr} + 2\dot{\varepsilon}_{\theta\theta}) + \partial_th] \ .
    \label{eq:stress-strain-rate-theta}
\end{eqnarray}

Here, 
\begin{equation}
     \ \ \ \ \dot{\varepsilon}_{rr} \approx \partial_r v_r + \tfrac{1}{2}\partial_t (\pd_r z)^2  \ ; \ 
\dot{\varepsilon}_{\theta\theta} \approx v_r/r
\label{eq:strain-disp}
\end{equation}
\end{subequations}
are the in-plane strain rate components.
Eqs.~\eqref{eq:stress-strain-rate}
feature 
the inherent geometric nonlinearity induced by rotational invariance of an in-plane tensor on a curved surface (in Eq.~\eqref{eq:strain-disp}), 
as well as thermodynamic and flow-dependent contributions to 
the thickness-integrated stress due to surface tension and 
viscous forces, respectively (Eqs.~(\ref{eq:stress-strain-rate-r},\ref{eq:stress-strain-rate-theta})). 
Assuming a nearly uniform thickness,  
\textit{i.e.} $h(r,t) \approx
h(t)+ \delta h(r,t)$ with $|\delta h|/h \ll 1$,
it is possible to express 
normal and in-plane force balance in terms of two scalar fields, 
$\Phi(r,t)$ and $z(r,t)$. 
Normalizing $\Phi$ by $2\gamma R_0^2$, 
$r$ and $z$ by $R_0$, and $t$ by $\alpha^2 \tau_{vc}$, where $\alpha$ is an arbitrary small parameter that characterizes the average initial slope $\pd_r z$ at $t\!=\!0$, and introducing a dimensionless parameter $\epsilon = (R_0\alpha)^{-1}h$, these two equations are (see SI):  
\begin{subequations}
\label{eq:Howell-FvK-sym}
\begin{align}
    \frac{1}{r}\frac{\partial}{\partial r}(\frac{\partial \Phi}{\partial r}\frac{\pd z}{\pd r}) &= -\Delta P(t) 
    + \epsilon^2 \hat{L}_r^2 \frac{\pd z}{\pd t}
        \label{eq:Howell-FvK-sym-1} \ \\
    {\frac{1}{3}}\nabla^4 \Phi &=
        \frac{1}{3}\hL_r^2\Phi=
        \rhodyn 
            \label{eq:Howell-FvK-sym-2} 
\end{align}  
\end{subequations}
where  
$\hL_r \!=\! 
r^{-1}\pd_r(r{\pd_r})$, and the dynamo-geometric charge density is a sum, 
\mbox{$\rhodyn \!=\!  \rhodyn^{(int)} \!+\!  \rhodyn^{(ext)}$,} with 
        \begin{align}   \rhodyn^{(int)}  = -\frac{\partial \GR}{\partial t}  \ \ , \ \ 
\rhodyn^{(ext)} = 
\frac{1}{3}\hL_r^2 (\GH \partial_t z) 
    \label{eq:Howell-FvK-sym-2b} 
    \end{align}  
Here, $\GR$ and $\GH$ are the intrinsic (Gaussian) and extrinsic (mean) curvatures of the mid-surface: 
$\GH \approx \tfrac{1}{2}\hat{L}_r^2 z$ and  
$\GR \approx \tfrac{1}{2r}\partial_r(\partial_r z)^2$, up to corrections of $O(|\nabla z|^2)$. We readily verify that the static (pressurized) state of the bubble, where the in-plane stress is just the 
surface tension ($\srr\!=\!\sqq\! =\!1$ in Eqs.~\ref{eq:stress-strain-rate}), is a trivial solution of Eqs.~(\ref{eq:Howell-FvK-sym}), reducing to the Laplace law, Eq.~
(\ref{eq:Laplace-law}). 
The appearance of 
$\pd_t\GR$ in $\rhodyn$
results from the geometric nonlinearity in Eq.~\eqref{eq:stress-strain-rate}, reflecting 
a source of stress due to areal shearing, whereas the appearance of $\GH$ in $\rhodyn$ results from the term $\partial_t h$ in Eq.~\eqref{eq:stress-strain-rate} and reflects a source of stress due to areal contraction/expansion 
\footnote{The original form of Eqs.~(\ref{eq:Howell-FvK-sym}) in Ref.~\cite{Howell1996} ignores the contribution $\rhodyn^{(ext)}$ in~\eqref{eq:Howell-FvK-sym-2}, whereas the version in Ref.~\cite{Mahadevan2010} is valid only for small perturbations of a hemispherical shape.}. Indeed, although the fluid is volumetrically-incompressible, areal compressibility, 
$\partial_r(r hv_r) \neq 0$, is enabled by temporal variation of the 
thickness, 
governed by the continuity (mass conservation) equation \cite{Stone1990}:
\begin{gather}
\pd_t h \approx  -r^{-1} \partial_r (r h v_r) - 2 h  \GH \partial_t z\ , 
\label{eq:continuity-00}
\end{gather}
Thus, the thickness variation implied by volumetric incompressibility is encoded in the term $\rhodyn^{(ext)}$,  
revealing that 
the only \textit{explicit} dependence of the
mid-surface dynamics on the film's thickness 
is in 
the effect of $\pd_t h$ on  
the thickness-integrated stress. 
In \textit{SI} we derive Eqs.~(\ref{eq:Howell-FvK-sym},\ref{eq:continuity-00}) in their general,  
non-axisymmetric version, 
and elaborate 
on the small-slope approximation and 
nearly-uniform thickness assumption.

%
%
%
%
%
%
%
Before moving to solve Eqs.~\eqref{eq:Howell-FvK-sym}, we note their formal 
similarity to the 
{F\"oppl-von K\'arm\'an} (FvK) 
equations of an elastic sheet with thickness $h$ and Young's modulus $E$.
This observation motivates us to employ the ``membrane theory'' approach of elasticity theory to analyze the axisymmetric dynamics,   Eq.~(\ref{eq:Howell-FvK-sym}), for $\epsilon\!=\!0$. Our numerical solution, to be discussed below, confirm that the omitted term is significant only at boundary layers describable by singular perturbation theory. More profoundly, 
the similarity to FvK equations reflects the SRA 
mapping 
between 
Newtonian hydrodynamics and Hookean elasticity,    \mbox{$\eta\!\cdot\! \partial_t \varepsilon_{ij} \  \leftrightarrow \ E \!\cdot\! \varepsilon_{ij}$}, which provides, in the absence of Gaussian curvature, a powerful tool to study 
viscous films and filaments   \cite{Buckmaster1975,ribe2001,LeMerrer2012, Slim2012,Pfingstag2011,Srinivasan2017,Okiely2019}. 
Nevertheless, the geometrical nonlinearity in Eq.~(\ref{eq:strain-disp}), which gives rise to  $\rhodyn$ in Eq.~(\ref{eq:Howell-FvK-sym-2}) \textit{versus} $-{\cal R}$ in the analogous FvK equation, violates SRA, rendering Eqs.~\ref{eq:Howell-FvK-sym} a genuine nonequilibrium dynamics.




\section*{Nucleation of a disclination-front pair}

\begin{figure*}   
    \begin{subfigure}[b]{0.2\hsize}
        \includegraphics[width=\hsize]{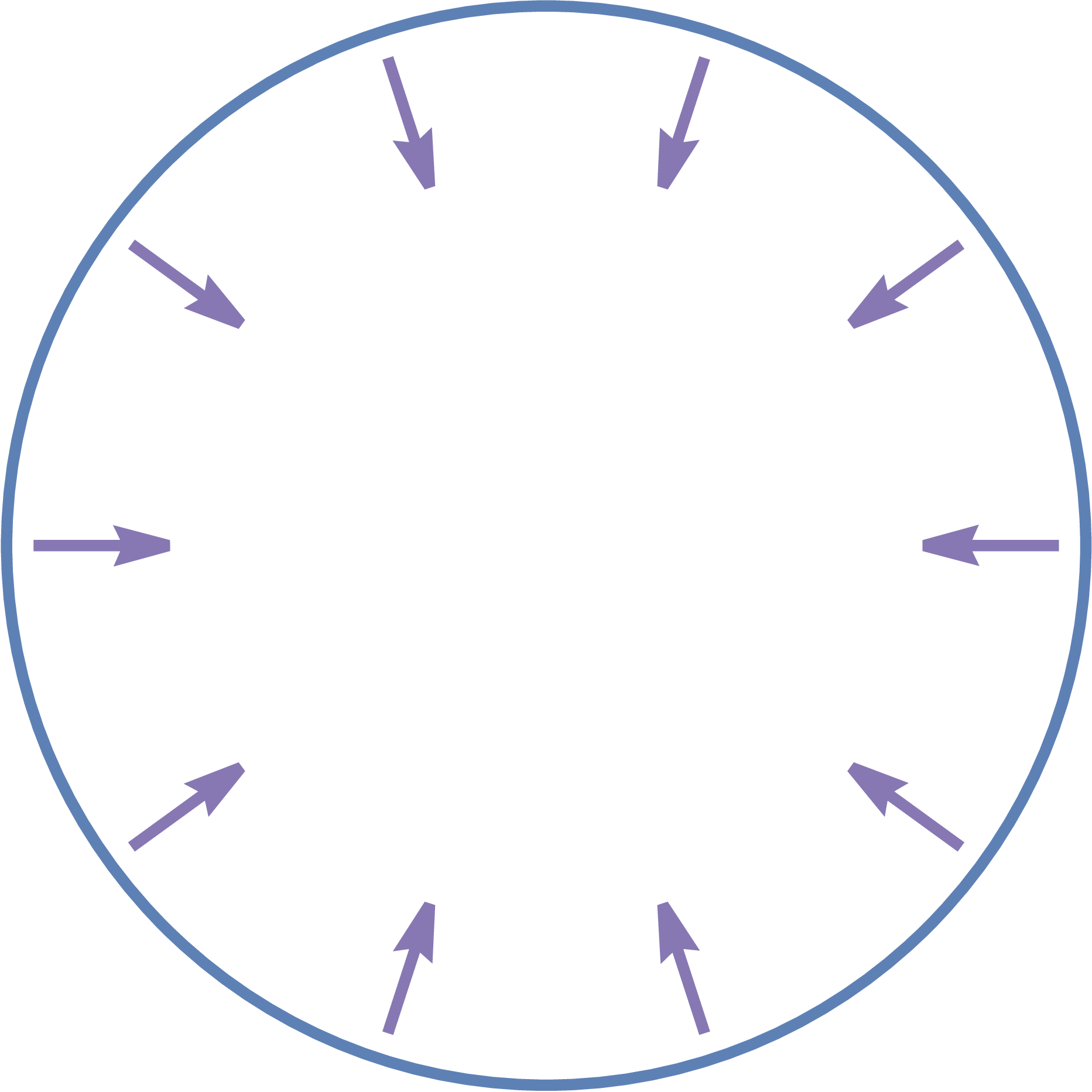}
        \caption{}
    \end{subfigure}\hfill
    \begin{subfigure}[b]{0.35\hsize}
        \includegraphics[width=\hsize]{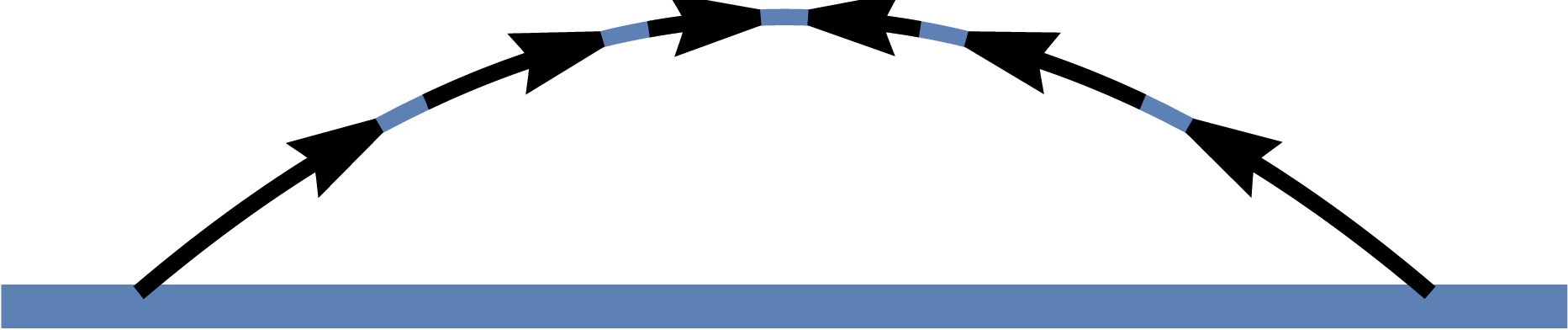}
        \vspace{0.2\hsize}
        \caption{}
    \end{subfigure}\hfill
    \begin{subfigure}[b]{0.35\hsize}
        \includegraphics[width=\hsize]{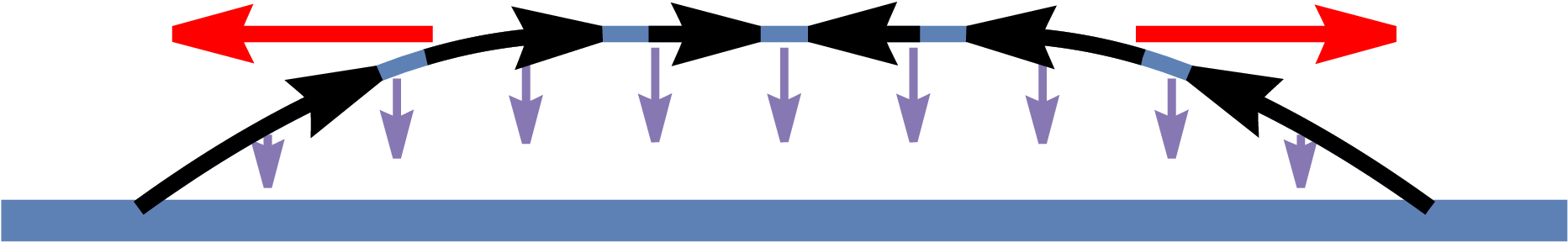}
        \vspace{0.2\hsize}
        \caption{}
    \end{subfigure}
    \caption{Schematic depiction of three possible solutions of 
    Eqs.~\eqref{eq:Howell-FvK-sym}. 
    (a) A spherically collapsing bubble (purple arrows) with no in-plane flow. Here the surface tension is compensated for by the viscous stress of a purely radial flow (thickening of the bubble). This solution is forbidden in the presence of a meniscus, which fixes the bubble's rim.  
    (b) A ``phantom bubble'' which does not collapse at all, and where tangential, in-plane flow (black arrows) creates viscous stress to compensate for surface tension. This exact solution is forbidden by thermodynamics, Eq.~\eqref{eq:1Thermo}. (c) The only physically-realizable  axisymmetric solution  -- a front separating a domain of flat surface from a curved surface. The front propagates outwards (red arrows), whereby the film experiences both in-plane flow and collapse \label{fig:schemSolutions}}
\end{figure*}

We proceed to analyze Eqs. (\ref{eq:Howell-FvK-sym}) with $\epsilon\!=\!0$, for an axisymmetric bubble evolution, where $\Delta P = 2e^{-t/T}$ (or other function that decays in time $T \!\ll \!1$), subject to initial conditions, where $z(r,t=0)$ is paraboloid (\textit{i.e.} hemi-sphere up to $O(\partial_rz)^2$), under isotropic uniform surface tension:
\begin{equation}
t=0: \ \  \ \ \ 
z \!= \!1\! -\! \frac{1}{2}r^2 \ \  \ , \ \ \  
\pd_r\Phi \!= \!r  
\label{eq:IC0}
\end{equation}
%
The boundary conditions (BCs) to Eqs. (\ref{eq:Howell-FvK-sym}) require some discussion. These PDEs for $z$ and $\Phi$ 
are $4^{th}$ order in $r$, requiring eight BCs. Six 
BCs are straightforward, arising from regularity and symmetry considerations (see SI):
\begin{align}
    &r\to 0: &\pd_r z \to 0 \ , \ \pd_r^3 z \to 0\ , \ \pd_r \Phi \to 0\ ,
    \label{eq:BC-hom-1}\\
    &r \to 1: &\Phi \to 0 \ , \ \pd_r^3 \Phi \to 0\ , \ z \to 0\ . \label{eq:BC-hom-2}
\end{align}
The next homogeneous BC results from the immobility of the meniscus discussed above, enforcing preservation of its shape at $r\approx 1$ while the rest of the film evolves (see SI):   
\begin{equation}
r \to 1:  \ \ \ \   
   \pd_t\pd_r^2 z 
   \to 0  . 
 \label{eq:BC-z}
\end{equation}
To understand the implication of Eq. \eqref{eq:BC-z} it is useful to integrate the normal force balance   Eq. \eqref{eq:Howell-FvK-sym-1}, 
\begin{equation}
    {\partial_r \Phi}\cdot \pd_r z = - \tfrac{r^2}{2} \Delta P(t)  \xrightarrow{t \gg T} 0.  
    \label{eq:Howell-FvK-sym-3aa} 
\end{equation}
Eq.~\eqref{eq:Howell-FvK-sym-3aa} implies that after the pressure drops, any portion of the film must be either completely flat, $\partial_rz \!=\!0$, or completely stress-free, $\partial_r\Phi\!=\!0$. 
One very simple solution of Eqs.~\eqref{eq:Howell-FvK-sym}
that obeys this condition is just a shrinking paraboloid, $z(r,t) = z_0(t)- r^2/2R(t)$, 
which is in fact nothing but the small-slope approximation of a uniformly shrinking sphere, see SI \cite{VanderFliert1995,HowellPrivate}.
In this solution (see Fig.~\ref{fig:schemSolutions}a), the film remains curved everywhere (and its curvature increases), in such a way that the    
charge densities, $\rhodyn^{(int)}$ and $\rhodyn^{(ext)}$ mutually cancel. 
Crucially, this solution is forbidden in our system due to the immobile meniscus which clamps the film boundary to $r=1$. 
By preventing the uniformly-shrinking paraboloid dynamics, 
the BC (\ref{eq:BC-z}) affects an imbalance of $\rho^{(int)}$ and $\rho^{(ext)}$. Indeed, we will show later that the dominant source of stress becomes the areal shearing, \textit{i.e.} $\rhodyn^{(int)}$, rather than areal contraction. 

The final, and only nonhomogenous BC, derives from a global constraint, imposed by the first law of thermodynamics:  
\begin{equation}
    P_{vis} = - \dot{E}_{surf}
    \label{eq:1Thermo}
\end{equation}
where $P_{vis}$ is the dissipation rate of kinetic energy in the viscous flow, and $E_{surf}$ is the surface energy, which is proportional to the film's area and its reduction drives the dynamics. In SI we give $P_{vis}$ and $\dot{E}_{surf}$ as area integrals of expressions that involve $\Phi(r,t)$ and $z(r,t)$.        

The importance of Eq. \eqref{eq:1Thermo} 
becomes evident by noticing an 
\emph{exact} yet \textit{non-physical} solution to Eqs. (\ref{eq:Howell-FvK-sym}), namely,  
$\pd_r\Phi = \tfrac{1}{2}r \Delta P(t), z = 1-\tfrac{1}{2}r^2$.   
This solution, which we dub ``phantom bubble'', 
describes a depressurized, yet immobile bubble, that 
retains the hemispherical shape of its mid-surface  
by sucking liquid from the bath through an inward radial velocity, $v_r \propto -r$. In such a state, 
$\dot{\varepsilon}_{rr},\dot{\varepsilon}_{\theta\theta}$, and $\partial_th$, 
given by  Eqs.~(\ref{eq:strain-disp},\ref{eq:continuity-00}), generate viscous stress in Eqs.~(\ref{eq:stress-strain-rate-r},\ref{eq:stress-strain-rate-theta})  
that counteracts surface tension, and yields the stress-free state necessary for normal force balance on a curved film, Eq.~(\ref{eq:Howell-FvK-sym-3aa}). 
Thus, this 
solution satisfies 
the hydrodynamic equations~(\ref{eq:Howell-FvK-sym}) and
BCs~(\ref{eq:BC-hom-1},\ref{eq:BC-hom-2},\ref{eq:BC-z}), but violates 
the thermodynamic constraint, Eq. \eqref{eq:1Thermo}, since it is characterized by  
$\dot{E}_{surf} = 0$ and  viscous flow that implies $P_{vis} >0$. Since this thermodynamically-inconsistent solution would necessarily emerge by
any set of completely homogeneous BCs, compatibility with Eq. \eqref{eq:1Thermo} requires
a \textit{non-homogeneous} BC, which we can  conveniently express as:
\begin{equation} 
  r \to 0:  \ \ \ \ \ \ 
     r^{-1}\pd_r \Phi - \pd_{rr} \Phi = q(t) \ , 
  \label{eq:BC-phi-2}
\end{equation}
such that $q\!=\!0$ at $t\!=\!0$, as implied by~Eq. \eqref{eq:IC0}, and at 
$t\!>\!0$ $q(t)$ is determined by requiring the solution of~Eq. \eqref{eq:Howell-FvK-sym} 
subject to  BCs~(\ref{eq:BC-hom-1},\ref{eq:BC-hom-2},\ref{eq:BC-z}) to satisfy the integral constraint, 
Eq. \eqref{eq:1Thermo}.

Notably, $q(t)$ in~Eq. \eqref{eq:BC-phi-2} is readily recognized as a ``point charge'' density,  
 $q(t) r^{-1} \delta(r) $, 
 added to $\rhodyn^{(int)}$ and $\rhodyn^{(ext)}$ in Eqs.~(\ref{eq:Howell-FvK-sym-2}). 
The elastic counterpart in FvK equations, called ``disclination'', is realized  
\textit{e.g.} 
by inserting an azimuthal sector of angle $-q/2\pi$ into a solid disk     
 \cite{LandauBook1986}. 
 For a film of volumetrically-incompressible liquid, an analogous effect 
 is associated with 
 temporal variations of the film's thickness, 
 governed by~Eq. \eqref{eq:continuity-00}.  
In analogy to the elastic case, the logarithmic divergence, $\Phi\sim r^2 \log r$, implied by the thermodynamically-consistent disclination,
 Eq.~\eqref{eq:BC-phi-2}, is regularized below a 
 cut-off, $\ellcut \!\sim\! \epsilon$ (where Trouton approximation 
 must be replaced by Stokes hydrodynamics in 3D), but whose presence does not affect the physics at $r \gg \ellcut$. 
 For numerical analysis, it is useful to solve Eqs.~(\ref{eq:Howell-FvK-sym}) in the interval $r \in [\ellcut,1]$, and replace~Eq. \eqref{eq:BC-phi-2} by  
$\pd_r \Phi = \ellcut \cdot [1+ q(t) \log \ellcut ]$ at $r\to \ellcut$. 

At 
$t \!\gg\! T$ the normal force balance Eq. \eqref{eq:Howell-FvK-sym-3aa} requires
any curved portion of the film to be stress-free, 
\textit{i.e.} 
$\partial_r\Phi \!= \!0$. However, 
Eq. \eqref{eq:BC-phi-2} with \mbox{$q(t)\neq 0$}, implies 
$\pd_r\Phi \neq 0$ near $r=0$. 
This
conflict is naturally 
resolved by 
a \textit{front-like}
dynamics, 
separating flattened-stressed and curved-unstressed regions: 
\begin{subequations}
  \label{eq:front-def}
\begin{gather}
  \Phi  \approx \Theta\left(r_f(t) - r\right)\Phi_{\dis} (r,t),   \\
    z  \approx z_f(t) + \Theta \left(r - r_f(t)\right)z_{\ann} (r,t) \ ,  
\end{gather}
\end{subequations}
where $\Theta(x)$ is the Heaviside function. The disclination potential, $\Phi_{\dis}$ and the 
peripheral shape, $z_{\ann}$ (both of which are smooth at $r \!\approx\! r_f(t)$), 
along with the location of the propagating front, $r\!=\!r_f(t), z\!=\!z_f(t)$, are discussed in the next section. Note that for $\dot{r}_f >0$ the surface area of the film is decreasing in time, such that the thermodynamic constraint~(\ref{eq:1Thermo}) can be satisfied for a suitable choice of $q(t)$.  
A numerical solution of Eqs.~(\ref{eq:Howell-FvK-sym},\ref{eq:BC-hom-1},\ref{eq:BC-hom-2},\ref{eq:BC-z},\ref{eq:BC-phi-2}) is shown in Fig.~\ref{fig:solutionPlots}a (see SI for numerical parameters). The numerical solution clearly shows the front dynamics, where the thickness parameter $\epsilon$ simply acts as a regularizer 
of the Heaviside functions of Eq. \eqref{eq:front-def}. More precisely, for $r \approx r_f(t)$: 
\begin{equation}
  z \approx z_f + 
  {z^*_{\ann}}
  z_{\BL}(\xi) \ \ 
 \ \   , \Phi \approx {\Phi^*_{\dis}}  \Phi_{\BL}(\xi)
  \label{eq:front-sim} 
\end{equation}
where $z^*_{\ann} = z_\ann(r_f,t) , \Phi^*_{\dis} = \Phi_{\dis}(r_f,t)$,  
$\xi = (r-r_f)/\ell_{\BL}$ and  $\ell_{\BL} \sim \sqrt{\epsilon}$. 
Here, $g_{\BL}(\xi,t)$ and  $\phi_{\BL}(-\xi,t)$ are scaling functions, which 
{approach $1$ for $\xi \to \infty$ and 0 for $\xi\to -\infty$.} 
%
%
In SI we give these functions and show that the rapid drop in gas pressure determines the initial location of the front, Eq.~(\ref{eq:front-def}):   
\begin{equation}
r_f(\Tstad) \equiv \lstad \sim \sqrt{\Tstad}
   \label{eq:ell-adi}
 \end{equation}
where $\Tstad \!\sim\! \max\{T,\epsilon\}$ such that both $\lstad,\Tstad\to 0$
 in the parameter regime of Eqs. \eqref{eq:regime-1} and \eqref{eq:T}. 
 
 \section*{Axisymmetric front propagation}
 We now turn to evaluate the curved shape, $z_{\ann}(r,t)$ at $r\!>\!r_f$, and stress function $\Phi_{\dis}(r,t)$ at the flat portion $r\!<\!r_f$, Eq. \eqref{eq:front-def}, along with 
the radial 
location of the front, $r_f(t)$. 
 
Inspecting 
Eq.~\eqref{eq:Howell-FvK-sym-2} we note 
that 
having a stress-free 
curved periphery requires not only 
vanishing of $\rhodyn(r,t)$ at $r>r_f(t)$, 
but also an overall dynamo-geometric charge neutrality,  
$2\pi \int_0^r r'dr' \rhodyn(r') \!=\! 0$, for any $r>r_f(t)$. This implies that the front dynamics~(\ref{eq:front-def}) is characterized by a 
charge distribution, Eq. \eqref{eq:Dyndensity}, such that the disclination charge is completely ``screened'' 
at the front, where the curvature changes abruptly. Thus, the formation of a front-disclination pair through rapid depressurization is 
understood as a \textit{topological instability}, akin to nucleating a charge-neutral excitation 
in electrostatic medium.  
The stress function in the flat part is readily found to be: 
\begin{equation}
\Phi_{\dis}(r,t) = 
\tfrac{1}{4}
q(t) r^2 \left(\log \frac{r}{r_f(t)} - \frac{1}{2}\right) 
    \label{eq:Phi-dis} \ ,
\end{equation}
such that the BC~\eqref{eq:BC-phi-2} is satisfied, and the radial stress, $\sigma_{rr} = r^{-1}\pd_r\Phi$ is continuous at the front, as implied by in-plane (radial) force balance. 

\begin{figure*}
   \begin{subfigure}[b]{0.45\hsize}
        \includegraphics[width=\hsize]{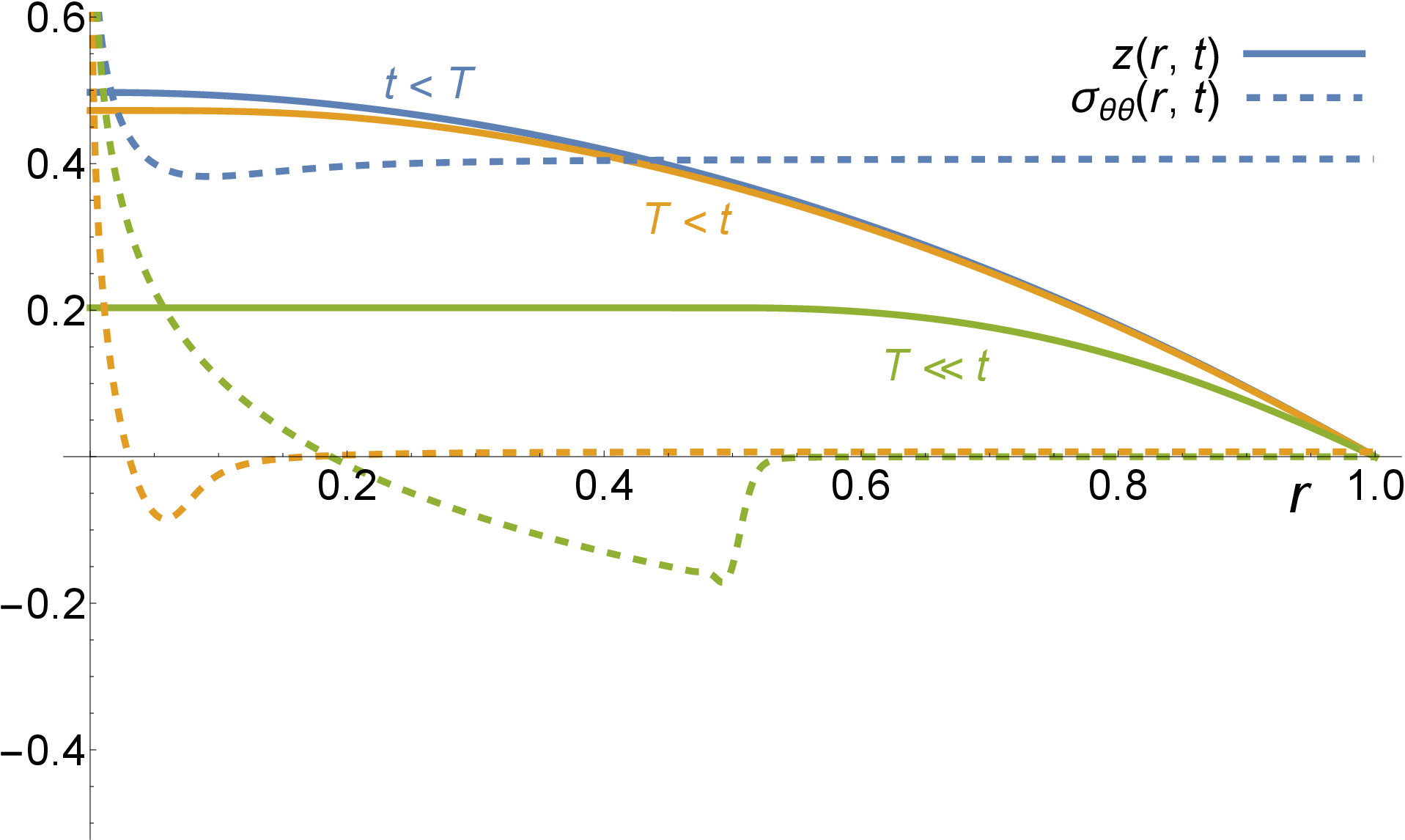}
        \caption{}
    \end{subfigure}\hfill
    \begin{subfigure}[b]{0.45\hsize}
        \includegraphics[width=\hsize]{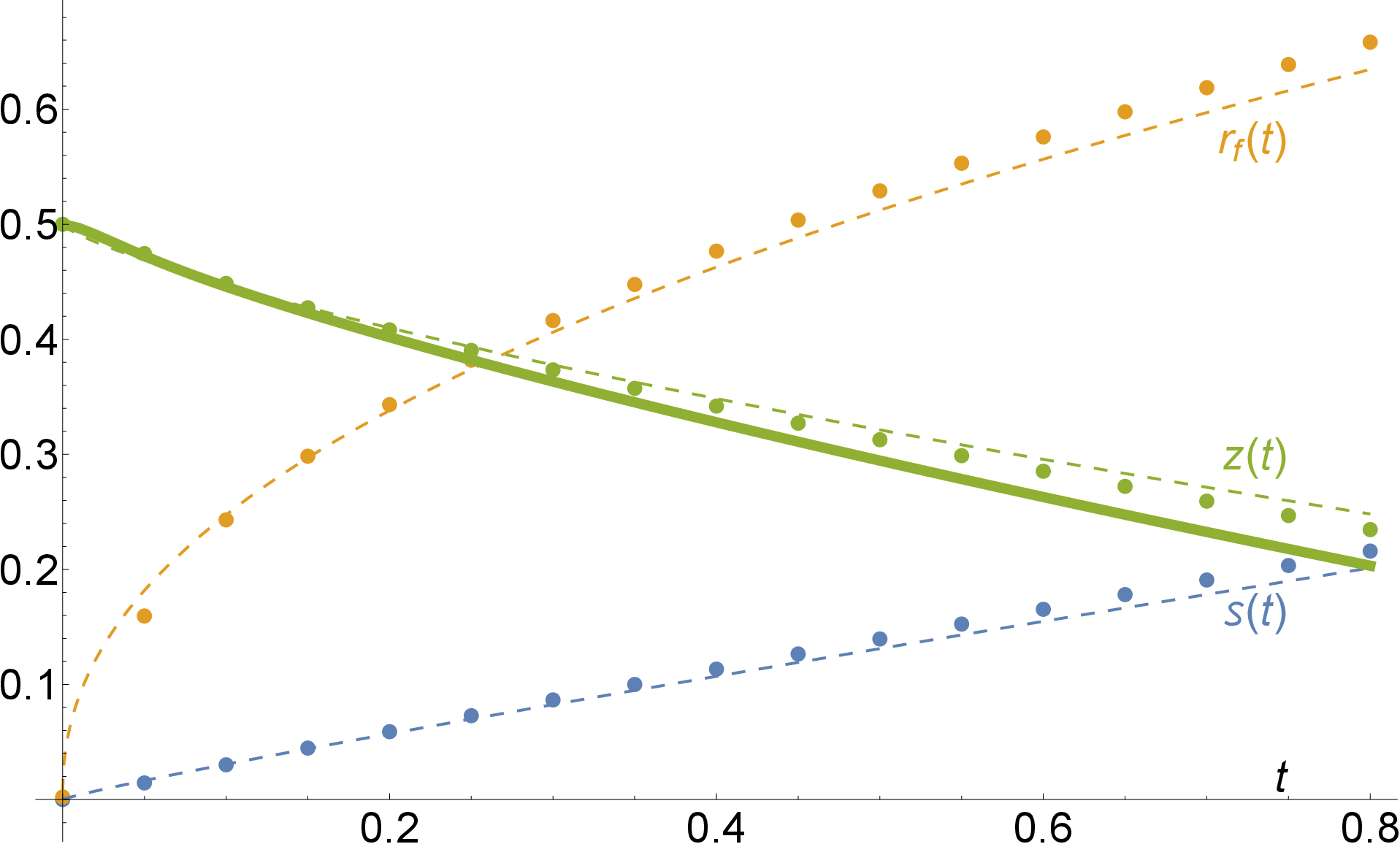}
        \caption{}
    \end{subfigure}
    \caption{Axisymmetric evolution of the surface dynamics. 
    (a) The profile of the bubble $z(r,t)$ and of the hoop stress $\sigma_{\theta\theta}(r,t) = \pd_{rr} \Phi$ at several times. As the bubble evolves after the decompression time $T$, a front forms, seen as a jump in the hoop stress, separating flattened vs. curved domains of the film. 
    The collapse of the bubble proceeds via the propagation of the front to the bubble rim. Directly behind the front, negative hoop stress (compression) drives the formation of radial wrinkles. (b) The evolution of the Seivert parameter $S(t)$, the front position $r_f(t)$ and the bubble apex position $z(r=0,t)= z_f(t)$. The dots and dashed lines represent the numerical and analytic solutions of the simplified \emph{intrinsic} model $\delta_{ext} = 0$ (see text), and show excellent agreement. The solid green line is the numerical solution of the full problem ($\delta_{ext} = 0$), showing that the simplified model captures the dynamics of the full model very well. See SI for numerical parameters. \label{fig:solutionPlots}}
\end{figure*} 
 The charge-neutral 
 nature of the front-disclination pair 
 points to a rather powerful analogy between in-plane force balance of a viscous film, Eq. \eqref{eq:Howell-FvK-sym-2}, and electrostatics in continuum media, which parallels a similar analogy between 2D elasticity and electrostatics \cite{Bowick2000,Irvine2010}.  
 Indeed, Eq. \eqref{eq:Howell-FvK-sym-2} can be recast as a conservation law, akin to Poisson equation: 
\begin{equation}
  \label{eq:elec-0}
 {\nabla\cdot \bER   = 
\rhodyn 
\ \  \Longleftrightarrow  \ \ \nabla\cdot \bJG = \rhodyn  
+  \nabla \cdot \bPG }
\end{equation}
where
\begin{equation}
  \label{eq:Er-def}
  \bER = \nabla V_\mathcal{R},~V_\mathcal{R} = \tfrac{1}{3} \nabla^2 \Phi =\tfrac{1}{3}{\rm Tr} \  {{\bf \sigma}}.  
\end{equation}
The ``polarization'' field $\bPG$ is divergenceless everywhere, $\nabla \cdot \bPG = 0$. 
%
%
and the ``curvature current'' $\bJG$ is analogous to 
Maxwell's electrostatic displacement, ${\bf D} = {\bf E} + {\bf P}$. Specifically, the curved-unstressed exterior, $r>r_f$, where $V_\mathcal{R}=0$ and $\rhodyn = 0$, is analogous to a conductor, while the planar-stressed interior part is analogous to 
an insulator, where a point charge at $r=0$ gives rise to potential 
$V_\mathcal{R} \sim \log (r)$. The front, 
$r\approx r_f$, is thus 
akin to a vacuum-conductor boundary, where charge must accumulate
to screen the point charge in the vacuum interior. In analogy to an electrostatic field, the potential gradient $\bER$ is discontinuous at the front, whereas the curvature current $\bJG$ remains continuous similarly to Maxwell's displacement.    
\\

Continuity of the curvature current 
implies $\bJG \!=\! \tfrac{1}{3}q(t) \hat{r}/r$ for any $0\!<\!r\!<\!1$, such that the polarization in the curved-unstressed ``conductor'', $r\!>\!r_f$, is $\bPG \!=\! \tfrac{1}{3}q(t)\hat{r}/r$, yielding:   
\begin{equation}
\tfrac{1}{3}q(t) =  
 -\frac{1}{2} \pd_t (\pd_rz_{\ann})^2  + \frac{r}{6} \pd_r [\pd_tz_{\ann} (\frac{1}{r} \pd_rz_{\ann} + \pd_{rr} z_{\ann}] \ . 
 \label{eq:pol-1}
\end{equation}
(Eq.~(\ref{eq:pol-1}) is obtained by integrating Eq.~\eqref{eq:Howell-FvK-sym-2} over a circle of radius $r\!>\!r_f$, using Eqs.~(\ref{eq:Dyndensity},\ref{eq:Howell-FvK-sym-2b})). For a given $q(t)$, 
Eq. \eqref{eq:pol-1} is a $3^{rd}$ order PDE in $r$ in a time-dependent interval $r\in (r_f(t), 1)$, whose advancement in time requires 3 BCs  
at $r=r_f$ and $r=1$, along with equation for advancing $r_f(t)$. 
Two of these 4 equations are obtained by specializing to the vicinity of $r\approx r_f(t)$.
A second integration of Eq. \eqref{eq:Howell-FvK-sym-2} across the front yields two 
equations 
at $r_f$. The first is  
continuity of the slope, 
and the second one relates discontinuities $\Phi_{dis}$ and 
$z_\ann$,
\begin{align}
    0 &=\pd_rz_{\ann}|_{r\to r_f},  
    \label{eq:BC-front-1}
    \\
0 &= [[\pd_{rr}\Phi]]_{r_f}+[[\frac{\dot{z}_f}{2} \pd_{rr}z]]_{r_f},
\label{eq:BC-front-2}
\end{align}
where the apical height $z_f$,  
defined in Eq. \eqref{eq:front-def}, 
is 
\begin{equation}
    z_f = -\int_{r_f}^1 \pd_r z_\ann(r',t) dr'.
\end{equation}
For any function $q(t) \!<\!0$ and initial conditions 
\mbox{$r_f(0) \!\approx\! 0$},  \mbox{$z_{\ann}(r,0) \!\approx\! 1\!-\!r^2/2$} (Eqs.~\ref{eq:ell-adi},\ref{eq:IC0}), we may use a
first-order 
integrator to compute 
$r_f(t)$ and the curved shape $z_{\ann}(r,t)$ in $r>r_f(t)$, by temporally integrating     
the nonlinear PDE (\ref{eq:pol-1}), along with Eqs.~(\ref{eq:BC-front-1},\ref{eq:BC-front-2}) at $r\!\!=\!\!r_f$ and $z_{\ann} \!=\! \pd_t\pd_{rr}z_{\ann}\!=\!0$ at $r\!\!=\!\!1$ (Eqs.~\ref{eq:BC-hom-2},\ref{eq:BC-z}). The physical $q(t)$ is then obtained by requiring Eq. \eqref{eq:1Thermo} to be satisfied at all times.    









While the prescription above fully solves the axisymmetric problem in principle, we can 
simplify the problem and obtain a valuable analytic approximation to the solution through 
an auxiliary version of Eqs. \eqref{eq:Howell-FvK-sym}, where the ``intrinsic'' and ``extrinsic'' contributions to $\rhodyn$ are explicitly separated out via an auxiliary variable $\delta_{ext}$,
\begin{equation}
    \rhodyn  \longrightarrow 
    \rhodyn^{(int)}  + \delta_{ext}  \rhodyn^{(ext)} \ .
    \label{eq:int_ext}
\end{equation}
Specifically, the ``intrinsic'' model, which is amenable to an analytic solution, corresponds to 
substituting $\pd_t h=0$ in Eq.~(\ref{eq:stress-strain-rate}), thereby ignoring the effect of the fluid's volumetric incompressibility and the associated areal contraction,~(Eq. \eqref{eq:continuity-00}, on the mid-surface dynamics \cite{Howell1996}.  
We show now that this analytic solution captures the behavior in the full model very well. 
{For $\delta_{ext}= 0$, the second term on the RHS of~Eq. \eqref{eq:pol-1} vanishes, with two ramifications. First, the solution of Eq. \eqref{eq:pol-1} reduces to
a Seivert surface of constant Gaussian 
curvature, $(\pd_r z_{per})^2 = r^2 - \frac{2}{3}\Sg(t)$, with $\dot{\Sg} \!= \! - q$. The solution evinces a jump in $\pd_r z$ at $r= r_f$. 
Second, the jump condition, Eq. \eqref{eq:BC-front-2}, reduces to an explicit equation of motion for $r_f$. Indeed, at the front, in the intrinsic model, one may switch to a comoving frame near the front, such that $\pd_t \approx -\dot{r}_f\pd_r$. Evaluating Eq. \eqref{eq:BC-front-2} in this case, one finds that $q(t) = -\frac{3\dot{r}_f}{2r_f}\left[(\pd_r z_\ann)^2\right]_{r_f}$.}
The dynamics are then completely dictated by $q(t)$, and one finds analytically {(see SI)},
\begin{align}
\label{eq:rf-S-intrinsic}
    r_f(t) &= \left[\frac{4}{3}\Sg(t)\right]^{1/2}, \\ \nn
    z_f(t) &= z(0,t) = r_f^2\left(\mathcal{F}(r_f^{-1}) - \mathcal{F}(1)\right),
\end{align}
where $\mathcal{F}(x)= \frac{1}{4}\left[2x\sqrt{x^2-\frac{1}{2}} - \log\left(2x+2\sqrt{x^2-\frac{1}{2}}\right)\right]$.  

Fig. \ref{fig:solutionPlots}b compares the numerical and analytic solutions of the intrinsic model, and also shows $z_b(t)$ for the full model $\delta_{ext} = 1$, with excellent agreement between the models. Indeed, as we show in the SI, when $\delta_{ext} > 0$, 
the jump in $\pd_r z_{per}$ at $r\!=\!r_f(t)$ is merely broadened over a transition zone of width $\delta_{ext}$, while the overall dynamics remain essentially intact. 
This observation highlights the crucial role of stress-induced areal shearing in Eq. \eqref{eq:Howell-FvK-sym}, on which we hinted 
earlier.

\section*{Wrinkling instability}

We started this paper with a basic question:  what is the origin of compression in the depressurized bubble? 
Our solution of the axisymmetric bubble collapse provides the answer immediately. A glance at Fig. \ref{fig:solutionPlots}a and Eq. \eqref{eq:Phi-dis} shows that the disclination induces hoop compression 
$\sigma_{\theta\theta} = \pd_{rr}\Phi <0$ in an annular zone, $r_f/e < r< r_f$, implying the film buckles along this axis, 
similarly to its 
elastic counterpart.
Thus, the physical mechanism underlying the wrinkling instability is the propagation of the front, which we showed to result from
the exclusion of a uniformly stress-free solution to Eq. \eqref{eq:Howell-FvK-sym}, 
as implied by the joint effect of   
the meniscus ``clamping'',  
Eq. \eqref{eq:BC-z}, and the thermodynamic law 
(\ref{eq:1Thermo}). 
We stress that our theory shows that a wrinkling instability for the collapsing bubble can \emph{only} arise from the front-induced hoop compression. 
This mechanism and the consequent quantitative predictions it entails, are conceptually different from previous proposals that attributed compression to contraction of the meniscus \cite{da2000rippling} or a highly nonuniform thickness profile \cite{Oratis2020}.  

Our axisymmetric front solution
enables us to carry out a quantitative linear stability analysis of the flattening film, and examine whether the wrinkle pattern  observed experimentally in Ref.~\cite{Oratis2020} is explained via this fundamental paradigm of pattern formation theory. 
We consider 
periodic,  infinitesimal deflections of the film's mid-surface from the axisymmetric dynamics in Eq. \eqref{eq:front-def}, namely, 
\begin{align}
    z(r,\theta,t) &= z(r,t) + \zeta_m(r,t) \cos(2\pi m \theta) , 
        \label{eq:s-mode}
        \\\nn
    \Phi(r,\theta,t) &= \Phi(r,t) + \phi_m(r,t) \cos(2\pi m \theta), 
\end{align}
where $m$ is an integer, and $z(r,t),\Phi(r,t)$ are the previously found solution. We expand
the general  (non-axisymmetric) form of Eq. \eqref{eq:Howell-FvK-sym} to linear order in $\zeta_m$ and $\phi_m$. Solutions of the linearized equation have the form: $\zeta_m(r,t) =e^{\omega_m t}\zeta_m(r),\phi_m(r,t) = e^{\omega_m t}\phi_m(r)$, 
such that positive values of $\text{Re}  (\omega_m)$ signal instability, and  ${\max}\{\text{Re} (\omega_m)\}$ defines the fastest-growing unstable mode $m^*$, which dominates
the observed pattern. 
After expanding 
we obtain four coupled PDEs. For our purpose it is enough to present one of them, which determines the exponential behavior and arises from normal force balance (see the SI for the other equations):
\begin{equation}
 (\pd_r^2\Phi)
\frac{m^2}{r^2}\zeta_m  
\!+\! 
\epsilon^2 \omega_m
\hat{{\cal L}}_m 
\left[\zeta_m\right]  =
 \label{eq:Howell-FvK-nonsym-lin-1} 
 \frac{3}{2}
      \pd_t\left(\pd_{rr}z
      \zeta_m \right) + \frac{1}{r}\pd_r
      (\pd_r \Phi_0 \pd_r \zeta)
\end{equation}
where $\hat{\cal L}_m = \hat{L}_r^2 - \frac{m^2}{r^4}\left(4-2r\pd_r+2r^2\pd_{rr}\right)+\frac{m^4}{r^4}$ is the 2D Laplacian.

Eq.~(\ref{eq:Howell-FvK-nonsym-lin-1}) is analogous to the normal force balance that determines stability of an axisymmetrically-confined  
elastic sheet. 
The first term on the LHS, with $\sigma_{\theta\theta} = \pd_{rr}\Phi <0$, is a destabilizing, compression-induced force, 
and the second term is ``viscous bending'' that restores axisymmetry.   
(In the elastic counterpart, the bending force is proportional to $\zeta_m$ rather than to $\pd_t \zeta_m$). The terms on the RHS of Eq. \eqref{eq:Howell-FvK-nonsym-lin-1} consist of radial  derivatives of $\zeta_m$ and $z$ (akin to ``tension-induced'' and ``curvature-induced'' restoring forces in the elastic counterpart), and are pronounced only at the vicinity of the front. 

For a given $m$, Eq.~(\ref{eq:Howell-FvK-nonsym-lin-1}) is an eigenvalue problem with a single eigenvalue $\omega_{m}$ and corresponding eigenfunctions $[\zeta_m(r), \phi_m(r)]$. Inspection of Eq.~(\ref{eq:Howell-FvK-nonsym-lin-1}) 
reveals that among the two terms on the RHS, the  
first one is non-vanishing only at the 
highly-curved front region, whereas the second one is smaller than the first term on the LHS (by a factor $m^{-2}$). Hence, ${\max}\{\text{Re} (\omega_m)\}$ is determined by balancing    
the two dominant terms 
on the LHS, yielding $\omega_m \sim (\epsilon m)^{-2}$ (up to pre-factor that depends on $r_f$), which is maximized at small, 
$\epsilon$-independent $m^*$. This observation is in sharp contrast with the experimental observations in Refs.~\cite{Debregeas1998,da2000rippling,Oratis2020} of a large and $\epsilon$-dependent $m^*$, see e.g. Fig. \ref{fig:expSetup}c.
%
Following a general argument by Howell \cite{Howell1996}, 
it was suggested in  Ref. \cite{Oratis2020} that the LHS of Eq.~(\ref{eq:Howell-FvK-nonsym-lin-1}) must be supplemented by another restoring force, ($Oh \cdot \epsilon)^{-2} \omega_m^2 \zeta_m$, which originates from the inertia of the fluid film. With this additional force, the fastest growing mode satisfies $m^* \sim \epsilon^{-1} Oh^{-1/3}r_f$ (see SI). 

\begin{figure}
    \centering
        \begin{subfigure}{0.48\hsize}
      \includegraphics[width=\hsize]{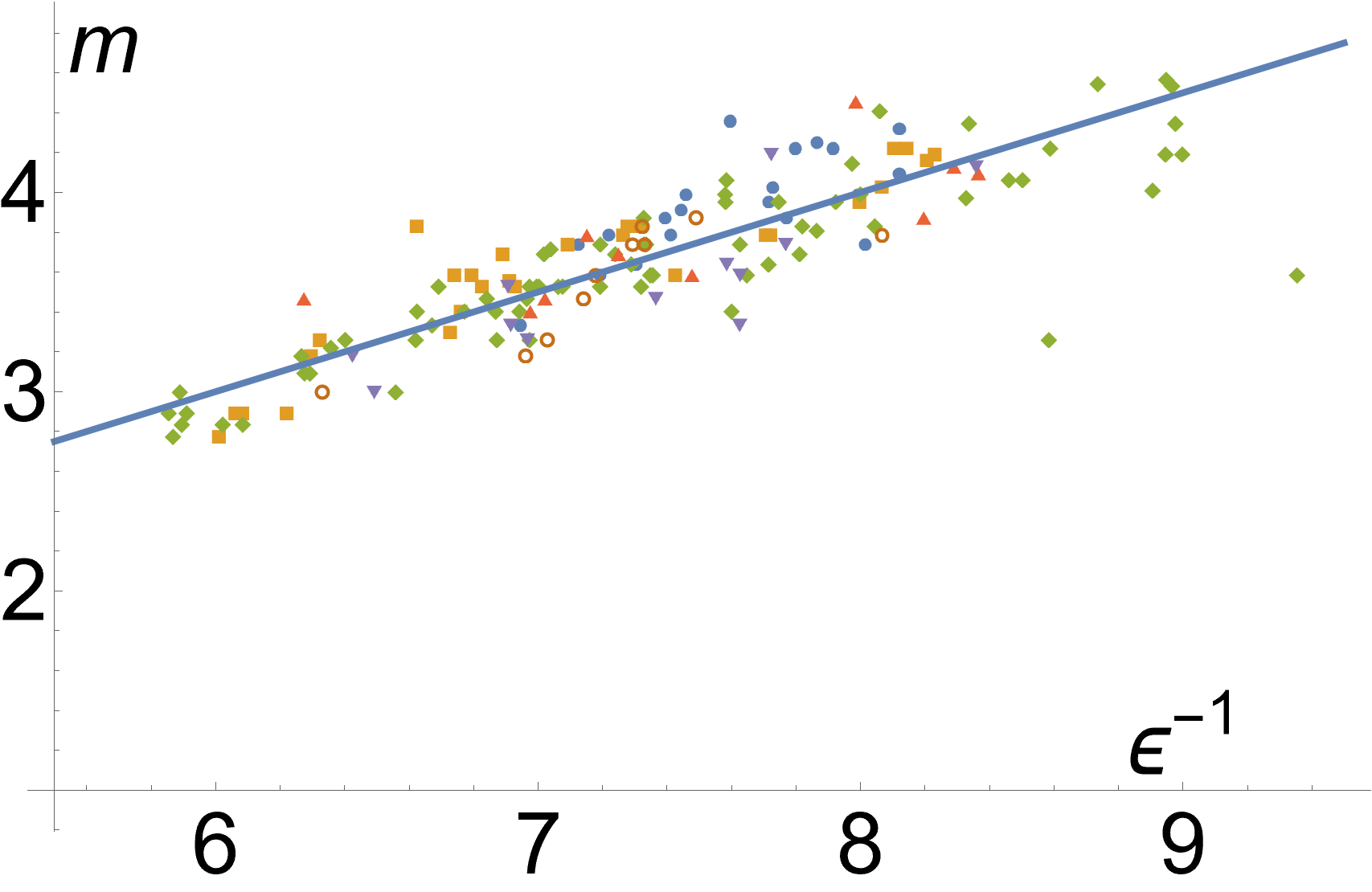}
      \caption{}
    \end{subfigure}\hfill
    \begin{subfigure}{0.48\hsize}
      \includegraphics[width=\hsize]{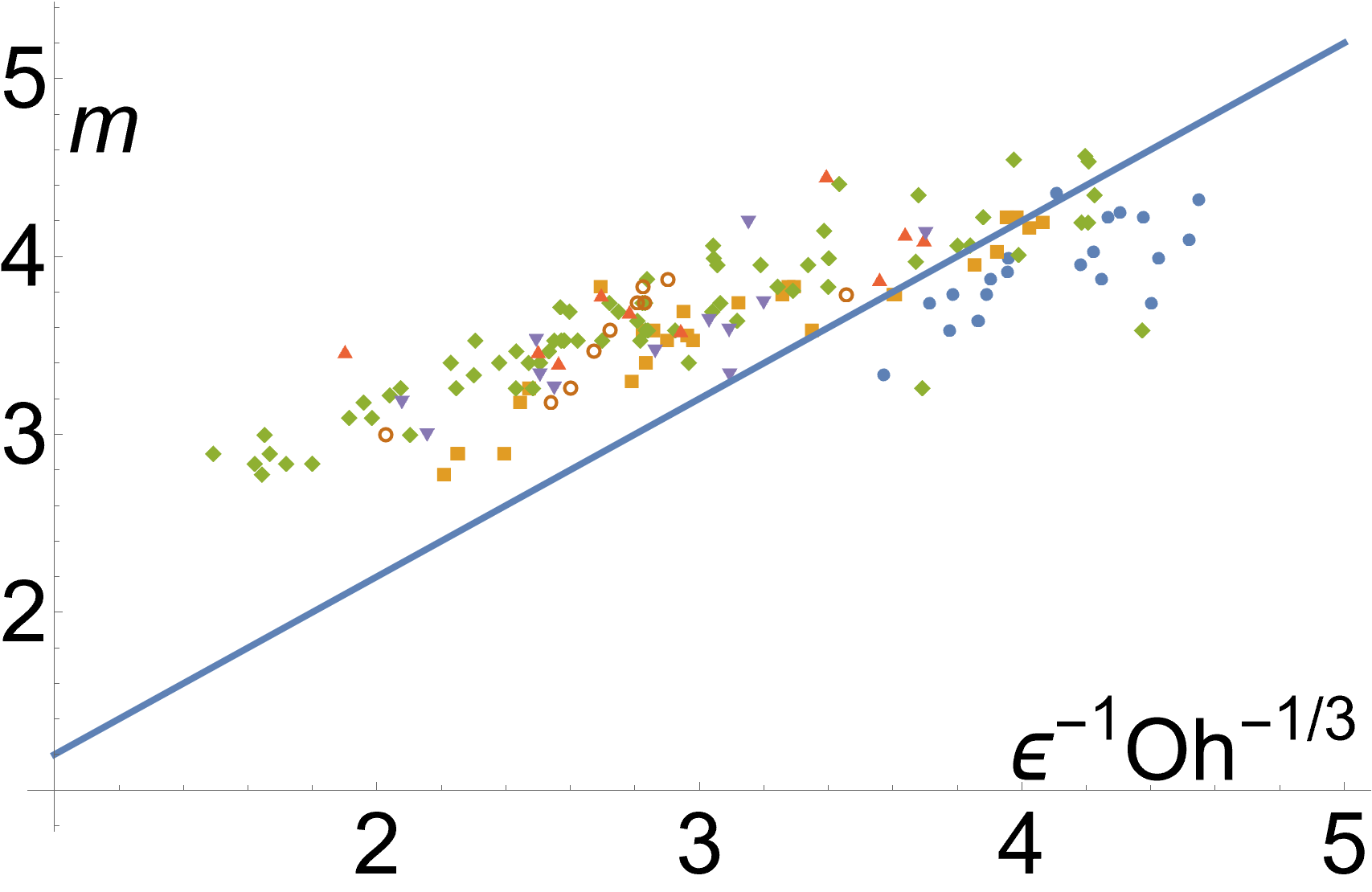}
      \caption{}
    \end{subfigure}
    \caption{
                Log-Log scaling of the wrinkle number $m^*$ in the experimental data, taken from Ref. \cite{Oratis2020}. Different colors represent different datasets from that work (see SI).
                (a) Scaling of $m$ with $\epsilon$ only. Good agreement is found with $m\sim\epsilon^{-1/2}$ (solid blue line). (b) Scaling of $m$ with the prediction of a linear stability analysis, with dependence on inertia. We note that changing the power law in $Oh$ (e.g. to the power law suggested in the above work) does not restore the scaling properties.}
    \label{fig:mscale}
\end{figure}

We now show that the experimental data appears to rule out \emph{any} significant $Oh$ dependence. 
Our re-analysis of the data of Ref.~\cite{Oratis2020},  
presented in Fig. \ref{fig:mscale}, reveals that the data scale very nicely with $m^* \sim \epsilon^{-1/2}$, and that introducing any $Oh$ dependence destroys the scaling.
%
This leads us to a conundrum.
On the one hand, an apparent irrelevance of inertia 
indicates that the dynamics is governed solely by viscous and capillary forces, even at the very 
onset of the wrinkling instability. On the other hand, the  dependence 
on the film's thickness contradicts 
linear stability analysis of the non-inertial theory. 
This conundrum suggests that unlike many examples of non-equilibrium pattern formation, the number of wrinkles in a depressurized viscous bubble is not determined when their amplitude is infinitesimal,  
but rather at a later, ``far-from-threshold`` stage, 
at which they grow 
to suppress the compressive hoop stress altogether.
Analysis of this stage 
is beyond the scope of this manuscript.
However, we note that in the front-dominated scenario, there is a finite hoop compression,
$\sigma_{\theta\theta} = \pd_{rr}\Phi<0 $,  
at the vicinity of the propagating front, even though compression is suppressed elsewhere in the film. We therefore conjecture that the optimal wrinkle amplitude is governed by curvature-induced 
terms at the front, akin to those on the RHS of 
Eq. \eqref{eq:Howell-FvK-nonsym-lin-1}.
A scaling analysis of the biharmonic operator $\hat{\mathcal{L}}_m$ shows that force balance indeed requires $m^* \sim \epsilon^{-1/2}$ with no $Oh$ dependence.

\section*{Discussion}

We offered here a
mechanism for 
the collapse of a highly viscous bubble following  
rapid depressurization. 
We showed that 
any curved portion of the film must become stress free, and that this condition cannot be realized by 
uniform shrinkage 
due to 
the  
immobile meniscus. 
Instead, 
depressurization 
triggers 
a topological instability: 
spontaneous nucleation of a
globally neutral dynamo-geometric 
charge distribution, which comprises a 
disclination-front pair. In this picture, the emerging surface dynamics is governed by 
curvature localized 
at a front, Eq. \eqref{eq:front-def}, propagating at a rate dictated by thermodynamics, ~Eq. \eqref{eq:1Thermo}, and separating 
an expanding planar core from a curved periphery, see Fig. \ref{fig:solutionPlots}.
The wrinkling instability is merely a byproduct of the hoop compression in the disclination-induced stress field at the expanding planar core. 

Our work reveals 
the bubble collapse problem as  
a peephole into an entire, 
almost unexplored, branch of classical fluid dynamics: non-inertial motion of curved, free-surface films, governed by a geometrically nonlinear and topologically nontrivial hydrodynamics.
The nonequilibrium nature of this hydrodynamics 
is encoded in a ``charge'' density $\rhodyn$, Eq. \eqref{eq:Howell-FvK-sym-2b}, that stems from temporal variation of the surface curvature, and the corresponding charge/stress relation, Eq. \eqref{eq:Howell-FvK-sym-2}, is akin to the charge/field relation in  
electrostatics of continuum media. Furthermore, the association of stress-generating ``charge'' with temporal variation of the curvature, rather than with the curvature itself, underlies a violation of SRA between Hookean elasticity and viscous hydrodynamics. \\

\begin{figure}
    \centering
    \includegraphics[width=0.6\hsize]{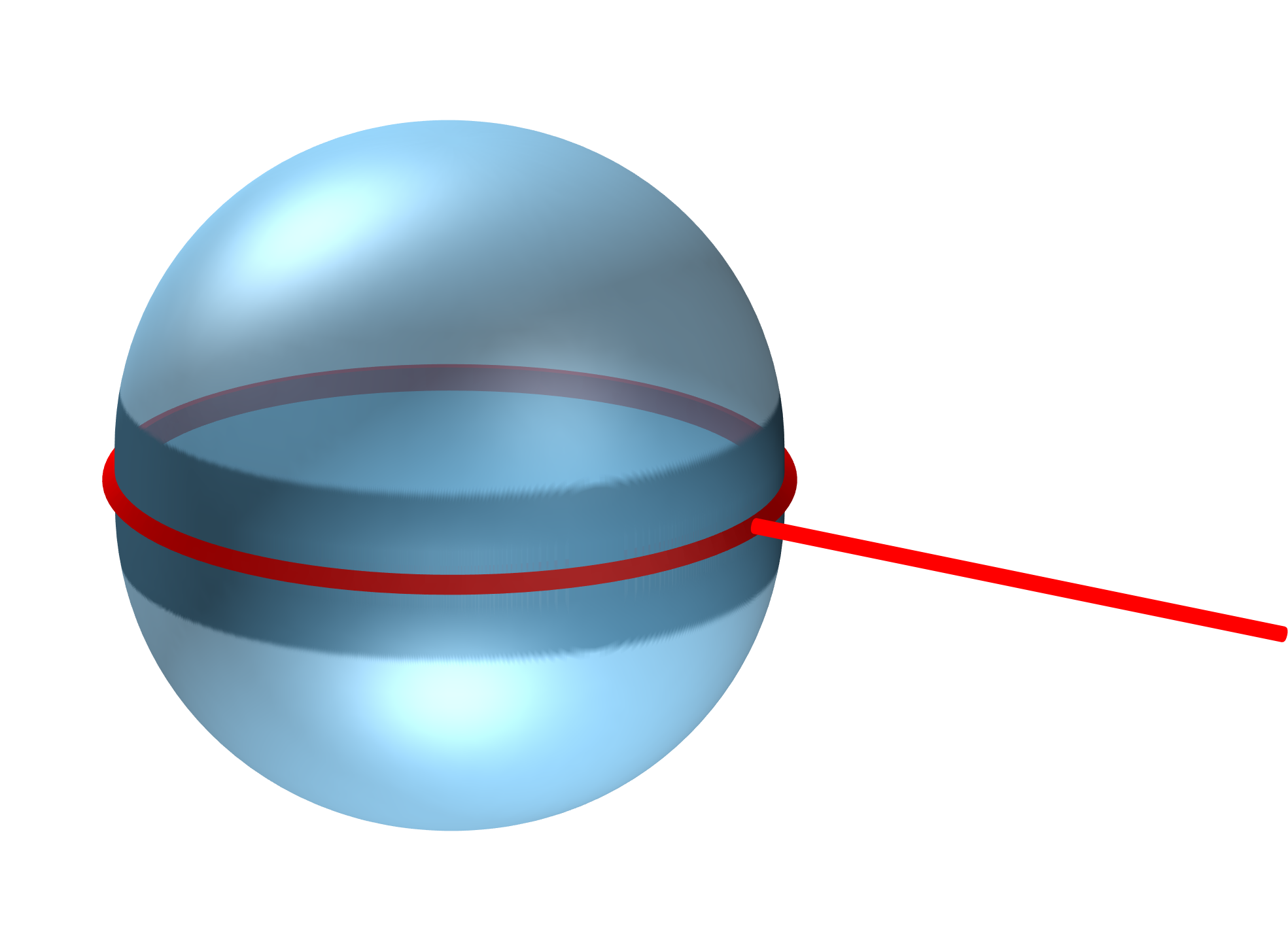}
    \caption{{A non-floating bubble configuration, where 
    two viscous films are stabilized on a solid ring. 
    Upon rapid depressurization, 
    the vanishing fluid velocity at the ring implies 
    flattening of each film that 
    propagates 
    \emph{inward}.}}
    \label{fig:expSuggest}
\end{figure}

Let us now 
discuss our  
results in view of 
experimental observations,
and elaborate on key theoretical and experimental challenges that remain towards a full understanding of the viscous bubble collapse. 

\textit{(i)}
A central theoretical prediction is that the surface evolves \textit{via} front propagation. Evidence for such surface dynamics, and experimental data for the propagation rate $\dot{r}_f(t)$, may be obtained by gathering and systematically analyzing side views of the collapsing bubble. 

\textit{(ii)} The front solution allows us to evaluate the collapse rate of the film, given by $\pd_t z(r=0,t)=\dot{z}_f(t)$, since by construction the bubble apex and the front have the same height, see 
Fig.~\ref{fig:solutionPlots}a and b. We find (see Eq. \eqref{eq:rf-S-intrinsic}) that at short times $\dot{z}_f$ is a constant, $\dot{z}_f \propto - q(T)$, up to logarithmic corrections, where the numerical value of the disclination charge is determined from Eq. \eqref{eq:1Thermo}. 
This result is in accord with experimental data 
(Fig.~2 of Ref.~\cite{Oratis2020})  that exhibits a constant speed, which is $O(1)$ in the natural speed units,  $R/\tau_{vc}$.

%
\textit{(iii)} 
Our analysis 
shows that the experimentally observed wrinkle number is determined solely by the aspect ratio $\epsilon$ of the film, in contradiction with predictions based on linear stability analysis of the axisymmetric state. This observation calls for finding a far-from-threshold, non-equilibrium mechanism that selects the wrinkle number.

%
\textit{(iv)} 
The simplest version of the hydrodynamics, represented in the current version of~Eq. \eqref{eq:Howell-FvK-sym}, 
assumes the initial state of the film is described by a simply-connected mid-surface and nearly uniform thickness profile. Incorporating rupture and spatially nonuniform thickness profile into this theoretical framework 
is a theoretical challenge, which may be crucial for quantitative comparison of theoretical predictions with experimental observations (\textit{e.g.} for the propagation rate of the front and the radial extent of the wrinkled region). 


\textit{(v)}
Our study highlights 
the central effect of the meniscus: ``clamping'' the film's edge and thereby forbidding uniform shrinkage, while allowing a free exchange of fluid mass with the bath. This dual role of the meniscus may be disentangled in 
a different set-up, depicted schematically in Fig. \ref{fig:expSuggest}:  
a thin solid ring of radius $r=1$ wets a bubble, 
and the interior gas is depressurized rapidly, {\emph{e.g.}} by pricking the bottom half. 
Similarly to the meniscus, the wetting ring effectively ``clamps''
the film, but unlike the meniscus it also  
arrests the fluid at $r\!=\!1$, 
thus prohibiting 
mass flux between the two halves of the film, 
and thereby replacing the homogeneous BC~(\ref{eq:BC-hom-2}) at $r=1$ 
by 
an \emph{inhomogeneous} one  (see SI).  
Since a non-homogeneous BC 
gives
rise to non-zero stress near $r=1$, which cannot be 
realized 
by a curved surface, we expect that 
after the rapid depressurization stage the film must be flattened there. As a result, 
yet another, inward-moving front must emerge at the periphery, $r_{p}(t)$, in addition to 
the outward-moving front at $r_f(t)$, 
such that 
the  
shrinking, curved portion of the film 
is buffered between the two fronts. \\

Beyond the Stokesian hydrodynamics of simply viscous films, our study raises questions concerning the effect of surface geometry on the hydrodynamics of viscous fluids that are also coupled  to elastic degrees of freedom. One basic example, of interest to the complex fluids community, is a free-surface film of viscoelastic polymeric suspension, whose stress is affected both by the solution viscosity and by the polymer elasticity.
Since the generator of stress in elastic sheets is the surface's curvature itself, 
one may ask to what extent 
the stress-generating ``charge'' density $\rhodyn$ in Eq. \eqref{eq:Howell-FvK-sym} is affected both by the rate of change of the surface's curvature and by the curvature itself. 

Another generalization of ~Eq. \eqref{eq:Howell-FvK-sym} is to the non-inertial hydrodynamics of purely 2D liquids that reside on a curved 2D surface such that their tangential  momentum is conserved. A physical realization of such a system is the recently discovered viscous hydrodynamics of strongly correlated  electrons in graphene and several other compounds~\cite{Gurzhi1963,Andreev2011,Hartnoll2015,Levitov2016,BandurinD.2016,Jesse2016,MollPhilipJ.2016}.  
For the collapsing viscous film 
the curvature current $\bJG$, Eq. \eqref{eq:elec-0}, is proportional to the gradient of the rate of change of surface area, $\dot{\varepsilon}$, occupied by a volumetrically-incompressible fluid, and the dynamo-geometric charge density $\rhodyn$ is required to vanish everywhere but at the front,  due to the stress-free condition implied by normal force balance, Eq. \eqref{eq:Howell-FvK-sym-1}.  
In contrast, 
for a 2D-incompressible fluid 
$\bJG$ is proportional to 
the gradient of the chemical potential (thus describing a flow of actual electric charge), and $\rhodyn$ is 
externally imposed by a time-dependent curvature of the embedding surface, \textit{e.g.} by appropriately vibrating the 2D sheet, through Eq. \eqref{eq:Howell-FvK-sym-2b}.       
These examples suggest that,
beyond 
the fascinating journey taken by depressurized bubbles, 
studies of curved liquid films may lead to new exciting insights on the interplay between hydrodynamics and surface geometry.
\acknowledgements{
We thank J. Bird, A. Oratis, H.A. Stone for many inspiring discussions on their paper \cite{Oratis2020} and beyond, and also thank J. Bird \& A. Oratis for the permission to present their data in Fig. 1c. We thank G. Grason, P. Howell, D. Kastor, L. Mahadevan, V. Mathai, N. Menon, J. Traschen, M. Moshe, D. Vella, J. Schmalian, J. Ruhman, and G. Weinstein for helpful discussions, and G. Grason, D. Vella, and H.A. Stone for their critical reading of the manuscript. We acknowledge support by the National Science Foundation under grant DMR 1822439 (BD), and by the Israel Science Foundation (ISF), and the Israeli Directorate for Defense Research and Development (DDR\&D) under grant No. 3467/21 (AK). This research was supported in part by the National Science Foundation under Grant No. NSF PHY-1748958 (AK). 
We gratefully acknowledge the hospitality of the Weizmann Institute of Science while this work was completed, and the support of the Weston Visiting Professorship Program (BD).
}
\newpage
\appendix

\onecolumngrid
\begin{center}
    {\huge \bf{Supplementary information (SI)}}
\end{center}
\section{Stokes (non-inertial) dynamics of curved films}
\label{app:EOM}
In this appendix we describe the derivation of the force balance equations~(\ref{eq:Howell-FvK-sym-1},\ref{eq:Howell-FvK-sym-2}) of the main text.
An early version of these equations, which does not include the `` extrinsic'' charge, $\rhodyn^{(ext)}$, was 
obtained by Howell in Ref.~\cite{Howell1996} for a thin film of volumetrically-incompressible fluid, by exploiting a formal similarity to FvK equations that describe the mechanical equilibrium of elastic sheets. In addition to ignoring fluid inertia and using small-slope approximation, which we discussed already in the main text, 
we elaborate here further on the negligible effect of spatial gradients in the film thickness on its viscous hydrodynamics. For this purpose, we start by extending Howell's derivation, using a more generic form of the stress tensor of a 2D isotropic fluid model
which does not depend explicitly on the film thickness and is not restricted to a film of incompressible liquid volume.
This allows us to evaluate (in a following section)  
when it is possible to ignore spatial variations of the film thickness. \\ 





\subsection{An invariant form of strain rate and stress}
Following Scriven ~\cite{Scriven1960}, we start with a generic, covariant formulation of the local mechanics and kinematics in a curved, 2D momentum-conserving viscous film. We consider a smooth surface 
$\vec{X}(u_1,u_2;t)$ made of viscous fluid {and described by two generalized coordinates $u_{1,2}$}, with metric and curvature tensors, $g_{ij},\kappa_{ij}$ ($i,j \in \{1,2\}$), respectively, such that $|\kappa_{ij}| \ll h^{-1}$.   
Denoting orthogonal unit vectors locally tangential to the surface 
by ${\bf t}_i$, and a normal vector by
$\hat{n}$, we decompose the 
fluid velocity $\vec{v}$ into tangent and normal components,  
${\bf v} = v_i {\bf t}_i$ and 
$v_n \hat{n}$, respectively. The mechanics of a Newtonian fluid is then expressed through a linear relation between the 2D tensors of  stress (force/length) and strain rate (inverse time): %
%
\begin{gather}
\sigma_{ij} =
\mu{|\!\det\stackrel{\leftrightarrow}{g}\!| \delta_{ij}} + 
\eta_{2d} (\dot{\varepsilon}_{ij} + g^{lm}\dot{\varepsilon}_{lm} 
\delta_{ij} ) \ ,
    \label{app:stress-strain-rate} 
    \end{gather}
    whereas the strain rate is given by: 
    \begin{gather}
 {\dot{\varepsilon}}_{ij} \!=\! 
        \tfrac{1}{2} ( {\bf t}_i \cdot {\cal D} \ {\bf v}_j  + {\bf t}_j \cdot {\cal D} \ {\bf v}_i)
    +  \tfrac{1}{2} \dot{g}_{ij}
    \ , 
    \label{app:strain-rate-vel}
\end{gather}
{where $\mathcal{D}$ denotes a covariant derivative}
In the above expressions, the viscous part of the stress tensor is proportional to the 2D shear viscosity $\eta_{2d}$ (pressure$\cdot$ length$\cdot$time), whereas an explicit dilatational viscosity is ignored. The thermodynamic part of the stress tensor is isotropic, proportional to a ``chemical potential'' $\mu$ (energy/area). Note that the strain rate is affected both by gradients of the tangential velocity, as well as by rate of change of the metric that does not involve tangential flow. 

For the problem we study here, the 2D surface is actually a mid-surface of a film of thickness $h(\vec{X},t)$ of a volumetrically-inompressible liquid. In this case, temporal variation of the mid-surface metric stems from the rate of change of film's thickness, $\pd_th$, as is evidenced in~\eqref{eq:stress-strain-rate} of the main text. Mass conservation is given by 
\cite{Stone1990}: 
\begin{gather}
\pd_t h = 
- \ddiv (h \vec{v})  = 
- \ddiv (h {\bf v})- h v_n (\ddiv \ \hat{n}) \ , 
\label{eq:continuity-00-sup}  \\
\text{where} \ \ 
\ddiv \vec{A} \equiv {\bf t_i} \cdot ({\bf t}_i \cdot {\cal D}) 
{\vec A} 
    \label{eq:ddiv}
    \end{gather}
is the surface covariant divergence of a vector field $\vec{A}$, 
and 
${\cal H} = \tfrac{1}{2}(\ddiv \ \hat{n})$ is the mean curvature of the surface. Equation~(\ref{eq:continuity-00}) of the main text is the axisymmetric version of~\eqref{eq:ddiv}.    \\

For a viscous film of finite thickness $h$, the thermodynamic part of the stress in Eq.~(\ref{eq:stress-strain-rate}) is determined by the surface energy of the two free surfaces, hence: 
\begin{equation}
\mu = 2\gamma \cdot (1 + h \nabla^2 h)  \ ,   
    \label{app:mu-gamma}
\end{equation}
(where $\nabla^2$ above is interpreted as Laplace-Beltrami operator).
The assumption of nearly-uniform thickness allows us to ignore thickness gradient in the above expression (see below), hence $\mu = 2\gamma$. More specifically, while temporal variation of the thickness, \textit{i.e.} the term $\pd_th$, may have a finite ($O(1)$) effect on the surface dynamics and is thus included in our analysis, spatial gradients in the thickness of a film, which is initially nearly uniform, remain $O(\epsilon)$ during the flattening process and can therefore be safely neglected.



\subsection{Tangential force balance}

Since our focus in this paper is on a thin film of volumetrically-inompressible fluid, we choose in this section not to follow the covariant 2D formulation introduced above, but instead the standard method in continuum mechanics, namely starting with the viscous stress of a 3D, volumtrically-incompressible liquid film and performing ``dimensional redution'' by integrating over the film's thickness. This method is analogous to the derivation of FvK equations in classical elasticity theory for a thin solid plate.  In this approach, we assume the fluid surface is described as $\vec{X}(x,y;t) = x \hat{x} + y\hat{y} + z(x,y;t) \hat{z}$, such that $|\nabla z|\!\ll \! 1$, and we can choose, up to corrections of $O(|\nabla z|^2)$, the two orthogonal unit tangent vectors as 
${\bf t}_1 \approx \hat{x} + \partial_x z\hat{z}$ 
and ${\bf t}_2 \approx \hat{y}+ \partial_y z \hat{z}$, and the normal vector is $\hat{n}\approx \hat{z}$. To obtain the 2D strain rate
and stress tensors
~\eqref{eq:stress-strain-rate}, we decompose the tangential velocity into corresponding components, ${\bf v} = v_x {\bf t}_1  + v_y {\bf t}_2$, and note that 
in the small-slope approximation the corresponding components of the strain rate tensor are: 
\begin{align}
\dot{\varepsilon}_{xx} &\approx \pd_x v_x \!+\! \tfrac{1}{2}\pd_t(\pd_x z)^2,\nn\\  
 \ \dot{\varepsilon}_{yy} &\approx \pd_y v_y \!+\! \tfrac{1}{2}\pd_t(\pd_y z)^2,      \nonumber \\
    \dot{\varepsilon}_{xy} &\approx \tfrac{1}{2}(\pd_yv_x + \pd_xv_y + \pd_t (\pd_xz\pd_yz)) \ .  
    \label{app:strain}
\end{align}
These expressions for the tangential (or ``in-plane'') components of the strain tensor are valid, up to corrections of $O(h/L)$, where $L$ is a characteristic lateral scale over which the dynamics varies, at any point in the film.  
Note that the quadratic terms in $\nabla z$ in the above expressions are analogous to the ``geometrically  nonlinear'' terms in FvK equations of elastic plates, and stem from rotational invariance of the strain rate tensor. Specifically, note that $\dot{\varepsilon}_{xx}$ can be written as $\partial_t [\partial_xu + \tfrac{1}{2}(\partial_xz)^2]$, where $u$ is a ``displacement'' of fluid element along $\hat{x}$, and the term in square brackets is the corresponding component of a strain tensor. Here, the presence of of the quadratic term expresses the fact that a simple rotation of a fluid element (where $\partial_x u \neq 0$) generates in fact no strain.                

In order to derive analogous expressions for the corresponding components of the thickness-averaged stress in the film, consider the $x$-component of the normal force balance inside the film: 
\begin{equation}-\partial_x p  + 2\eta \partial_x \dot{\varepsilon}_{xx} + 2\eta  \partial_y  \dot{\varepsilon}_{xy} = 0
\label{app:q11}
\end{equation}
where fluid inertia is ignored, and $\eta$ is the dynamic shear viscosity of the liquid (pressure$\cdot$time).  
Note that, similarly to the tangential strain rate components in Eq.~(\ref{app:strain}), the pressure $p$ is also constant across the film's thickness, up to $O(h/L)$. The pressure value is readily obtained by noticing that the free surfaces imply its balancing with a viscous force: 
$$ p = 2\eta \partial_z v_n \ $$ 
whereas the volumetric inompressibility of the liquid implies that (similarly to Eq.~(\ref{eq:ddiv}):
$$\partial_z v_n  = - [v_n (\nabla \cdot \hat{n}) + \nabla \cdot {\bf v}]  = - [2 v_n {\cal H} + (\pd_x v_x + \pd_y v_y]  \ . $$
Plugging the pressure into Eq.~(\ref{app:q11}), and performing the analogous manipulation for the force balance in the $\hat{y}$ direction, we note that the two tangential force balance equations can be written as: 
\begin{equation}
\nabla \cdot \stackrel{\leftrightarrow}{\sigma} = 0
    \label{app:q12}
\end{equation}
where $\stackrel{\leftrightarrow}{\sigma}$ is a thickness-integrated stress, whose components are: 
\begin{align}
{\sigma}_{xx} &\approx 2\gamma  + 
2\eta h (2\dot{\varepsilon}_{xx}  + 
{\varepsilon}_{yy} + 2 v_n {\cal H})  \nonumber \\
{\sigma}_{yy} &\approx 2\gamma  + 
2\eta h (2\dot{\varepsilon}_{yy}  + 
\dot{\varepsilon}_{xx} + 2v_n {\cal H}) \nonumber  \\
\dot{\sigma}_{xy} &\approx 2
\eta h \ 
\dot{\varepsilon}_{xy}  \ .  
\label{app:stress}
\end{align}

Conversely, the strain rate tensors are given by:
\begin{align}
\dot{\varepsilon}_{xx} &\approx  
\frac{1}{6\eta h}
(2\sigma_{xx}  - 
\sigma_{yy} 
-
2\gamma) - 
v_n {\cal H} \ ,  \nonumber \\
\dot{\varepsilon}_{yy} &\approx  
\frac{1}{6\eta h}
(2\sigma_{yy}  - 
\sigma_{xx} 
-
2\gamma) -
v_n {\cal H} \ ,  \nonumber \\
\dot{\varepsilon}_{xy} &\approx \tfrac{1}{2\eta h}
{\sigma}_{xy}  \ .  
\label{app:strain-2}
\end{align}


It is well known (\textit{e.g.} from linear elasticity \cite{LandauBook1986}) that Eq.~(\ref{app:q12}) is automatically satisfied by the celebrated Airy stress potential, $\Phil(x,y,t)$, such that:  
\begin{equation}
  \sigma_{xx} = \partial_{yy} \Phi\ , \ \sigma_{yy} = \partial_{xx} \Phi  \ , \sigma_{xy} = 
  - \partial_{xy} \Phi 
  \label{app:Airy}
\end{equation}
Substituting \eqref{app:Airy} for the stresses in ~\eqref{app:strain-2}, adding the second derivatives of the first line with respect to $y$ and the second line with respect to $x$ and subtracting the mixed second derivative of the third line, we obtain Eq.~(\ref{eq:Howell-FvK-sym-2})of the main text: 
\begin{equation}
\nabla^4 \Phi 
= \eta h \cdot [-3 \partial_t {\cal R}  + \nabla^2(v_n {\cal H})]
%
    \label{app:Stokes-Poisson}
\end{equation}
where $\nabla^2 \!= \! \partial_{xx} + \partial_{yy}$. %

\subsection{Normal force balance}
The normal force balance on the film is expressed by Eq.~(\ref{eq:Howell-FvK-sym-1}) of the main text. 
This equation
was obtained by Howell \cite{Howell1996} as a generalization of the {\emph{viscida}} \cite{Buckmaster1975}, which is itself analogous to the celebrated Euler's {\emph{elastica}} that describes the normal force balance at mechanical equilibrium of a solid sheet subjected to developable deformation ({\emph{i.e.}} $\GR = 0$). A reader who is interested in a derivation of Eq.~(\ref{eq:Howell-FvK-sym-1}) of the main text through ``dimensional reduction'' of the Navier-Stokes equation of a non-inertial, volumetrically incompressible liquid film (Trouton's approach) is referred to Refs.~\cite{Buckmaster1975} and Secs.~2,4 of Ref.~\cite{Howell1996}. Here we will employ the similarity to the {\emph{elastica}} in order to physically motivate this equation. 

\paragraph*{{{Elastica:}}} 
In the absence of external tangential forces, the planar deformation of a naturally flat elastic sheet, $\vec{X}(x,y) = (x,y,z(x))$ with $|z'|\ll 1$, is characterized by a single non-vanishing stress component that is spatially constant, $\sigma_{xx}(x,y) = \sigma$, and the normal force balance at mechanical equilibrium is given by:  
\begin{equation}
    -\sigma \kappa + B \kappa{''} = f_n \ , \ \ \kappa \approx z{''}  \ , 
    \label{app:elastica}
\end{equation}
where $B$ is the elastic bending modulus and $f_n$ is an external force/area exerted normally to the sheet.   One may readily recognize the first term in this simplified version of the {\emph{elastica}} as the resistance of a tensed string (corresponding to $\sigma >0$) to deflections from flatness, whereas the second term originates from the resistance of a naturally-planar sheet to curvature. Alternatively, Eq.~(\ref{app:elastica}) is obtained as the Euler-Lagrange equation that minimizes the elastic energy $\approx \tfrac{1}{2} (\sigma \cdot (z')^2 + B \cdot (\kappa)^2)$, with respect to variations of the shape $z(x)$.  

\paragraph*{{{Viscida:}}}
The small-slope version of the {\emph{viscida}} \cite{Buckmaster1975}, which describes the normal force balance in planar deformations of a free-surface viscous film, is now obtained from Eq.~(\ref{app:elastica})
by expressing the stress $\sigma$ through its capillary and viscous contributions, $\sigma  = 2\gamma + 4\eta h \dot{\varepsilon}_{xx}$, where the strain rate $\dot{\varepsilon}_{xx}$ is given in Eq.~(\ref{app:strain}), and replacing the elastic bending force by a ``viscous bending'', namely, $B \to \eta h^3 \pd_t$. We thus obtain: 
\begin{equation}
    -\sigma \kappa + \eta h^3 \pd_t \kappa{''} = f_n \ ,     \label{app:viscida}
\end{equation}
Physically, the reason that the 
viscous bending force does not depend on the curvature $\kappa$ (but only on its gradients) is identical to the reason that the viscous stress depends on gradients of the tangential velocity but not on the velocity itself. In both cases, it is the mutual shearing of liquid layers that generates mechanical resistance rather than net (rigid-body like) motion of the liquid body. 

\paragraph*{{$1^{st}$ FvK equation:}} Turning back to the case of an elastic sheet, we note that Eq.~(\ref{app:elastica}) 
is readily generalized to non-planar deformations, {\emph{i.e.}} $\vec{X}(x,y) = (x,y,z(x,y))$ with $|\nabla z|\ll 1$, where the stress and curvature are now second rank tensors, rather than scalar functions, defined on a surface. The product $\sigma \kappa$ becomes a scalar product of two tensors, and the second derivative of the curvature is replaced by the Laplacian of the trace of the curvature tensor. Consequently, normal force balance is: 
\begin{equation}
    -\sigma_{ji} \kappa_{ij} + B \nabla^2 \kappa_{ii} = f_n \ \ ,  \ \ \text{where:}   \ 
    \kappa_{ij} \approx \pd_{ij} z  \ , 
    \label{app:1FvK}
\end{equation}
and the stress tensor $\sigma_{ij}$ is obtained by solving the $2^{nd}$ FvK equation (that describe in-plane force balance, akin to~\eqref{app:Stokes-Poisson}).  

\paragraph*{Equations~(\ref{eq:Howell-FvK-sym-1}):}
The formal relation between Eqs.~(\ref{eq:Howell-FvK-sym-1}) of the main text and (\ref{app:1FvK}) is identical to the relation between the {\emph{viscida}}, Eq.~(\ref{app:viscida}) and the {\emph{elastica}}, Eq.~(\ref{app:elastica}). Namely, the elastic stress tensor $\sigma_{ij}$ is replaced by the liquid stress, which consists of capillary and viscous contributions according to Eq.~(\ref{eq:stress-strain-rate}), and is found by solving Eq.~(\ref{eq:Howell-FvK-sym-2}) of the main text. Additionally, the elastic bending force is replaced by a viscous bending force, $\eta h^3 \pd_t \nabla^2 \kappa_{ii}$. The resulting equation can be written as: 
\begin{equation}
    -\sigma_{ji} \kappa_{ij} + B \pd_t \nabla^2 \kappa_{ii} = f_n(t) \ \ ,  \ \ \text{where:}   \ 
    \kappa_{ij} \approx \pd_{ij} z  \ , 
    \label{app:1FvK-n}
\end{equation}
Equation~(\ref{eq:Howell-FvK-sym-1}) of the main text, where $f_n(t) = -\Delta P(t)$, is the axisymmetric realization of this equation.\\ 

\subsection{Comparison with previous versions}
To our knowledge, the first attempt to derive equations of motion for the mid-surface of a thin film of viscous liquid (Stokes) flow, in the presence of Gaussian curvature, was done by Howell~\cite{Howell1996}. The difference between \eqref{app:Stokes-Poisson} and its counterpart in Ref.~\cite{Howell1996} is the second term in the square brackets on the RHS. As we pointed our above, this difference is related to our inclusion of the effect of the temporal variation of thickness, $\pd_t h$, implied by mass conservation (Eq.~\ref{eq:ddiv}), on the in-plane strain rate tensor.

Another related paper is Ref.~\cite{Mahadevan2010}, which addressed the viscous dynamics of a pressurized bubble slightly perturbed from its equilibrium hemispherical shape. 
In this case, 
the shape of the mid-surface is $z(r,t) = z_{sph}(r) + \delta z(r,\theta,t)$, with $|\nabla \delta z| \ll |\nabla z_{sph}|$, and the surface dynamics in Ref.~\cite{Mahadevan2010} is obtained by expanding  Eqs.~(\ref{app:1FvK-n},\ref{app:Stokes-Poisson}) to linear order in $\delta z$.      



\subsection{Spherically symmetric solution}
Assume now a spherical, free-standing bubble, where the film's mid-surface has radius $R_0$ and thickness $h_0 \!\ll \! R_0$, and assume that the interior gas's pressure is suddenly suppressed at $t=0$ from $P_a + 4\gamma/R_0$ to the ambient pressure $P_a$. We consider  a non-inertial, volumetrically-incompressible, spherically symmetric solution to this dynamics.

Following Ref.~\cite{Vandefliert1995}, one may solve this problem by considering spherically-symmetric, volumetrically-incompressible  flow in the \textit{bulk} of the film, $\vec{v} = \dot{R}\rho^{-2} \hat{\rho}$ (where $\rho \in [R(t)-h(t),R(t) + h(t)]$ is the distance of a fluid element in the film from the bubble's center) and determine the mid-surface shrinking rate, $\dot{R}$, by employing free-surface boundary conditions at the  
internal/external surfaces, $\rho = R(t) \pm h(t)$.  (We thank P. Howell for pointing out this simple solution of spherically symmetric dynamics). 
For our purpose it suffices
to solve this problem within 
a 
small-slope description of an axisymmetric mid-surface dynamics, as given by  Eqs.~(\ref{eq:Howell-FvK-sym}) of the main text. 
We can consider a small patch of an evolving spherical 
surface 
of radius $R(t)$ 
as a paraboloid, $z(r,t)  = a(t) + r^2/2R(t) + O(r/R)^2$, where $r$ is now the distance from the axis that connects the sphere center to the center of the small patch (such that $r \!\ll\! R(t)$). The small-slope approximation $|\nabla z| \!\ll \! 1$ now amounts to $r/R(t) \! \ll \!1$. The Gaussian and mean curvatures, evaluated through the expressions in~\eqref{eq:Howell-FvK-sym-2b} of the main text, are, respectively, ${\cal R} = R^{-2}$, 
and ${\cal H} = 
R^{-1}$,  
and the normal velocity is: $v_n = \pd_t z \sqrt{1+ (\pd_r z)^2}  = \dot{a} +\tfrac{1}{2}(\dot{a} - \dot{R}) (r/R)^2 + O(r/R)^4$, such that   $\rhodyn = \rhodyn^{(ext)} + \rhodyn^{(int)} = R^{-3} (4\dot{R} + 2\dot{a})/3  \ + \ O(r/R)^2$.
Since
a necessary condition for a stress-free dynamics (as implied by normal force balance, see main text), is $\rhodyn=0$, we obtain the axisymmetric dynamics of the paraboloid, $z(r,t)  = -2R(t) + r^2/2R(t)$, as the small-slope version of the stress-free spherically-symmetric solution. We remind the reader that such a ``uniformly shrinking'' paraboloid does not satisfy the effective clamping at the meniscus.     




\section{Boundary conditions and thermodynamics}
\label{app:BC}
The BCs for the surface dynamics, Eqs.~(\ref{eq:Howell-FvK-sym}) of the main text, consist of a subset imposed at $r \to 0$, Eq.~\eqref{eq:BC-hom-1} of the main text, 
%
and at the meniscus, $r \to 1$,Eqs.~\eqref{eq:BC-hom-2} and~\eqref{eq:BC-z} of the main text. Additionally, we have the thermodynamic constraint, Eq.~\eqref{eq:1Thermo},  that determines disclination charge $q(t)$ in the non-homogenous BC~\eqref{eq:BC-phi-2} of the main text.   
   

\subsection{Homogeneous BCs at $r \to 0$} 
At $r\to 0$ we have one non-homogeneous BC~\eqref{eq:BC-phi-2}, on whose rationale we elaborated in the main text, and   
the 3 homogeneous BCs in Eq.~\eqref{eq:BC-hom-1} of the main text, which we discuss here. 

Consider first the BCs on the shape $z(r,t)$ at $r\to 0$. Regardless of  depressurization rate, the absence of an external,  localized distribution of normal force anywhere on the film implies that for any finite thickness ({\emph{i.e.}} $\epsilon >0$) the rate of change of the curvature components is finite anywhere.
Specifically, integrating Eq.~(\ref{eq:Howell-FvK-sym-1}) over a small, 
vicinity of $r=0$ implies that $z(r,t) \approx z_0(t) + b_2(t) r^2 + b_4(t) r^4 +\cdots $, such that two natural BCs are: 
$\pd_r z = 0$ and 
$\pd_{rrr} z = 0$.  

Consider now the stress potential, $\Phi(r,t)$ at the vicinity of $r\to 0$, and express it formally as a sum of 4 solutions to the homogeneous (axisymmetric) bi-Laplacian equation, $\Phi \sim \ cst  \ , \ r^2 \ , \ r^2 \log r \ , \ \log r$ (in addition to possible  non-homogeneous contribution from $\rhodyn (r\to 0)$). 
Among the 4 homogeneous solution only the last one is forbidden, since 
$ \Phi \sim \log r \ \Longrightarrow \ |\srr| \ , \ |\sqq| \ \sim \ r^{-2}$, which imply an infinite rate of kinetic energy dissipated by viscous flow (see below). The BC $\pd_r \Phi \to 0$ allows the 3 other homogeneous solutions and eliminates only the non-physical one.           


\subsection{BCs at $r\to 1$}
At $r\to 1$ we have 4  homognous BC's, given by Eqs.~(\ref{eq:BC-hom-2},\ref{eq:BC-z}) of the main text. Two of them are rather obvious: The condition $\Phi \to 0$ at $r\to 1$ simply sets an arbitrary constant to the potential (gauge invariance, similarly to electrostatic potential), and the condition $z \to 0$ at $r \to 1$ derives from the immobility of the meniscus, on which we elaborated in the main text. Next we discuss the remaining two BCs at $r \to 1$.


For the thin film of a floating bubble the ``boundary'' $r=1$ is in fact a meniscus -- a nearly symmetric elevation of the liquid bath, of size $\sim \ell_c \ll 1$, in both sides of the circle $r=R_0=1$ (see schematic Fig.~\ref{fig:expSetup}a), as if the bubble would have been replaced by a thin, perfectly wetting solid wall protruding vertically from the liquid bath at $r=1$. 
This free-surface structure is dominated by gravity and surface tension, and is thus described by the non-linear Young-Laplace equation \cite{Anderson2006} (assuming viscous stresses due to flow have negligible effect in this region). 
Specifically, the height of the ideal Young-Laplace meniscus is $z_{top} \approx \sqrt{2}\ell_c$, and its width near the top is characterized by a parabolic profile, $w(z) \sim  (z-z_{top})^2/\ell_c$. Since the real meniscus must terminate at a finite  
width $h_0 \!\ll\! \ell_c$, a continuous transition from the film to the meniscus occurs 
through a {\emph{boundary layer}}, of length 
$\ell_{\BL} \sim \sqrt{h_0\ell_c}  \sim \sqrt{\epsilon/Bo}$.
In addition to providing a ``pinning ring'' to the film that resists changes of its initial radius $R_0$, the meniscus acts as a vertically-oriented ``funnel'', through which liquid can flow from/into the the bath. 
Even though one may employ methods of singular perturbation theory to 
carry out a quantitative analysis of the transition between the film and meniscus of large bubbles ({\emph{i.e.}} $Bo \gg 1$), following an analogous study for small bubbles 
(Ref.~\cite{Howell1999}), we will not pursue this approach here. Instead, we will obtain BCs for the film at the {\emph{interior vicinity}} of the meniscus, ({\emph{i.e.}} $|r-\ell_c| \sim  \ell{\BL}$) by assuming that vertically-oriented funneling of liquid into/from the film in the vicinity of the initial meniscus (of radius $R_0=1$) persists throughout the depressurization process, even if the meniscus size may deviate from $\ell_c$ (due to viscous stress) and so does the precise structure of the meniscus profile. Experimental observations clearly support this assumption (see {\emph{e.g.}} SI movies 1-3 in Ref.~\cite{Oratis2020}).

As long as liquid can flow freely through the meniscus, the momentum flux at $r\to 1$, that is $2\pi r \sigma_{rr}({r\to 1})$, 
is not restricted by the liquid bath. 
Mathematically, we express this physical condition by a
\textit{homogeneous} equation that involves derivatives of the stress potential $\Phi$ at $r\!\to \!1$ and is satisfied by the
initial (pressurized) state of the bubble, where the stress is given everywhere by the isotropic surface tension. 
A general BC of this type is: 
\begin{equation}
\text{at} \ r \to 1: \ \ c_1\left(\pd_r \Phi - \pd_{rr} \Phi \right)  + c_2 \cdot \pd_{rrr} \Phi  = 0   \ , 
\label{app:general-BC}
\end{equation}
where $c_1,c_2$ are arbitrary (and may be in principle functions of $t$, and depend on the dimensionless parameters $\epsilon$ and/or $T$). 
The BC $\pd_{rrr}\Phi \to 0$ in Eq.~\eqref{eq:BC-hom-2} of the main text   
is realized by choosing 
$c_1=0$. 
We
verified that the evolution of the bubble collapse is insensitive to other choices. 

The remaining BC at $r\to 1$ pertains to the flux of angular momentum, that is the torque exerted by the meniscus on the liquid film. 
Such a torque exists since the steady, vertical orientation of the meniscus implies a sharp transition in the tangent direction of the mid-surface at the vicinity of $r\approx 1$. We emphasize that this is a real physical effect and not merely a mathematical artifact of the small-slope approximation that we employed intensively for describing the surface dynamics. Indeed, one may view this situation as a viscous analogue of peeling a thin elastic sheet off a rigid adhessive substrate, in which case the torque exerted by the substrate on the sheet gives rise to discontinuity of the curvature, $[[\kappa]] \propto \sqrt{\sigma/B}$, at the peeling front, such that the tangent direction undergoes a finite, $O(1)$ variation over a ``bendo-capillary'' distance, $\sqrt{B/\sigma}$ \cite{Obreimoff1930}. 
%
Focusing on the normal force balance, Eq.~(\ref{eq:Howell-FvK-sym-1}) of the main text, we can 
pursue further the elastic analogy by noticing that in the vicinity of the meniscus, where the radial component of the curvature tensor is large, force balance must be dominated by terms associated with it. 
Consequently, at the boundary layer, $|r-1|\lesssim \ell_{BL}$, where we expect the stress components to vary smoothly,  
it is possible to replace Eq.~(\ref{eq:Howell-FvK-sym-1}) by the \textit{viscida}, Eq.~(\ref{app:viscida}), where $\sigma_{rr} \to \sigma(t)$,  $\kappa_{rr} \to \kappa(s,t) = \pd_s \cos^{-1} [[{\bf t}_r \cdot \hat{r}]]$, and $s$ is an arclength parameter of the surface along the tangent direction ${\bf t}_r$. Importantly, the one-dimensional \textit{viscida} does not rely on a small-slope approximation and one may thus employ it to interpolate between the bulk of the film (where  $|{\bf t}_r -\hat{r}| \sim O(g^2) \ll 1$) and the meniscus 
(where  $|{\bf t}_r \to -\hat{z}$). 
Integrating over the boundary layer, we obtain a relation between the jumps incurred by the curvature and tangent angle upon transitioning from the film to the meniscus:
\begin{equation}
 r\to 1: \ \  
 \epsilon^2 \cdot \partial_t \pd_s {\cal H}
 \ + \ 
 c_3 \ell_{\BL} \sigma \cdot [[\cos^{-1}({\bf t}_r \cdot \hat{r})]] \approx 0
  \label{app:BC-22}  
  \end{equation}
where the numerical value of $c_3$ must be determined by carrying out an asymptotic matching of the film and the meniscus.   
Going back to the small-slope approximation for the inner part of the bubble, Eq.~(\ref{app:BC-22}) becomes:
\begin{equation}
    \mbox{at} \ r \to 1:~
     \epsilon^2 \partial_t (r^{-1} \pd_{rr}z + \pd_{rrr} z)   \ + \  c_3  \ell_{\BL}\pd_r\Phi  \cdot (\pd_r z+1) \ \approx \ 0.
    \label{app:general-BC-2}
\end{equation}
While the second term in~\eqref{app:general-BC-2} may dominate during the rapid depressurization period (at which time it implies a ``freezing'' of the meniscus in its original shape), the fact that the film must become stress-free makes this term irrelevant after this short period. Similarly, since the third derivative of a parabolic shape vanishes identically, we retain only the first term in the temporal derivative (associated with variation of the azimuthal curvature at the vicinity of the meniscus),           yielding the BC~\eqref{eq:BC-z} of the main text. {We verified that the evolution of the bubble collapse is insensitive to modifying this choice, see Fig. \ref{fig:bcComp2}.}

For the example of a non-floating bubble (schematic Fig.~\ref{fig:expSuggest} of main text), the static ring at $r=1$ implies that the tangential fluid velocity vanishes there (where we assume standard ''no-slip'' at the solid-liquid contact). 
With the aid of Eqs.~\eqref{eq:stress-strain-rate} and \eqref{eq:continuity-00} of the main text, this condition can be expressed as an \textit{inhomogeneous} BC on the stress potential:
\begin{equation}
    \text{at} \  r \to 1: 
    \pd_{rr} \Phi  = 1 \ , 
\end{equation}
which replaces the homogeneous BC~\eqref{app:general-BC}. 
\subsection{Heat production and release of surface energy} 
\label{appsubsec:heat-surface}

In the main text we invoked the $1^{st}$ law of thermodynamics through~\eqref{eq:1Thermo},which determines the disclination charge $q(t)$ in~\eqref{eq:BC-phi-2} of the main text. Here we express the heat production rate $P_{vis}$, and the release of surface energy $\dot{E}_{surf}$ in terms of the functions $\Phi(r,t)$ and $z(r,t)$ that define the axisymmetric dynamics.    

Consider a small volume element inside a free-surface film, $-h/2<z<h/2$, of   
viscous, volumetrically-incompressible, axisymmetric flow. Ignoring the heat production due to temperature gradients, the heat production rate in this volume an be written as    
$ 2\eta (\dot{\varepsilon}_{rr}^2 +\dot{\varepsilon}_{\theta\theta}^2  + \dot{\varepsilon}_{zz}^2)  \ ,   
$
(see \textit{e.g.} Eq. 49.6 of Ref.~ \cite{landau1987course}). Substituting $\dot{\varepsilon}_{zz} = - (\dot{\varepsilon}_{rr} + \dot{\varepsilon}_
{\theta\theta})$ and integrating over the thickness we obtain the heat prodution rate per area of the film: 
\begin{equation}
  p_{vis} = 4\eta h  (\dot{\varepsilon}_{rr}^2 +\dot{\varepsilon}_{\theta\theta}^2  + \dot{\varepsilon}_{rr}\dot{\varepsilon}_{\theta\theta})  \ , 
  \label{app:heat-prod-1}
\end{equation}
The heat production per area of a liquid film in a free-surface film with curved mid-surface ($z(r;t) \neq 0$) is readily obtained from~\eqref{app:heat-prod-1} by using the corresponding expressions for an axisymmetric strain rate components in~\eqref{app:strain}, namely, $\dot{\varepsilon}_{rr} = \pd_rv_r + \tfrac{1}{2} \pd_t(\pd_r z)^2 \ ; \ \dot{\varepsilon}_{\theta\theta} = v_r/r$. Incorporating the viscous flow due to temporal variation of the thickness requires us to revise the above calculation by substituting: 
$\dot{\varepsilon}_{zz} \to \dot{\varepsilon}_{nn} = - (\dot{\varepsilon}_{rr} + \dot{\varepsilon}_{\theta\theta} + 2v_n{\cal H})$, such that we obtain:
\begin{equation}
  p_{vis} = 4\eta h  \left(\dot{\varepsilon}_{rr}^2 +\dot{\varepsilon}_{\theta\theta}^2  + \dot{\varepsilon}_{rr}\dot{\varepsilon}_{\theta\theta} + 2(v_n{\cal H})^2 + 2 v_n{\cal H}(
  (\dot{\varepsilon}_{rr} + \dot{\varepsilon}_{\theta\theta} 
  \right)  \ , 
  \label{app:heat-prod-2}
\end{equation}
where again, $\dot{\varepsilon}_{rr} = \pd_rv_r + \tfrac{1}{2} (\pd_r z)^2 \ ; \ \dot{\varepsilon}_{\theta\theta} = v_r/r$. The total heat production rate is obtained upon integrating over the whole surface: 
\begin{equation}
    P_{vis} = 2\pi \int_{r=0}^{1} p_{vis} \ rdr \ .  
    \label{app:heat-prod-3}
\end{equation}
The strain rate components, $\dot{\varepsilon}_{rr}$ and $\dot{\varepsilon}_{\theta\theta}$, along with the mean curvature ${\cal H}$ and normal velocity, $v_n \approx \pd_t z$, are given in terms of $\Phi(r,t)$ and $z(r,t)$ through Eqs.~(\ref{eq:stress-strain-rate},\ref{eq:Howell-FvK-sym-2b}) of the main text. 

The surface area due to the (two faces of) an axisymmetric surface $z(r,t)$ with surface energy $\gamma$, is given by  
\begin{equation}
\dot{E}_{surf} = 2\gamma \dot{A} \ \ ; \ \  
\dot{A} = 
2\pi \ \pd_t \int_{r=0}^1 \sqrt{1 + (\pd_rz)^2} \ r dr \approx \pi \  \pd_t \int_{r=0}^1 (\pd_rz)^2 \ rdr 
\label{app:Surf-1}
\end{equation} \\

We note in passing that Eqs.~(\ref{app:heat-prod-1}-\ref{app:heat-prod-3}) rule out the homogeneous mode of the stress potential, $\Phi \sim \log r$ at $r\to 0$, thus justifying the BC $\pd_r \Phi \to 0$ at $r \to 0$ on which we elaborated above (part A of this section). Indeed, for $\Phi \sim \log r$ we would have $|\dot{\varepsilon}_{rr}|,|\dot{\varepsilon}_{\theta\theta}|  \sim r^{-2}$, and consequently $P_{vis} =\infty$.     
 
\section{Detailed description of the axisymmetric front dynamics}
\label{app:detailed}
In this section we provide additional details regarding the front propagation. We begin by deriving the general set of equations for the ``auxiliary'' model, as we named it in the main text, with $\delta_{ext} > 0$. Then, we construct the solution for the ``intrinsic'' model with $\delta_{ext} = 0$, see \eqref{eq:int_ext}. Finally, we give further analytic and numerical arguments as to why the intrinsic model captures well the dynamics of the ``full`` model, with $\delta_{ext} = 1$, by analyzing the auxiliary model at short times and small $\delta_{ext}$.

Our starting point is the front solution, Eq. \eqref{eq:front-def}, reproduced here for clarity:
\begin{subequations}
  \label{eq:front-def-sup}
\begin{gather}
  \Phi  \approx \Theta\left(r_f(t) - r\right)\Phi_{\dis} (r,t),   \\
    z  \approx z_f(t) + \Theta \left(r - r_f(t)\right)z_{\ann} (r,t).
\end{gather}
\end{subequations}
Eq. \eqref{eq:front-def-sup} is valid at the limit $\epsilon\to 0$, which we take throughout this section. The fundamental electrostatic-like 
equation of our system, namely the in-plane force balance Eqs. \eqref{eq:elec-0} - \eqref{eq:Er-def}, as well as the disclination form of the potential, \eqref{eq:Phi-dis}, retain this form 
in both the intrinsic and auxiliary models, including 
the ``full'' model $\delta_{ext}=1$. For convenience, we rewrite it here,
\begin{equation}
\label{eq:gauss-sup}
    \frac{1}{r}\pd_r r E_\mathcal{R} - \rhodyn^{(int)}-\delta_{ext}\rhodyn^{(ext)} = 0,
\end{equation}
where we recall that $\rhodyn^{(int)} = -(2r)^{-1}\pd_r\pd_t(\pd_r z)^2, \rhodyn^{(ext)} = (3r)^{-1}\pd_r r \pd_r \mathcal{H}(\pd_t z)$. We now integrate Eq. \eqref{eq:gauss-sup} twice to obtain equations for the electric field $E_\mathcal{R}$ and electrostatic potential $V_\mathcal{R}$, which are dictated by $\Phi$. Integrating once we find,
\begin{equation}
\label{eq:gauss-q-sup}
    \frac{q(t)}{3 r} =  + E_\mathcal{R}(r,t) + P_\mathcal{R}(r,t),  
\end{equation}
where the polarization obeys,
\begin{equation}
    \label{eq:Pr-sup}
    P_\mathcal{R} = P_\mathcal{R}^{(int)}+ P_\mathcal{R}^{(ext)} \equiv\frac{1}{2r}\pd_t (\pd_r z)^2 - \frac{\delta_{ext}}{3}\pd_r\left(\mathcal{H}(\pd_t z)\right).
\end{equation}
Integrating again we obtain,
\begin{equation}
    \label{eq:Vr-sup}
    V(t) + \frac{q(t)}{3}\log(r) = V_\mathcal{R}(r,t) + V_P(r,t),
\end{equation}
where $V_P$ is the contribution to the potential from the polarization, such that

\begin{subequations}
\label{eq:Vp-sup}
\begin{equation}
   V_P = \left\{\begin{array}{ll}
        0 & r < r_f \\
        V_P^{(int)}+ V_P^{(ext)} & r > r_f 
   \end{array}\right.,
\end{equation}
where
\begin{equation}
    V_P^{(int)}= \int_{r_f^{-}}^r \frac{\pd_t(\pd_r z(r',t))^2}{2r'}dr',\quad V_P^{(ext)} = - \frac{\delta_{ext}}{3}\mathcal{H}(\pd_t z).
\end{equation}
\end{subequations}
Here $r_f^{-}$ is situated at $r$ infinitesimally less than $r_f$, i.e. the integral covers the front itself. The constant $V(t)$ is found from the requirement (due to radial force balance at the front) that $\sigma_{rr}=0$ at $r=r_f$. 
Eqs. \eqref{eq:gauss-sup}-\eqref{eq:Vp-sup} 
are akin to
the 
equations 
of Maxwell electric displacement field and associated potential in continuous media. They must be identically obeyed throughout the entire film, and should be insensitive to the ``jump'' at the front. Specializing to the 
front and using Eq.~\eqref{eq:Phi-dis} of the main text sets the potential to 
\begin{equation}
\label{eq:Vt-sup}
    V(t) = 
    \frac{q(t)}{6} \cdot 
    \left\{1  - 2 \log \left(r_f(t)\right) \right\}.
\end{equation}
Furthermore, at the front, $r=r_f$, $V_\mathcal{R}$ jumps from a value of $q/6$ to zero, and hence $V_P$ must \emph{also} jump from zero to $q/6$,
\begin{equation}
    [[V_P]]_{r\to r_f} = \frac{q(t)}{6}.
    \label{eq:Vp-jump-sup}
\end{equation}
As we shall show, Eqs. \eqref{eq:gauss-q-sup}-\eqref{eq:Vp-jump-sup} are enough to obtain the dynamics of the front.

\subsection{The intrinsic model}
\subsubsection{Solution of the intrinsic model}
Let us now simplify to the intrinsic model, $\delta_{ext} = 0$. In that case, Eq. \eqref{eq:Pr-sup} simplifies to the following,
\begin{equation}
\frac{q(t)}{3} =  
 -\frac{1}{2} \pd_t (\pd_rz_{\ann})^2.
 \label{eq:pol-1-sup}
\end{equation}
In addition, the jump in $V_P$ implies
a jump in $(\pd_r z)^2$. This immediately gives the solution of Eq. \eqref{eq:gauss-q-sup} with
\begin{equation}
\label{eq:zAnn-int-sup}
    \pd_r z_{\ann} = -\sqrt{r^2-\tilde{\Sg}(t)}, z_{
    \ann}(r_f) = 0,
\end{equation}
which describes a Seivert surface of constant Gaussian curvature. For convenience, we define: 
\begin{equation}
    \tilde{\Sg} = f_0\Sg, \quad f_0 = \frac{2}{3},
    \label{eq:def2-sup}
\end{equation}
and find that Eqs.~(\ref{eq:gauss-q-sup},\ref{eq:Pr-sup}) imply 
that 
the surface evolves according to
\begin{equation}
\label{eq:Sdot-sup}
    \dot{\Sg}= -q.
\end{equation}
Next, consider Eq. \eqref{eq:Vp-jump-sup}. Using the piecewise form of $z$, Eq. \eqref{eq:front-def-sup}, and evaluating the jump condition, Eq. \eqref{eq:Vp-jump-sup} in an infinitesimal region around the front, we obtain,
\begin{equation}
    \frac{q}{6} = \frac{\dot{r}_f}{2 r_f} (r_f^2 - f_0 \Sg),
    \label{eq:qq-sup}
\end{equation}
which yields 
the solution in the main text, Eq. \eqref{eq:rf-S-intrinsic}. Thus, the jump condition completely determines the dynamics. Fig. \ref{fig:appNumPic1}
depicts the evolution of the intrinsic model. It shows how the slope of $z_\ann$ jumps at the front, and evolves exactly according to the analytic expression, Eq. \eqref{eq:front-def-sup}.

\subsubsection{The disclination charge} 
In the above calculation for the intrinsic model, we considered the disclination charge $q(t)$ as a given temporal function, and determined the surface dynamics, Eqs.~(\ref{eq:zAnn-int-sup},\ref{eq:def2-sup}) through the  
two $1^{st}$ order ODE's~(\ref{eq:Sdot-sup},\ref{eq:qq-sup}). 
Now we turn to determine the disclination charge $q(t)$, and thereby fully determine the dynamics.   

In the main text, we noted that $q(t)$ is determined by the thermodynamic Eq. \eqref{eq:1Thermo}, where the rates of heat production rate, $P_{vis}$, and surface energy release, $\dot{E}_{surf}$, are given in Subsec.~\ref{appsubsec:heat-surface} of the SI.    
As we pointed out in the main text, in the intrinsic model the effect of temporal variation of the film thickness on the mechanics is ignored altogether, and therefore we use $p_{vis}$, Eq.~\eqref{app:heat-prod-1}, to compute the total heat production rate, Eq.~\eqref{app:heat-prod-3}.  The strain rates $\dot{\varepsilon}_{rr},
\dot{\varepsilon}_{\theta\theta}$, are computed from $\Phi_{dis}$, \eqref{eq:Phi-dis}), and the stress-strain rate relation, Eq.~\eqref{eq:stress-strain-rate} of the main text, with $\srr = r^{-1}\pd_r\Phi$ and $\sqq = \pd_{rr}\Phi$, yielding: 
\begin{equation} \dot{\varepsilon}_{rr}  = 
\tfrac{1}{6}\left\{-2 + \Theta(r-r_f)(\log \tfrac{r}{r_f} -1) \right\}  \ \ ; \ \ 
\dot{\varepsilon}_{\theta\theta}  = 
\tfrac{1}{6}\left\{-2 + \Theta(r-r_f)(\log \tfrac{r}{r_f} +2) \right\}   \ , 
\label{app:strain-int}
\end{equation} 
where $\Theta(x)$ is the Heaviside function. For the variation rate of surface energy $\dot{E}_{surf} = 2\gamma \dot{A}$, Eq.~\eqref{app:Surf-1}, we find using Eq.~\eqref{eq:zAnn-int-sup}: 
$$\dot{A}  \approx \pi 
\left((\tfrac{2}{3} S - r_f^2)r_f\dot{r}_f +\tfrac{1}{3}(1-r_f^2) q\right)  .  
$$ 
Substituting the above expressions in Eqs.~(\ref{app:heat-prod-3},\ref{app:Surf-1}) and Eq.~\eqref{eq:1Thermo} of the main text, we obtain: 
\begin{equation}
  \label{eq:q-thermo}
  q(t) = (r_f)^{-2}\left(\sqrt{1-4r_f^2} - 1\right) 
\end{equation}
At short times, where $r_f \!\ll \!1$, we obtain $q \approx -2$. \\
For the full model, we do not have an analytic expression, but the similarity of the dynamics (see below) 
suggests that Eq. \eqref{eq:q-thermo} remains a reasonable approximation.

However, Eq. \eqref{eq:1Thermo} 
is difficult to implement numerically, due to its inherently nonlocal form. Hence, in our numerical analysis we did not evaluate it explicitly, but rather replaced it by a \emph{local} constraint in which $q(t)$ is determined at the core, 
$r = \ellcore$, namely
\begin{equation}
  \label{eq:finite-stress}
  \sigma_{rr}(\ellcore) = 1 \Rightarrow q(t) = -2\log(r_f/\ellcore).
\end{equation}
Numerically, we find $\ellcore = 3a_0$, where $a_0$ is the numerical mesh constant (see Sec. \ref{app:numerics}). Clearly, Eq. \eqref{eq:finite-stress} and \eqref{eq:q-thermo} yield essentially the same dynamics, 
as long as $r_f$ is small but larger than $\ellcore$ (which condition is always satisfied). The figures in the manuscript all use the local constraint above.
\subsubsection{The velocity field}
It is useful to employ the analytic solution we obtained for the intrinsic model, $\delta_{ext}=0$, to address also corresponding velocity field. 

Inspecting the strain rate components that we evaluated above, Eq.~\eqref{app:strain-int}, we note that the presence of a disclination-like singularity at $r\to 0$ must come in tandem with a generalized version of the kinematic relation between velocity and strain rate, Eq.~\eqref{eq:strain-disp} of the main text, which includes 
a {\emph{velocity independent}} contribution to the hoop strain rate rate: 
\begin{gather}
\dot{\varepsilon}_{rr} = \pd_r v_r + \tfrac{1}{2} \pd_t (\pd_z^2)  \ \ ; \ \  
\dot{\varepsilon}_{\theta\theta} = v_r /r  + D \cdot q 
\label{eq:strain-field} \\
\text{where:} \ \ D = \left\{ 
\begin{array}{cc}
   {8}/{3}  \ &  \ r<r_f    \\
    {1}/{3}  \ & \  r>r_f  \ 
\end{array}
\right.  
\label{eq:D-0}
\end{gather}
such that the fluid velocity is: 
\begin{gather}
    v_r = - \frac{1}{3} r \cdot \left\{ 
    \begin{array}{cc}
     1 -  2 q  (\log \tfrac{r}{r_f}-2)  \   &  \ r<r_f   \ , \\
       (1+q)   \ & \ r> r_f 
    \end{array}
    \right.
    \label{eq:vel-field}
\end{gather}
Notice that the  
velocity field is discontinuous at $r_f$ ({\emph{i.e.}} undergoes an $O(1)$ variation over 
{the scale
$\ell_{\BL} \sim \sqrt{\epsilon}$, 
such that its average value $\langle v_r \rangle_{r_f}$ and jump $[[v_r]]_{r_f}$ are: 
\begin{equation}
\langle v_r \rangle_{r_f} = r_f (-\frac{1}{3} + \frac{1}{2} q) \ \ ; \ \ [[v_r]]_{r_f} = q r_f   \ .
\label{eq:shock}
\end{equation}
While the above expressions are specific to the analytically-tractable intrinsic model, $\delta_{ext} = 0$, we note that a discontinuity of $v_r$ across the front, $r=r_f$, does not stem from the discontinuity of the slope $\pd_r z$ (which is smoothed our for any $\delta_{ext}>0$), but rather from the discontinuity of $\sigma_{\theta\theta} = \pd_{rr}\Phi$, and consequently $\dot{\varepsilon}_{\theta\theta}$, which characterizes also the physical model, $\delta_{ext}=1$. Consequently, we expect that a sharp variation in the tangential fluid velocity across the front is a real effect.        

We note that the ``revision'', Eq.~\eqref{eq:strain-field}, of the common relationship between velocity and strain rate in a free-surface fluid film is analogous to the revision of the displacement-strain relationship implied by disclination in an elastic sheet. The elastic analogue of a disclination is realized, for instance, by inserting an azimuthal sector of opening angle $\propto q$ 
into a solid disk (Volterra construction \cite{LandauBook1986}), and features an analogous  contribution to the hoop strain, proportional to the disclination charge $q$, similarly to Eq.~(\ref{eq:strain-field}). 
In our surface dynamics, the dynamo-geometrical charge $q(t)$ reflects an axisymmetric ``excess" strain rate of hoops that is not captured by the 2D velocity field $v_r$
\subsubsection{Spatial variation of thickness}
Gradients in the film thickness $h$ imply corrections to Eq.~\eqref{eq:Howell-FvK-sym-2} of the main text. These corrections stem from two sources. The first type, related to Eq.~
\eqref{app:mu-gamma}, originates in the thermodynamic (capillary) stress, 
in Eq.~(\ref{app:stress}), 
$2\gamma \to 2\gamma(1 + h\nabla^2h)$, where $\nabla^2h$ is a small-slope approximation to the curvature of the thickness profile. 
Including this
correction in the above analysis 
yields through the double differentiation of~\eqref{app:strain-2}, an additional term, $2\gamma 
\nabla^4 h$ in Eq.~(\ref{app:Stokes-Poisson}). 
The second type of thickness correction is the viscous stress in~\eqref{app:stress}. 
Upon double differentiation of Eq.~\eqref{app:strain-2}, 
one finds additional terms in Eq.~\eqref{app:Stokes-Poisson} that are proportional to $h^{-1} \nabla^2 h \ \nabla^2 \Phi$ and  
$h^{-1}\nabla h \cdot \nabla(\nabla^2 \Phi)$.
%
%
Employing our solution of the intrinsic model, we will show below that, as long as the initial thickness profile is smooth, with $|\nabla h| \!\ll\! 1$, these thickness gradient terms have negligible effect on the surface dynamics. Strictly speaking, variation of the film's thickness from its initial state occurs only in the true physical model, $\delta_{ext} =1$, which incorporates fluid mass consrvation through Eq.~(\ref{eq:continuity-00}) of the main text. Nevertheless, even for the intrinsic model, $\delta_{ext}=0$, for which analytic expressions of the radial velocity $v_r(r,t)$ and mid-surface shape $z(r,t)$ are given in the preceding sections, we can study the evolution of the thickness $h(r,t)$ through Eq.~\eqref{eq:continuity-00} of the main text as a totally passive scalar function. We argue that the upper bound estimate we obtain below through this analysis remain valid also for the physical model, $\delta_{ext}=1$.      

Consider then the evolution of the thickness field $h(r,t)$, assuming the film thickness is initially uniform, 
$h(r,t\!=\!0) \!= \!\epsilon$ in the dimensionless convention used in the main text, or more generally, $h(r,t\!=\!0) = \epsilon + \Delta h_0(r)$, where
{$\epsilon\nabla^2 \Delta h_0 \! \ll \!1$. }
%
Substituting in Eq.~\eqref{eq:continuity-00} of the main text analytic the expressions for $v_r$ and $\pd_tz$ we obtained above,   
we note that $\pd_t h \!\sim\!O(\epsilon)$, exhibits a logarithmic divergence $\!\sim\! \log r$, which terminates at the 
disclination core, $r\!\sim\! \ellcore$, but is otherwise continuous everywhere except at
the vicinity  
of the moving front, $r\!=\!r_f(t)$, such that $\nabla^2 h \!\sim \!O(\epsilon)$ both at $r\!<\!r_f$ and $r\!>\!r_f$.
Turning to the vicinity of the front, we note that the mid-surface's slope, $\pd_r z$, and fluid velocity $v_r$, undergo
$O(1)$ jump 
across the front, as indicated by 
Eqs.~(\ref{eq:zAnn-int-sup},
\ref{eq:shock}). The continuity equation,Eq.~(\ref{eq:continuity-00}) of the main text, thus yields $\pd_t [[h]]_{r_f} \approx -\tfrac{5}{3}q \epsilon$
and consequently a jump in the thickness across the front, $[[h]]_{r_f} \!\sim\! \epsilon t$. This implies that 
$\nabla^2 h \! \sim \! \epsilon t /\ell_{\BL}^2 \!\sim \! O(t)$ at the front, where we {again} used  
the front width $\ell_{\BL} \!\sim\! \sqrt{\epsilon}$.

We conclude that for an initially smooth profile of the film thickness, the 
corrections to due thickness gradients in the capillary and viscous terms in the stress tensor (\ref{app:stress})
 remain small 
throughout the flattening process ($t \sim O(1)$) and do not affect the surface dynamics. 
We note though that thickness gradients must be incorporated into Eqs.~(\ref{eq:Howell-FvK-sym}), at least locally, into 
in order to describe several aspects of the surface dynamics that are not addressed in this manuscript. An important example, 
is rupture-induced depressurization, where the thickness gradient at the vicinity of the hole's edge must be considered, as was done for a planar film in Refs.~\cite{Brenner1999,Bush2009}. Another important example is in the experiment of Oratis \textit{et al.} \cite{Oratis2020}, where it was shown in \cite{Bartlett2023} that the initial thickness profile is nonuniform due to the slow drainage that occurs when the gas bubble rises to the interface.
\subsection{The auxiliary model}
We now generalize to the case $\delta_{ext}>0$. In contrast to the intrinsic model ($\delta_{ext} = 0$), we do not have explicit analytic expressions for $z_{\ann}$ in this 
case, nevertheless we are able to show that the qualitative behavior of the front remains unchanged. Specifically, 
we find that the
external terms (associated with mean curvature ${\cal H}$) simply act 
to broaden the jump in $\pd_r z_{\ann}$ at $r\approx r_f$ into a boundary layer of width $\sim \delta_{ext}$, such that the form of Eq. \eqref{eq:zAnn-int-sup} remains valid, but only in a ``far-field'' region of the interval $r>r_f$.  
This smoothing of the jump in $\pd_r z_{\ann}$ means that  
the jump in $V_P$, which is dictated by Eq. \eqref{eq:Vp-jump-sup} and is insensitive to the value of $\delta_{ext}$, is now taken up by the external contribution, $V_P^{(ext)}$ in Eq.~\eqref{eq:Vp-sup}.

To see this, let us consider
a front dynamics,  Eq.~\eqref{eq:front-def-sup}, for a given $q(t)$, with $\Phi_{\dis}$ given by Eq.~\eqref{eq:Phi-dis} of the main text, and the 
the following ansatz for $z_\ann$:
\begin{equation}
\label{eq:z-ann-aux-sup}
    (\pd_r z_\ann)^2 = (r^2 - \tilde{\Sg})\Theta(r - r_f^0) +   r G_\delta\left(\frac{r - r_f^0}{\delta_{ext}}\right) \ , 
\end{equation}
where $r_f^0$ the location of the front in the intrinsic model (defined formally through Eq.~\eqref{eq:qq-sup}), 
the first term on 
the RHS is formally identical to the solution of the intrinsic model (Eq.~\ref{eq:zAnn-int-sup}), 
and the 
second term on the RHS desribes a localized boundary layer function $r G_\delta$ of width $\delta_{ext}$, which vanishes at the front, $G_\delta(r_f - r_f^0) = 0$. In addition to 
$G_\delta$ 
the ansatz~(\ref{eq:front-def-sup},
\ref{eq:z-ann-aux-sup}) is characterized by two temporal functions, $r_f(t)$ and $\tilde{\Sg}(t)$. We will show now that for a given value of $\delta_{ext}$ there exist  $\{r_f(t) \ , \tilde{\Sg}(t)\}$ for which the surface dynamics is properly described by the ansatz~\eqref{eq:z-ann-aux-sup}. Furthermore, we will show that
$\{r_f(t) \ , \tilde{\Sg}(t)\}$ can be computed perturbatively by treating $\delta_{ext}$ 
as an 
(artificial) expansion parameter around the analytic solution of the intrinsic model $\delta_{ext} = 0$. 
%
Note that as defined, $G_\delta$ has a discontinuous jump as it crosses $r_f^0$, rather than $r_f$.

Employing the ansatz~(\ref{eq:front-def-sup},\ref{eq:z-ann-aux-sup}), we now analyze the problem at short times, but still well 
after the formation of the front, $T \ll t \ll 1$, where Eq.~\eqref{eq:z-ann-aux-sup} implies that $\tilde{\Sg} \!\ll \!1$.  
We assume, and confirm in a self-consistently manner, that for $\tilde{\Sg} \ll 1$ the vicinity of the perimeter, $r\lesssim 1$, is desribed by the first (``bare'') term on the RHS of Eq.~\eqref{eq:z-ann-aux-sup}.   
%
%
To see this, begin with the jump condition at the front for $V_P$, noticing that Eqs.~\eqref{eq:Vp-sup} and \eqref{eq:Vp-jump-sup} imply,
\begin{equation}
\label{eq:jump-delta-ext-sup}
    [[V_P]]_{r\to r_f} = -\frac{\delta_{ext}}{3}\dot{z}_f [[\pd_{rr}z]]_{r\to r_f},
\end{equation}
yielding,
\begin{equation}
\label{eq:jump-delta-ext-2-sup}
    \left(\pd_{rr}z_\ann\right)_{r\to r_f} = - \frac{q}{2\dot{z}_f\delta_{ext}}.
\end{equation}
In Eq. \eqref{eq:jump-delta-ext-sup} we kept only the highest order spatial derivative, which appears only in $V_P^{(ext)}$. This implies that, in contrast to the intrinsic model ($\delta_{ext} = 0$)
, the jump no longer 
determines the evolution of $r_f$.
Instead, Eq. \eqref{eq:jump-delta-ext-2-sup} implies 
a second BC for $G_\delta$ (in addition to the aforementioned condition, $G_\delta(r_f - r_f^0) = 0$). 

Next we consider the polarization in the vicinity of $r \lesssim 1$, \textit{i.e.} far away from the front. 
By our assumption, $G_\delta$ is negligible there, 
hence we take the intrinsic contribution only. Recalling Eq.~\eqref{eq:gauss-q-sup} with $\ER = 0$ at $r>r_f$, and with $P_\mathcal{R}$ given by \eqref{eq:Pr-sup} with $P_\mathcal{R}^{(int)}$ of similar form as in the intrinsic model, and 
\begin{align}
    P_\mathcal{R}^{(ext)} &= -\frac{\delta_{ext}}{3}\left[(\pd_t z)\pd_r\mathcal{H}+\mathcal{H}\pd_t(\pd_r z)\right] \nn\\
        &\approx +\frac{\delta_{ext}}{3}\pd_t(\pd_r z),
\end{align}
where we used $\mathcal{H}\approx -1$, expanding to leading order in $\Sg(t)$. This leads again to Eq. \eqref{eq:Sdot-sup}, with
\begin{equation}
\label{eq:Sg-delta-ext-sup}
    \tilde{\Sg} = f_\delta \Sg,\quad f_\delta = \frac{2}{3-\delta_{ext}} \ , 
\end{equation}
thereby self-consistently confirming our assumption, that for $\Sg \ll 1$, $G_\delta$ vanishes near $r=1$.

Eq.~(\ref{eq:Sg-delta-ext-sup}) defines $\Sg$ for any $\delta \geq 0$, so that we need only address now $r_f$. 
To this end, 
we consider again the electrostatic-like  potential. Plugging in our ansatz, Eq. \eqref{eq:z-ann-aux-sup}, into Eq. \eqref{eq:Vp-sup}, and  integrating up to the vicinity of $r=1$, we find that the smooth $\log r$ dependence in Eq. \eqref{eq:Vr-sup} is provided by the terms in $V_P$ that depend explicitly on $\Sg$, leaving an equation that involves only $q(t), r_f(t), r_f^0(t)$ (and an additional unknown time-dependent constant from $G_\delta$). Explicitly, the equation for $r_f$ is given by: 
\begin{equation}
    \label{eq:rf0-vp-sup}
    \frac{q(t)}{6} + 
    \frac{q(t)}{3} \log \frac{1}{r_f} = \frac{q(t)}{2}f_\delta \log\frac{1}{r_f^0} - \frac{\dot{r}_f^0}{2 r_f^0} ((r_f^0)^2 - f_\delta \Sg) + \delta_{ext}\int  G_\delta(x) dx,
\end{equation}
where in the last term we used the $\delta_{ext}$ dependence of $G_\delta$ and its localized nature to bound its integral.

Clearly, Eq. \eqref{eq:rf0-vp-sup} determines the dynamics of $r_f$. At 
zeroth order in $\delta_{ext}$, it reduces to the 
intrinsic model's equation of motion for $r_f^0$.
Expanding the equation in powers of $\delta$ (where we ignore a possible weak dependence of $q(t)$ on $\delta_{ext}$),
we obtain a formal expression for 
$r_f$ as successive orders of $\delta_{ext}$: 
\begin{equation}
    r_f = r_f^0 + \delta_{ext}r_f^1 + \cdots
\end{equation}
One may readily verify that the next order term sets $r_f^1 = O(1)$, implying an $O(\delta_{ext})$ correction to $r_f$ as expected.

Fig. \ref{fig:supDeltaSnap} provides numerical confirmation of the arguments we gave above. Fig. \ref{fig:supDeltaSnap}a shows $\pd_r z_\ann$ for different $\delta_{ext}$ at a specific time. The formation of the boundary layer is clear, and remarkably, all the lines cross at a single point, confirming that $\dot{r}_f^0$ is independent of $\delta_{ext}$. Figs. \ref{fig:supDeltaSnap}b and \ref{fig:supDeltaSnap}c compare $\Sg(t)$ and $r_f(t)$ for different $\delta_{ext}$, confirming the form of Eq. \eqref{eq:Sg-delta-ext-sup} and showing that the boundary layer merely shifts $r_f$ relative to $r_f^0$ by approximately a constant of order $\delta_{ext}$.



\section{Singular perturbation theory for the microscopic quantities $T,\epsilon$} \label{app:pert-theory}
In this section we address two elements in the axisymmetric surface dynamics of a rapidly depressurized bubble, where the 
viscous bending term is pronounced  -- the radius $\lstad$ of the flattened nucleus, Eq.~(\ref{eq:ell-adi}) of the main text, and the width $\ell_{\BL}$ of the moving front that separates the flattened zone from the curved periphery, Eq.~(\ref{eq:front-sim}).
In both calculations we employ the analytic solution of the intrinsic model ($\delta_{ext} = 0$).

\subsection{Adiabatically depressurized core}
The size $\lstad$ of the initial ``nucleus'', which marks the zone flattened while the pressure drops, and thereby defines the initial value of $r_f$, is set by the two small parameters, $\epsilon$ and $T$ (Eqs.~(\ref{eq:regime-1},\ref{eq:T}). 
To see this, let us start by following our ``membrane limit'' approach of the main text, setting $\epsilon \to 0$ in Eq.~\eqref{eq:Howell-FvK-sym-1} of the main text. Assuming that at the core zone, $r<\lstad$, the film remains close to 
its initial, uniformly-tensed state (in contrast to the periphery, which remains curved and hence becomes stress-free when the pressure drops), we expand Eqs.~\eqref{eq:Howell-FvK-sym} around the uniformly-tensed initial condition, Eq.~\eqref{eq:IC0} of the main text, to leading order in the deviation, $r^2 \delta(r,t)$, from the initial stress potential. This expansion yields:          
\begin{gather}
  \pd_r \Phi \approx r\left(1 + 2\delta(r,t)\right) \  
  ,   \ \pd_rz \approx - e^{-\tfrac{t}{T}}  r\left(1+\delta(r,t)\right) \ , 
\label{app:no-flow} \\
\text{where} \ \ 
\delta(r,t)  = \tfrac{3}{16}\tfrac{r^2}{T} e^{-\tfrac{2t}{T}}  \ , \nonumber \end{gather}
%
Eq.~\eqref{app:no-flow} shows that the perturbation to the stress remains small (\textit{i.e.} $|\delta| \ll 1$) only for $r\ll \lstad \sim \sqrt{T}$, yielding the initial core size for sufficiently small $\epsilon$.    

Considering depressurization rate, $T$, we apply a self-consistency test to determine whether neglecting 
the viscous bending term in Eq.~\eqref{eq:Howell-FvK-sym-1} of the main text, which underlies Eq.~\eqref{app:no-flow} is justified. 
Estimating the time derivative during the depressurization period as $\pd_t \sim T^{-1}$, and the spatial derivative and slope in the adiabatic region by $\pd_{rrr} \sim \lstad^{-3}$ and $\pd_rz \sim \lstad$, respectively, with $\lstad \sim \sqrt{T}$, we find that viscous bending is negligible in comparison to pressure (which is $O(1)$ at $t <O(T)$), as long as:    
\begin{equation}
    \epsilon^2 |\pd_t \pd_{rrrr} z| \ll |\Delta P|  \ \Longrightarrow \ \epsilon \!\ll\! T \!\ll\! 1 \ . 
    \label{app:ineq-1}
\end{equation}
In the complementary parameter regime, $ T \!\ll\!\epsilon\!\ll\! 1$, the resistance of viscous bending to the formation of a flattened disk in the spherically-shaped film is dominant, and assuming this process occurs over a time longer than $T$ (which assumption must be verified self-consistently) we may neglect now the pressure. 
Assuming again that the stress within the initial flattening core of the film remains close to the original (uniformly tensed) state,  $\sigma_{rr} \approx 1$, Eq.~\eqref{eq:Howell-FvK-sym-1} of the main text now becomes: 
\begin{align}
    \frac{1}{r}\pd_r(rg) &\approx \epsilon^2 \pd_t \frac{1}{r}\pd_r (r \pd_r (r^{-1} \pd_r (rg)))   \nn\\
                           \Rightarrow~g &\approx \epsilon^2 \pd_t \pd_r(r^{-1} \pd_r (rg)) ,  
    \label{app:inv-diff}
\end{align}
where we denote $g = \pd_rz$. 
Interestingly, the formal structure of this equation is an ``inverse diffusion'' (that is, $\pd_t \to \int dt$). This observation
suggests we can approximate the flattening disk via a similarity solution,
in analogy to diffusion of a scalar field: 
\begin{equation}
g(r,t) = -  t^{-1/2}\epsilon F(\xi)  \ \ , \ \ \xi \equiv r t^{1/2}\epsilon^{-1} \ , 
    \label{app:similarity} 
\end{equation}
such that Eq.~\eqref{app:inv-diff} reduces
to an ODE for the similarity function $F(\xi)$: 
\begin{align}
    &(\xi^{-1}-2\xi)F - F' + 2\xi F'' +\xi^2 F'''=0.
              \label{app:similarity-2} 
\end{align}
Eq.~(\ref{app:similarity-2}) has analytic solutions in terms of hypergeometric functions. We identify the physical solution as having $F(0) = 0$, corresponding to $g(r=0,t)$ and a finite $F(\xi\to \infty)$, corresponding to a flattening region, see Eq. (\ref{app:similarity}). Fig. \ref{fig:hypergeo} depicts the result.

  To obtain the scaling of  
  $\lstad,\Tstad$ in this parameter regime (complementary to~\eqref{app:ineq-1}) we note that the above similarity solution of Eq.~\eqref{eq:Howell-FvK-sym-1} of the main text implies one scaling relation: $\lstad \Tstad^{-1} \epsilon^{-1} \sim 1$. A second scaling relation is obtained by requiring compatibility with a solution of Eq.~\eqref{eq:Howell-FvK-sym-2} of the main text. For such a solution the LHS of Eq.~\eqref{eq:Howell-FvK-sym-2} describes an $O(1)$ spatial variation of the stress (between the tensed core and the stress-free periphery) over a length scale $\lstad$, whereas the RHS describes $O(1)$ temporal variation of the Gaussian curvature ${\cal R}$ (from a spherical shape to flat disk) over a time scale $\Tstad$. Matching the RHS with the LHS (and noting that the stress itself is a second derivative of the stress potential), this consideration yields the scaling relation: $\lstad^{-2} \sim \Tstad^{-1}$. Taken together, these two conditions show that $\lstad \sim \sqrt{\epsilon} \ , \ \Tstad \sim \epsilon$, and combining with the scaling obtained above for the parameter regime~\eqref{app:ineq-1}, we obtain 
Eq.~\eqref{eq:ell-adi} of the main text. 

\subsection{Inner structure of the moving front}
As we noted in the main text, 
the viscous bending term becomes dominant at the propagating front, $r \approx r_f(t)$, where the large gradient in $\pd_rz(r,t)$ enables a balance between the two sides of Eq.~(\ref{eq:Howell-FvK-sym-1}) of the main text, thus allowing both $\pd_r\Phi$ and $\pd_rz$ to attain finite values (even through $\Delta P=0$) that interpolate across the Heaviside functions, Eq.~\eqref{eq:front-def} of the main text.

The method of asymptotic matching posits that, when analyzed in terms of the ``inner'' variable $\xi = [r-r_f(t)]/\ell_{\BL}$, the ``inner'' functions, $\Phi_{\BL}$ and $z_{\BL}$, must match the respective ``outer'' functions, $\Phi_{dis}$ 
and $z_{\ann}$, at $\xi \to \pm \infty$. Consequently, Eqs~(\ref{eq:front-def},\ref{eq:Phi-dis})
of the main text indicate that 
$\pd_r\Phi = \sqrt{\epsilon}\Psi_{\BL}$, where $\Psi_{\BL}$ and $z_{\BL}'$ are both $\sim O(1)$. Changing variables from $(r,t) \to \xi$, and recalling that $\ell_{\BL} \ll 1$, we readily note that a balance between the two sides of Eqs.~(\ref{eq:Howell-FvK-sym}) of the main text implies that $\ell_{\BL}\sim \sqrt{\epsilon}$. This scaling is confirmed by the numerical solution, see \textit{e.g.} Figs. \ref{fig:appNumPic1}
and \ref{fig:appNumPic2},
and the discussion regarding them in Sec. \ref{app:numerics}.
In the front vicinity, the PDEs~(\ref{eq:Howell-FvK-sym}) of the main text reduce to a coupled set of ODEs for the functions $\Psi_{\BL} (\xi,t)$ and $g_{\BL} (\xi,t) = z_{\BL}' (\xi,t)$, where the explicit time $t$ acts merely as a parameter (and $(\cdot)'$ indicates derivation w.r.t. $\xi$):  
\begin{subequations}
    \label{app:front-eqs}
\begin{gather}
    r_f^{-1}\cdot (\Psi_{\BL} g_{\BL})' = -\dot{r}_f \cdot g_{\BL}'''  \\
     \Psi_{\BL}''' =  \frac{3}{2} \ \dot{r}_f \cdot (g_{\BL}^2)'   \  , 
\end{gather}
The BCs for these equations, obtained by matching with $\Phi_{dis}$ 
and $g_{\ann}$, Eq.~\eqref{eq:front-def} of the main text, are: 
\begin{eqnarray}
   \xi \to \infty&:& \ g_{\BL} \to -\sqrt{r_f^2 - \tilde{\Sg}} \  \ , \ \ \Psi_{\BL} \to 0 \nonumber \\
    \xi \to -\infty&:& \ g_{\BL} \to 0 \ \  ,\  \ \Psi_{\BL}' \to q/2  
\end{eqnarray}
\end{subequations}
where $r_f,\dot{r}_f,\tilde{\Sg}$ are totally determined by the ``outer'' analysis that was described in the previous section of the SI. 
Thus, for a given set of outer parameters, $r_f,\dot{r}_f,S,q$, a solution of ODEs~\eqref{app:front-eqs}, which one may obtain numerically, fully determines the inner functions, $\Phi_{\BL}$ and $z_{\BL}$. For the sake of brevity, and since the exact solution of these equations is of marginal importance for the purpose of this manuscript, we will not describe it here.


\section{Dynamics of the wrinkled state} 
\label{app:wrinkle}
Relaxing the assumption of axial symmetry (but retaining a small slope approximation), the force balance Eqs.~(\ref{eq:Howell-FvK-sym}) of the main text acquire azimuthal dependence upon replacing $\hL^2_r \!\to\! \hL^2 \!=\!\hL^2_r \!+\! \tfrac{1}{r^2}\pd_\theta^2$, and 
$\tfrac{1}{r} \pd_r (\pd_r \Phi \pd_r z) \!\to\! \sigma_{ij} \pd_{ij} z$, where the stress components $\sigma_{ij}$ are given in terms of the stress potential $\Phi$ 
and the Gaussian and mean curvatures are expressed through the determinant of the curvature tensor. Below we write the equations (written here for the full model, {\emph{i.e.}} 
$\delta_{ext}=1$).     

Substituting the single-mode ansatz~
Eq.~\eqref{eq:s-mode} 
of the main text in this generalized version of the force balance equations, and retaining only $\theta$-independent terms and terms proportional to $e^{im\theta}$, one obtains from Eq.~\eqref{eq:Howell-FvK-sym-1} of the main text two equations 
for the normal force balance: 
\label{app:singlemode}
\begin{flalign}
    r^{-1}\pd_r\left((\pd_r z_0) (\pd_r \Phi_0)\right) +
    \frac{1}{2r}\pd_r(\pd_r\varphi\pd_r\zeta) -
    \frac{m^2}{2r^2}\left[\pd_r^2(\zeta\varphi) - 2\pd_r(r^{-1}\varphi\zeta)\right] =  -\Delta P(t)  +\epsilon^2 \hat{L}_r^2(\pd_t z_0) \label{eq:normalA}\\
    r^{-1}\pd_r\left((\pd_r z_0) (\pd_r \varphi)\right) + r^{-1}\pd_r\left((\pd_r \zeta) (\pd_r \Phi_0)\right) - \frac{m^2}{r^2}(\pd_r^2 z_0 \varphi + \zeta \pd_r^2 \Phi_0) = -\delta P_m(t) + \epsilon^2 \hat{\mathcal{L}}_m(\pd_t \zeta).\label{eq:normalB}
\end{flalign}
Here,
\begin{equation}
    \hat{\mathcal{L}}_m[f] = \left[\hat{L}_r^2 - \frac{m^2}{r^4}\left(4 -2r\pd_r +2r^2\pd_r^2\right) + \frac{m^4}{r^4}\right]f,
\end{equation}
and $\delta P_m(t)$ is an ``angular'' force, which breaks the axial symmetry explicitly, and represents some microscopic fluctuations, {\emph{e.g.}} thermal fluctuations. 
The equations for tangential force balance are: 
\begin{flalign}
\hat{L}_r^2 \Phi_0 = 
\frac{1}{r}\pd_r
\left\{
-\frac{3}{2}\pd_t 
\left(
(\pd_rz_0)^2 + \frac{1}{2}(\pd_r\zeta)^2  - \frac{m^2}{r^2} \pd_{rr}\frac{\zeta^2}{r^2}
\right) 
+\frac{r}{2} 
\left(
\pd_tz_0 
\frac{1}{r}\pd_r(rz_0)
+ \frac{1}{2}\pd_t \zeta
\left(
\frac{1}{r}\pd_r (r\zeta)
-\frac{m^2}{r^2}\zeta
\right)\right)
\right\}
\label{eq:tangentA} \\
    \hat{\mathcal{L}}_m \varphi = 
    -\frac{3}{r} \pd_t 
    \left\{
    \pd_{rr}z_0 \pd_r\zeta + \pd_{rr}\zeta \pd_r z_0 - \frac{m^2}{r^2}
\zeta \pd_{rr}z_0     \right\}
+ \frac{1}{2} 
(-\frac{m^2}{r^2} + \frac{1}{r}\pd_r r \pd_r)
\left[ \pd_t \zeta  \frac{1}{r}\pd_r(rz_0)  + \pd_tz_0 (\frac{1}{r}\pd_r(r\zeta) - \frac{m^2}{r^2} \zeta) 
\right] 
\label{eq:tangentB}.
\end{flalign}
Analogous set of equations have been analyzed for elastic sheets \cite{Davidovitch2012,Davidovitch2019} and shells \cite{Taffetani2017} that undergo wrinkling instabilities above a threshold level of confinement. Even though such equations are derived through an uncontrolled truncation of higher-order harmonics (\textit{i.e.} terms $\propto e^{ikm\theta}$ with $k>1$), it has been shown in these studies that this method 
captures quantitatively the non-perturbative effect of radial wrinkles on the stress in the confined body and the consequent surface dynamics.


\section{Numerical methods}
\label{app:numerics}

In this section we describe the numerical algorithm we used in this work. We also provide details of the analysis and parameters used to generate the quantitative figures in this paper.

\subsection*{The numerical algorithm}

We obtained  all the numerical results reported in this paper by implementing an exact numerical solution of the axisymmetric equations, Eqs. \eqref{eq:Howell-FvK-sym-1} and \eqref{eq:Howell-FvK-sym-2}. 
The equations form a set of nonlinear coupled differential-algebraic equations (DAEs) for the fields $z(r,t),\Phi(r,t)$.

Our equations of motion are prescribed in terms of the scalars $\Phi,z$ rather than the currents or velocities. In addition, Eqs. \eqref{eq:normalA}-\eqref{eq:tangentB} only involve even powers of spatial derivatives. Hence it is natural to define the numerical difference equations on the numerical lattice sites, rather than on the bonds as is common in hydrodynamic difference schemes. We discretized all differential operators symmetrically, balancing forward and backwards difference operators. For example, the operator $\hat{L}_r^2$ is implemented as,
\begin{equation}
    \hat{L}_r^2[f_n] \to r_n^{-1}D_f[r_{n-1/2}D_b(r_n^{-1}D_b(r_{n+1/2}D_f f_n))],
\end{equation}
where 
\begin{equation*}
    D_f(X_n) = \frac{X_{n+1}-X_n}{a_0}, D_b(X_n) = \frac{X_{n}-X_{n-1}}{a_0}
\end{equation*}
are the usual forward/backward differences, $X_{n+1/2} = 0.5(X_n+X_{n+1})$ is the averaged field value on the bond position, and 
\begin{equation}
    r_n = a_0 n, n = 0\ldots N-1, a_0 = (N-1)^{-1}
\end{equation} 
defines the numerical lattice from 0 to 1, where $N$ is the mesh size.


Because we are treating a system of DAEs, it is not possible to integrate the equations directly (either explicitly or implicitly). We solved the equations using the method of lines via Mathematica's DAE solver implemented in the NDSolve function. As a general rule, the equations were stable as long as $\sqrt{\epsilon}/T \lesssim 1$, as can be expected from the $\epsilon^2\pd_t$ structure in Eqs. \eqref{eq:normalA}-\eqref{eq:normalB}.
We did not systematically investigate the stability of the equations. 

\subsection*{Boundary conditions}

As discussed in the main text, the development of the front-disclination pair and the following dynamics of the bubble collapse require non-homogeneous boundary conditions for the Airy stress function. Table \ref{tab:numericBC} details the numerical boundary conditions we used in our simulations.

The Neumann boundary condition for $\Phi_0$ at $r=0$ is denoted as $n_0(t)$, given by
\begin{equation}
    n_0(t) = \frac{a_0}{2}e^{-t/T}.
\end{equation}
The reason for this choice is to interpolate smoothly from the initial conditions, which do not obey Neumann boundary conditions on the discrete lattice, to the final one which does. Leaving the original $O(a_0)$ condition does not change the numerics qualitatively but did affect the stability of the numerics and the behavior near $r=0$. This is true for other boundary conditions as well: changing the boundary conditions did not qualitatively change our results, with the exception that, as discussed in the main text, changing the $r=0$ boundary conditions for $\Phi_0$ to fully homogeneous ones leads to formation of a phantom bubble, rather than a propagating front. As a general rule, changing the conditions by e.g. modifying the constants $1,-1$ to other numbers, or by changing the second derivative constraint on $z(r=1)$ to a first order one, just resulted in the formation of a boundary layer of order $\ell_{BL}$.



\subsection*{Detailed description of the figures and the analysis used to generate them}

We now provide details of our figure production. For each figure, we first give a sketch of the technique used (if relevant), and then provide the relevant parameters.
\subsubsection*{Figure \ref{fig:expSetup}}

In this figure, only panel (b) is a quantitative solution of our equations. The parameters used to generate it are:
\begin{equation}
    \epsilon = \frac{1}{3}\times 10^{-4}, T = 0.01, N=2^{10}, \delta_{ext}=1
\end{equation}
and it is a snapshot of the solution at $t=0.8$
\subsubsection*{Figure \ref{fig:schemSolutions}}
This figure does not use quantitative data, except for the bubble profile in panel (c), which is a snapshot from the same solution as Fig.  \ref{fig:expSetup}b, at $t=0.35$.

\subsubsection*{Figure \ref{fig:solutionPlots}}

In order to generate this figure, we solved the equations numerically for the set of parameters given below.
Panel (a) depicts the solution for the full model, $\delta_{ext}=1$. To generate panel (b), we solved the auxiliary model with $\delta_{ext}=10^{-4}$, i.e. a solution very close to the intrinsic one.
Then, we extracted the positions of $r_f(t),\Sg(t)$ by fitting the numerical $(\pd_r z)^2(r,t)$ to the relevant Sievert surface expression in Eqs. \eqref{eq:front-def-sup} and \eqref{eq:zAnn-int-sup} of the SI and \eqref{eq:Phi-dis} of the main text (we regularized the Heaviside functions by a boundary layer of width $\sqrt{\epsilon}\sim \ell_{BL}$). Fig. \ref{fig:appNumPic1} depicts some typical results of the fitting process, showing excellent agreement with the analytic expressions and the resulting high accuracy of the fit.

The analytic expressions are just those appearing in the text. They do not depend on any fitting parameter except for $\ellcore$. In our fitting we took
\begin{equation}
    \ellcore = 3a_0, \label{eq:ellcoreNum}
\end{equation}
which is exactly the size of the numerical core defined by the boundary conditions above. This choice also appeared, in a qualitative analysis, to give a best fit to the data.

The parameters used to generate this figure are:
\begin{equation}
    \epsilon = 1\times (1/3)\times 10^{-4}, T = 0.01, N = 2^{10}.
\end{equation}

\subsubsection*{Figure \ref{fig:mscale}}

The data in this figure was received directly from the authors of Ref. \cite{Oratis2020}. The values of $R, a_0, \gamma, \eta, \rho$ necessary for the scaling analysis were all taken from the experimental data and can be obtained either from the work itself or from the authors. The four colors blue, yellow, green and red represent different viscosities $\eta = 10, 100, 800, 300$ Pa$\cdot$s, respectively. We obtained each panel in the figure by rescaling the horizontal axis differently; the scaling is given in the label of each panel. We then fit the data to a linear equation using Mathematica's FindFit procedure.

\subsubsection*{Figure \ref{fig:bcComp2}}

The parameters used to generate this figure are:
\begin{equation}
    \epsilon = 1\times 10^{-5}, T = 0.01, N = 2^9, \delta_{ext} = 0.
  \end{equation}

\subsubsection*{Figure \ref{fig:appNumPic1}}
The parameters used to generate panel (a) of this figure are:
\begin{equation}
    \epsilon = 1\times 10^{-5}, T = 0.01, N = 2^9, \delta_{ext} = 0.
\end{equation}
The parameters used to generate panels (b,c) of this figure are:
\begin{equation}
    \epsilon = (1/3)\times 10^{-4}, T = 0.01, N = 2^{10}, \delta_{ext} = 10^{-4}.
\end{equation}
The reasons for using two different datasets were purely technical.

\subsubsection*{Figure \ref{fig:supDeltaSnap}}

In order to generate this figure, we solved the equations numerically for the set of parameters given below, using the $\delta_{ext}$ values shown in the figure panels. Panel (a) shows snapshots at $t=0.6$. To find the parameters $r_f, \Sg$ shown in panels (b,c) we fit first fit $\sigma_{\theta\theta}$ to the analytic expression, as discussed above for Fig. \ref{fig:solutionPlots}. Then we fit $(\pd_r z)^2$ to the intrinsic analytic expression, broadened by a boundary layer with a width obtain as an additional fitting parameter. Fig. \ref{fig:appNumPic2} shows an example of the fitting process, which is clearly very successful, proving the accuracy of our analysis in Sec. \ref{app:detailed}.

The parameters used to generate this figure are:
\begin{equation}
    \epsilon = 1\times (1/3)\times 10^{-4}, T = 0.01, N = 2^{10}, \delta_{ext} = 10^{-4},0.5,1.
\end{equation}


\begin{figure}
    \centering
    \begin{subfigure}{0.49\hsize}     \includegraphics[width=\hsize]{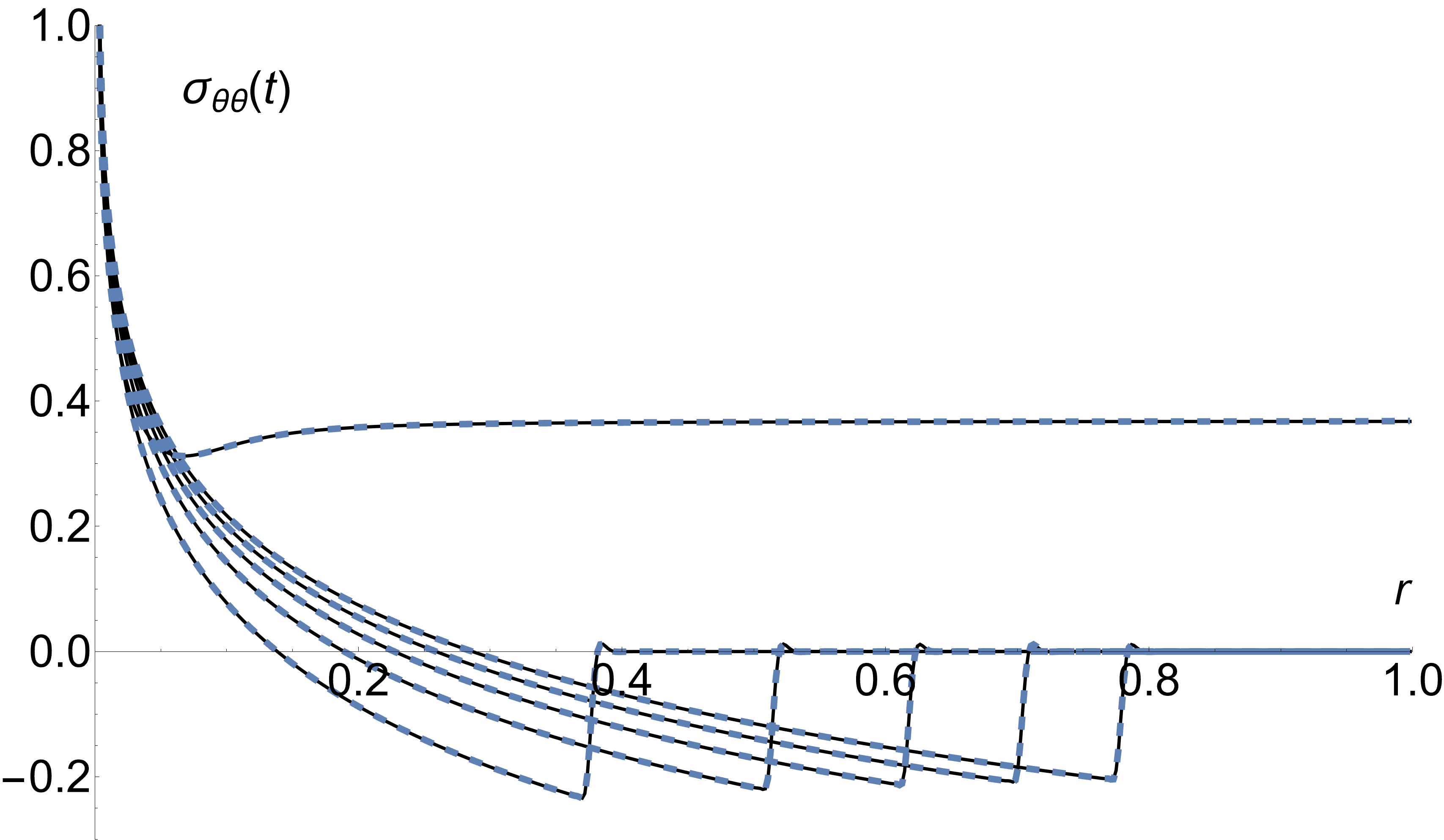}
    \end{subfigure}
    \begin{subfigure}{0.49\hsize}
        \includegraphics[width=\hsize]{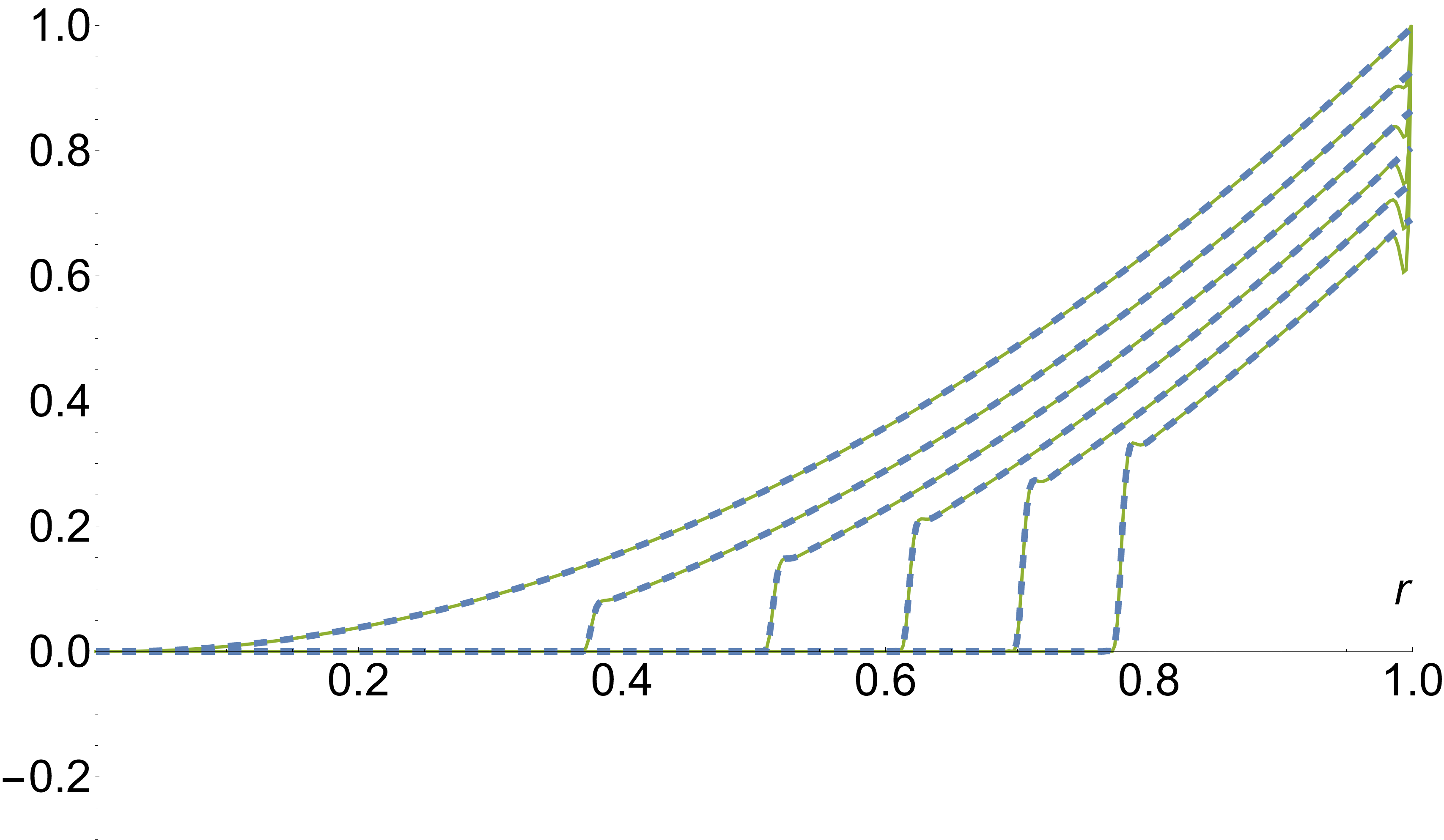}
    \end{subfigure}
    \caption{Insensitivity of the bubble collapse to details of the BCs \eqref{app:general-BC} and \eqref{app:general-BC-2} at $r\to1$  (using the intrinsic model for clarity). (a) The  traces depict the evolution of $\sigma_{\theta\theta}(t)$ for various times in a solution of the axissymmetric equations. The dashed blue lines correspond to $c_1 = 0, c_2 = 1$, which is the choice we made for the numerics in this paper. The solid black line corresponds to $c_1=1, c_2 = 0$. The two sets of data are indistinguishable within the numerical accuracy of the solution. (b) The  traces depict the evolution of $(\pd_r z)^2$ for various times in a solution of the axissymmetric equations The dashed blue lines correspond to $\pd_{rr} z = 0$, equivalent to choosing $c_3=0$, which is the choice we made for the numerics in this paper. The solid green line corresponds to $\pd_r z=-1$, equivalent to choosing $1 \ll c_3$. The two sets of data are indistinguishable within the numerical accuracy of the solution, except for a small boundary layer near $r=1$.}
    \label{fig:bcComp2}
\end{figure}

\begin{figure}
    \begin{subfigure}{0.32\hsize}
        \includegraphics[width=\hsize,clip,trim=0 0 0 50]{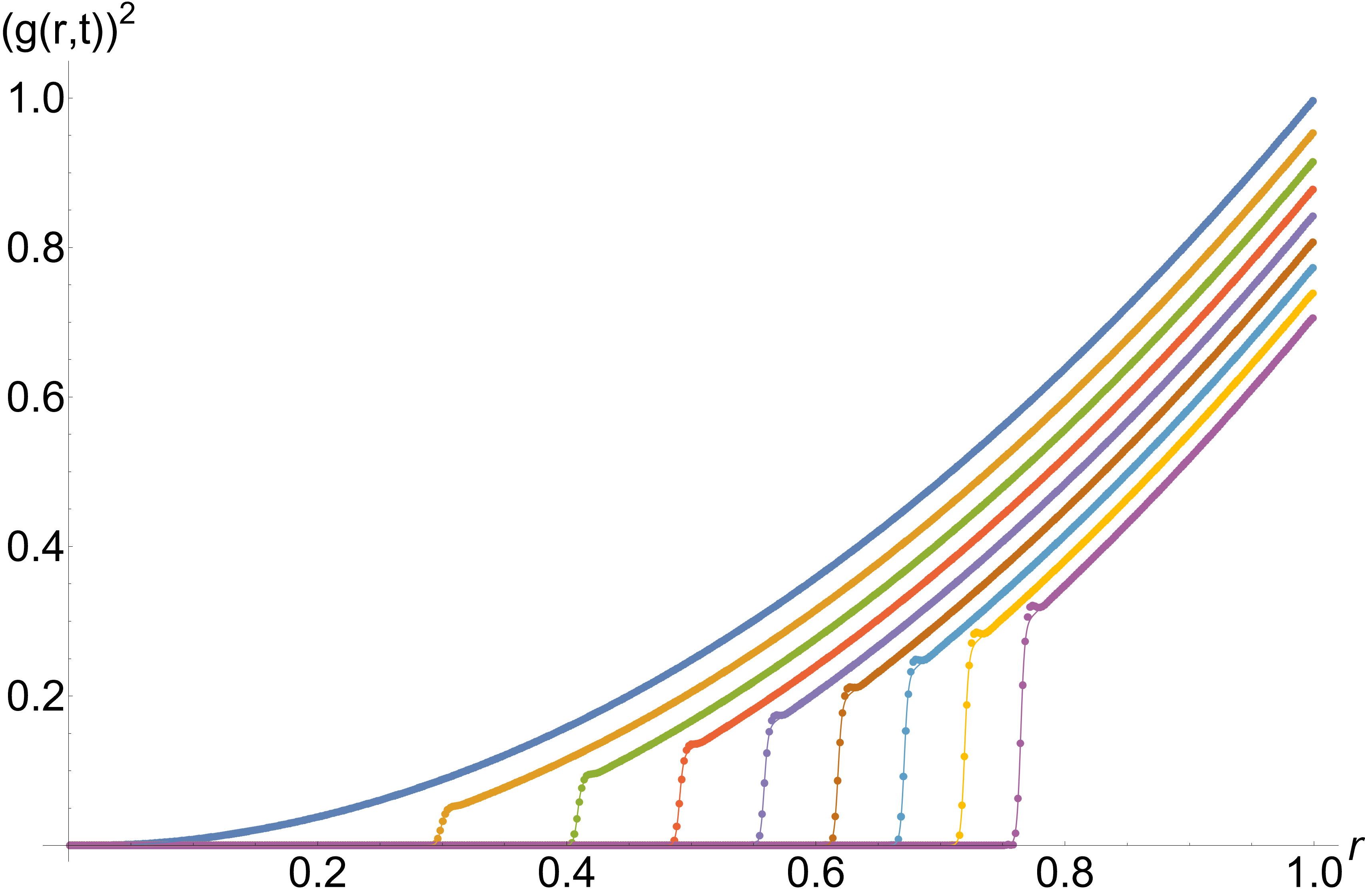}
       \caption{}
    \end{subfigure}
    \begin{subfigure}{0.32\hsize}
        \includegraphics[width=\hsize]{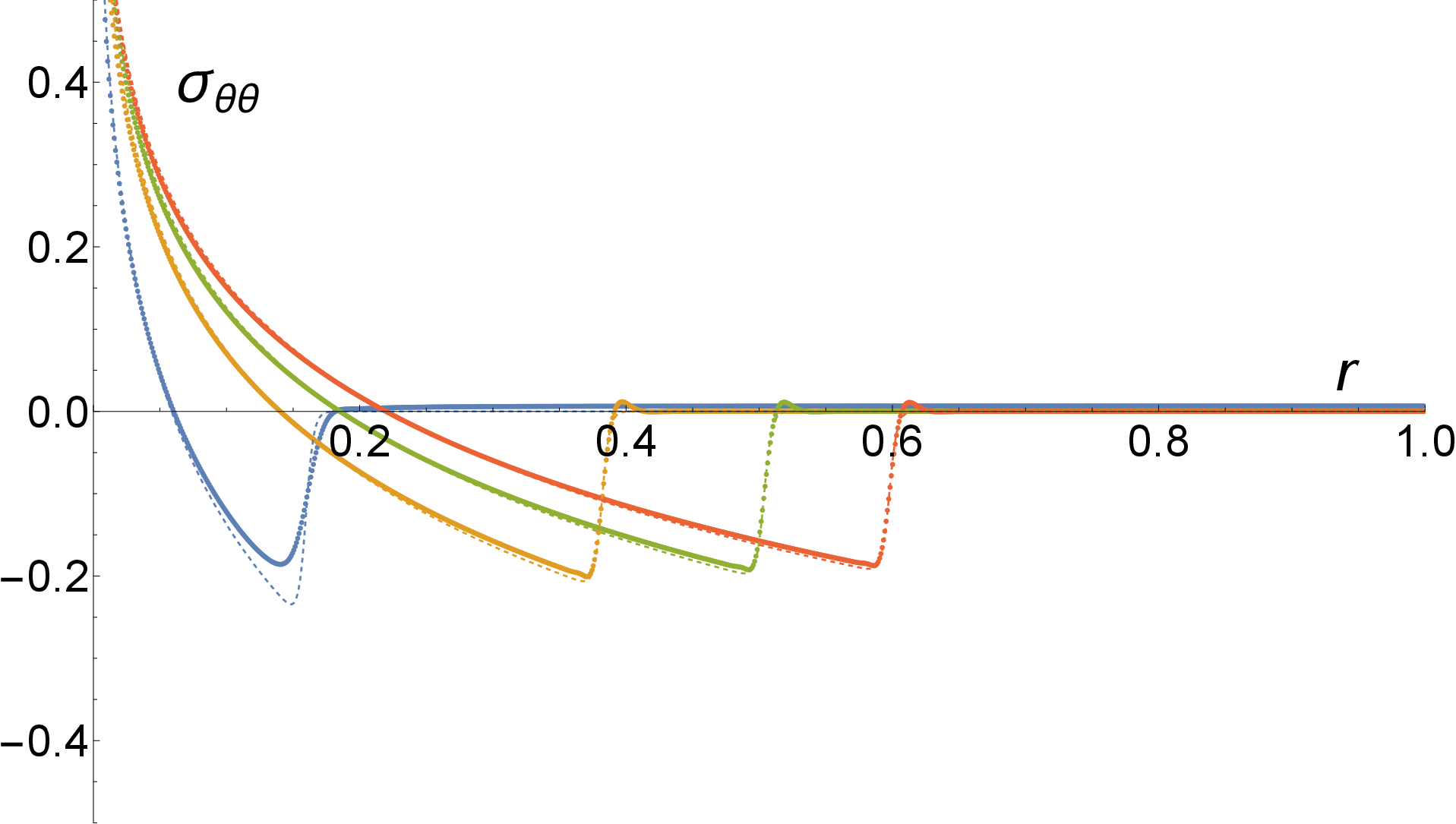}
       \caption{}
    \end{subfigure}
    \begin{subfigure}{0.32\hsize}
        \includegraphics[width=\hsize]{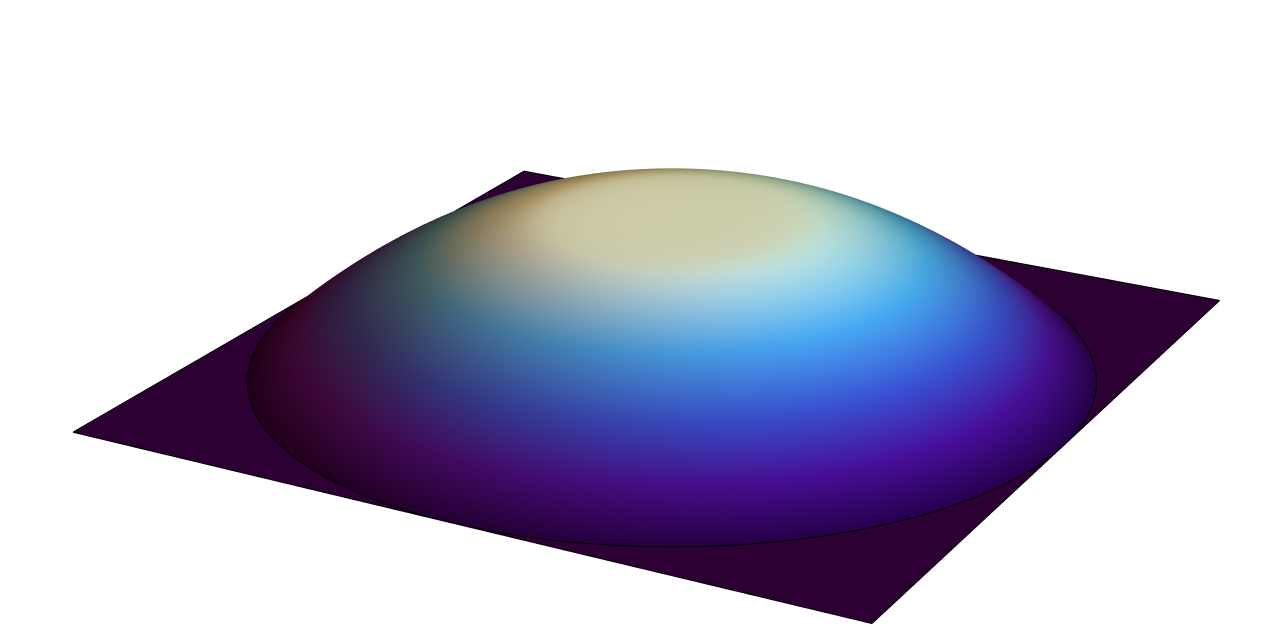}
        \caption{}
    \end{subfigure}
    \caption{Evolution of the intrinsic solution. (a,b) Evolution of $(\pd_r z)^2$ and $\sigma_{\theta\theta}$. The dots are the numerical solution at different times, and the thin lines are the analytic expression, Eq. \eqref{eq:front-def-sup}. (c) A snapshot of the intrinsic model evolution. Notice the abrupt change of slope at the front, compared to Fig. \ref{fig:expSetup}b.}
    \label{fig:appNumPic1}
\end{figure}

\begin{figure}
    \centering
    \begin{subfigure}{0.3\hsize}
        \includegraphics[width=\hsize]{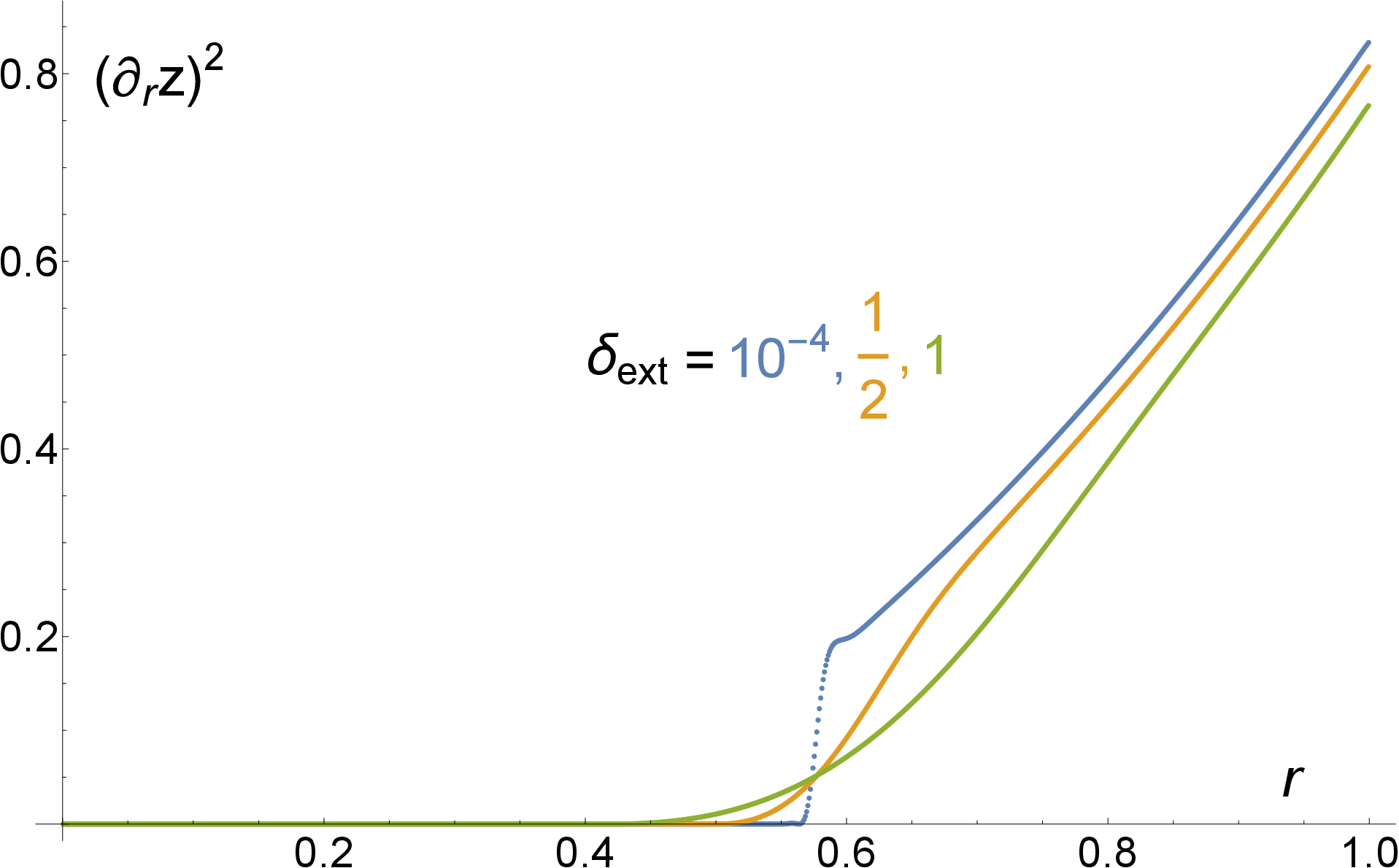}
        \caption{}
    \end{subfigure}\hfill
    \begin{subfigure}{0.3\hsize}
        \includegraphics[width=\hsize]{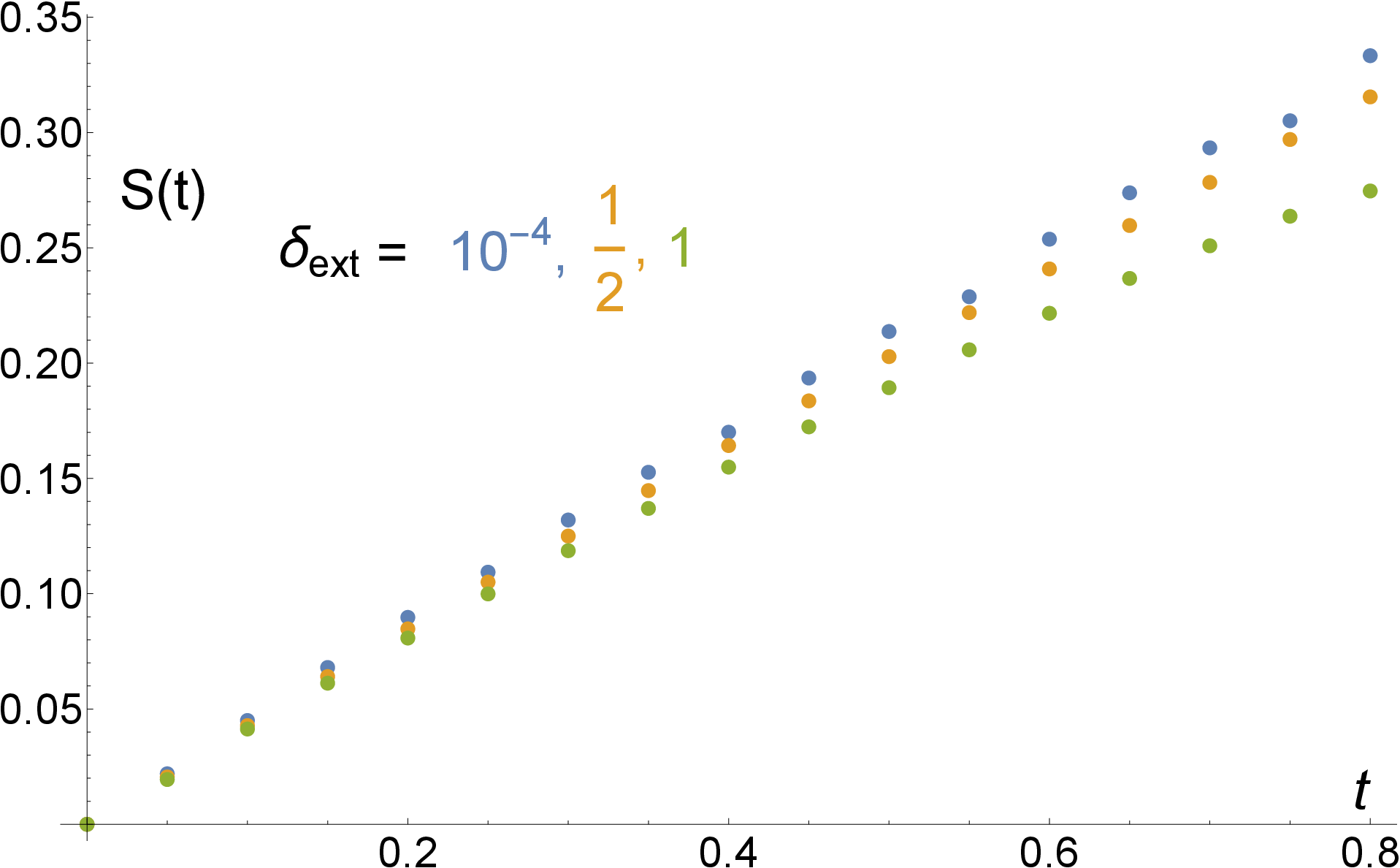}
        \caption{}
    \end{subfigure}\hfill
    \begin{subfigure}{0.3\hsize}
        \includegraphics[width=\hsize]{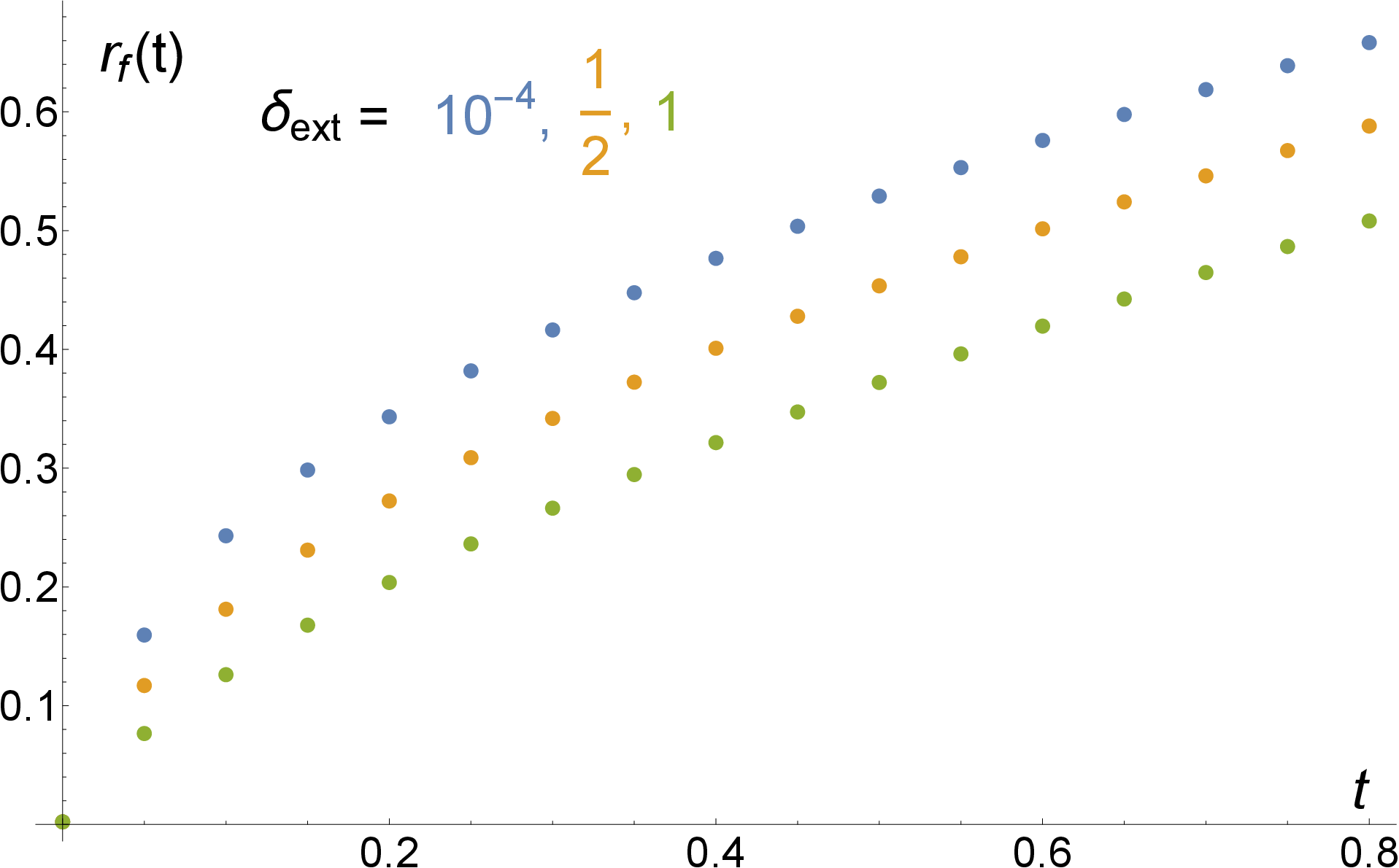}
        \caption{}
    \end{subfigure}
    \caption{Solution of the auxiliary model for various $\delta_{ext}$. The blue traces depict an extremely small $\delta_{ext}$, approximately conforming to the intrinsic model predictions as show in the main text. The yellow and green traces depict finite $\delta_{ext}$, with the green begin the full model $\delta_{ext} = 1$. (a) A snapshot of $(\pd_r z_\ann)^2$ for different $\delta_{ext}$, showing the formation of a boundary layer around the jump in $\pd_r z_\ann$ in the intrinsic model. The centerline of the boundary layer, which is the front in the intrinsic model, $r_f^0$, remains approximately unchanged, as seen by the crossing of the lines. (b) $\Sg(t)$ for different $\delta_{ext}$. The independence on $\delta_{ext}$ at short times confirms the validity of Eq. \eqref{eq:Sg-delta-ext-sup}. (c) $r_f(t)$ for different $\delta_{ext}$, showing that increasing $\delta_{ext}$ merely forms an almost constant boundary layer.}
    \label{fig:supDeltaSnap}
\end{figure}

\begin{figure}
    \centering    \includegraphics[width=\hsize]{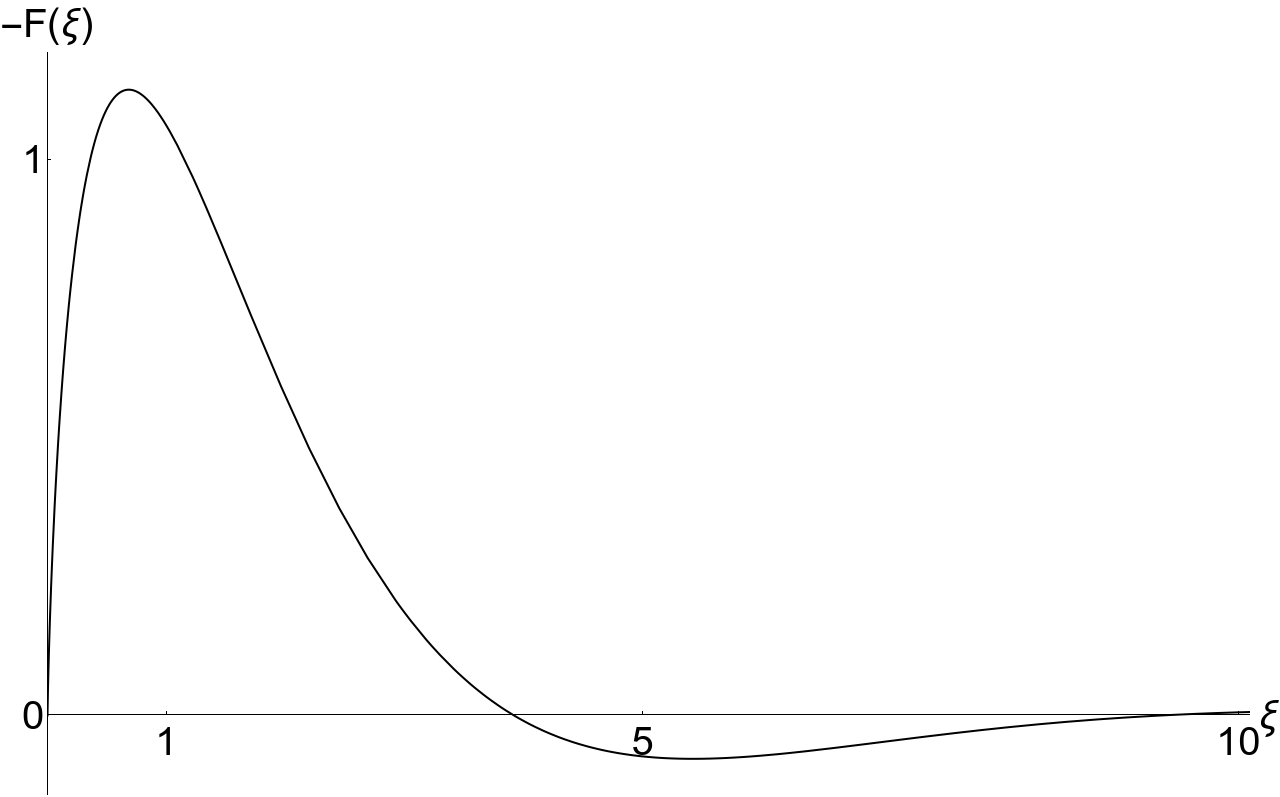}
    \caption{Similarity solution of the ``inverse diffusion'' equations~\eqref{app:inv-diff} and \eqref{app:similarity-2}.}
    \label{fig:hypergeo}
\end{figure}

\begin{figure}
    \centering
    \begin{subfigure}{0.49\hsize}
        \includegraphics[width=\hsize]{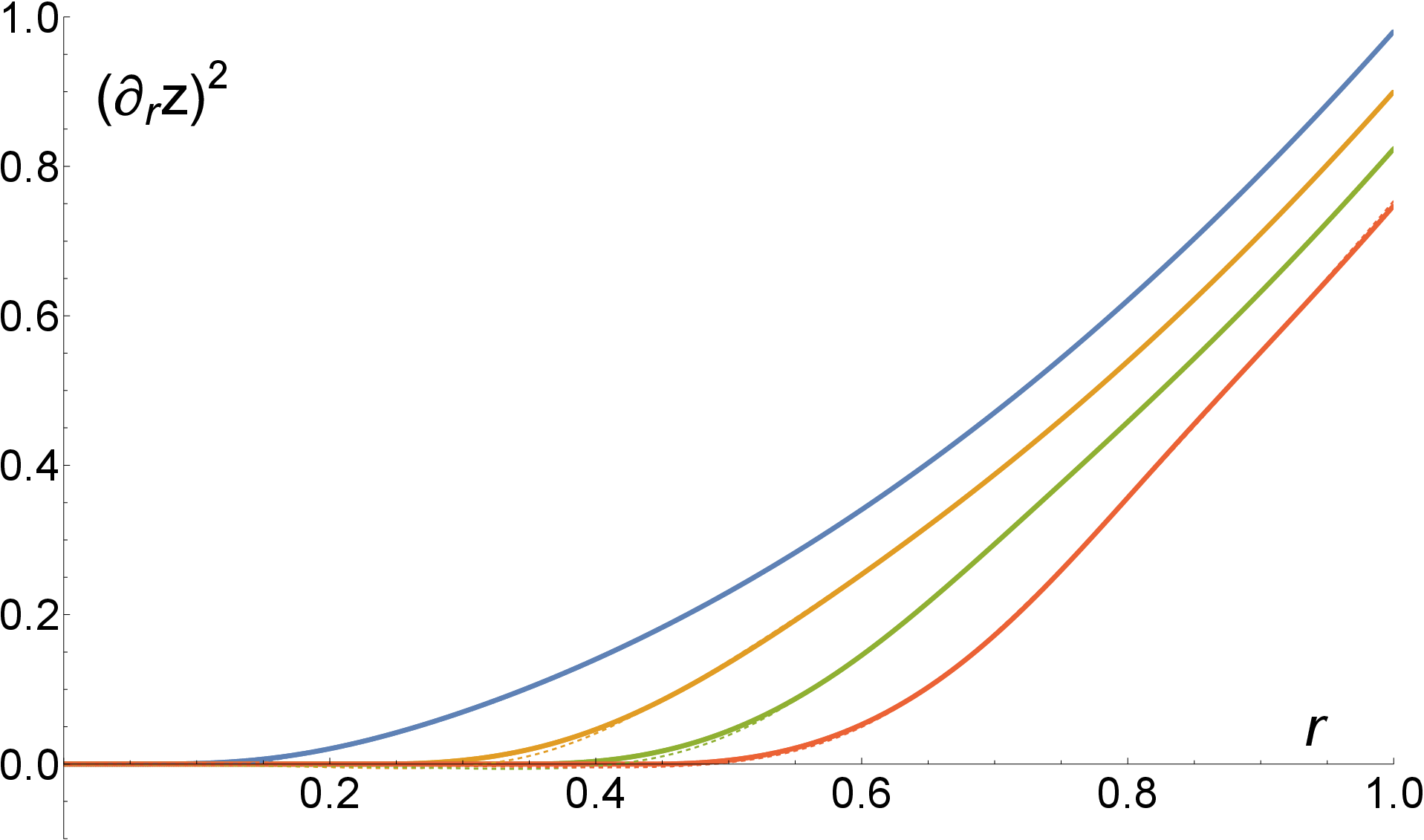}
    \end{subfigure}
    \begin{subfigure}{0.49\hsize}
        \includegraphics[width=\hsize]{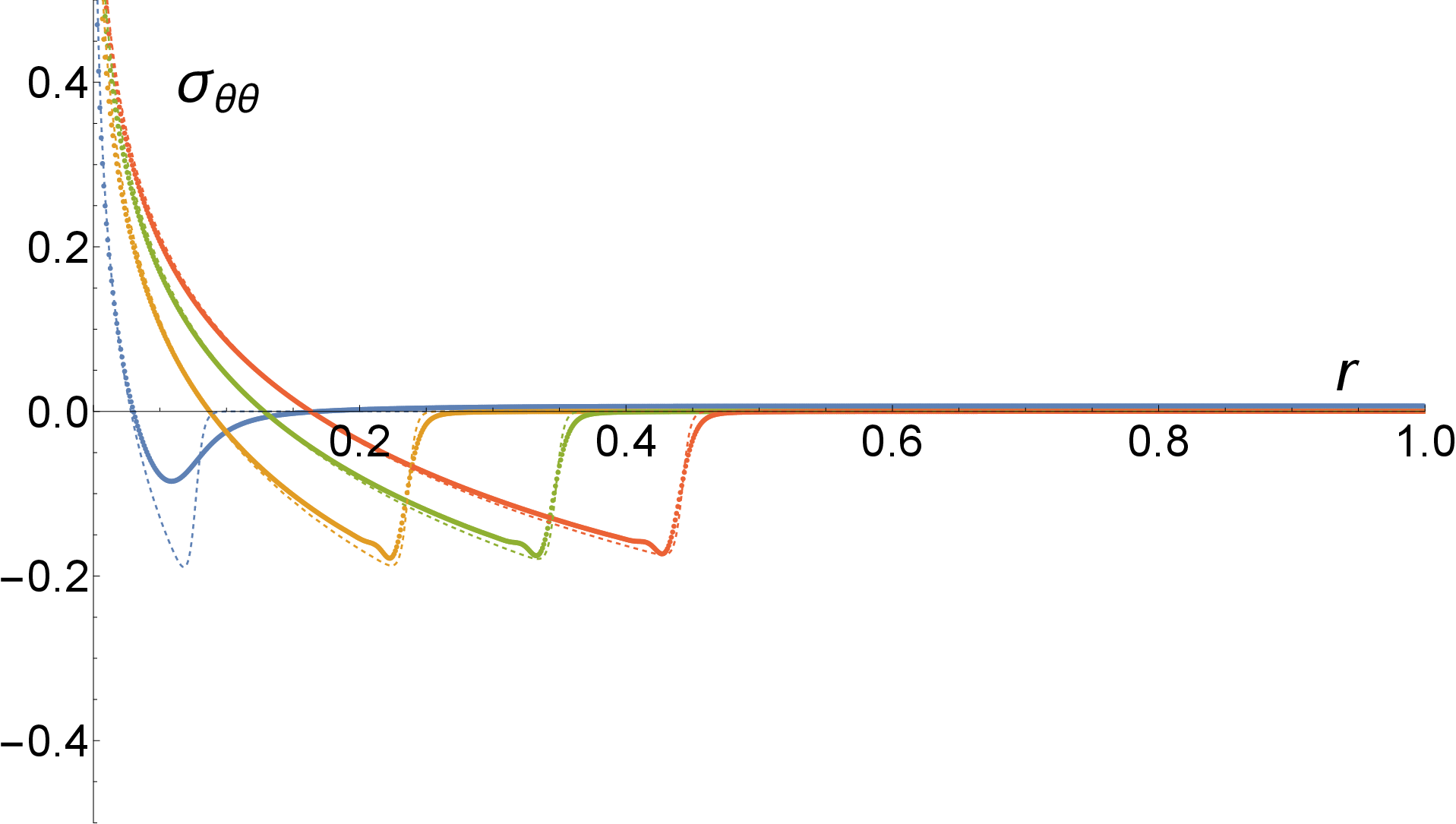}
    \end{subfigure}
    \caption{Depiction of the fitting process for $\delta_{ext} = 1$. The figure depicts $(\pd_r z)^2$ (a) and $\sigma_{\theta\theta}$ (b). The dots are the numerical solution at different times, and the thin solid lines are the analytic expression, Eq. \eqref{eq:front-def-sup}. For panel (a), the analytic expression is broadened by a boundary layer as discussed in the text regarding Fig. \ref{fig:supDeltaSnap}.}
    \label{fig:appNumPic2}
\end{figure}
\begin{table}[]
  \caption{Numerical implementation of boundary conditions for $\Phi_0$ and $z_0$} \label{tab:numericBC}
    \centering
   \begin{tabular}{|c|c|c|}
\hline
Boundary & $\Phi_0$&$z_0$ \\
\hline
    $r_0=0$ & 
    \begin{tabular}{l}
        $D_f (\Phi_0)_0 = n_0(t)$  \\
        $r_{3/2}^{-1}D_f(\Phi_0)_1 = 1$ 
    \end{tabular} 
    &
    \begin{tabular}{l}
        $D_f(z_0)_0 = -\frac{1}{2}h$  \\
        $r_{3/2}^{-1}D_f(z_0)_1 = 1$
    \end{tabular} 
       \\\hline
    $r_{N-1}=1$ & 
    \begin{tabular}{l}
        $D_b^3(\Phi_0)_{N-1} = 0$  \\
        $(\Phi_0)_{N-1} = 0$ 
    \end{tabular} 
    &
    \begin{tabular}{l}
        $D_b^2 (z_0)_{N-1} = -1$  \\
        $(z_0)_{N-1} = 0$
    \end{tabular}\\\hline
\end{tabular}
  
\end{table}

\bibliography{pnas-sample}

\end{document}